\documentclass[11pt]{article}

\usepackage{graphicx}
\usepackage{epsfig}
\usepackage{bbm}
\usepackage{amsmath, amssymb}
\usepackage[sort&compress, square, numbers]{natbib}
\usepackage{subfig}
\usepackage{float}
\usepackage{braket}
\usepackage{resizegather}
\usepackage{multirow}
\usepackage{tikz}
\usepackage{longtable}

\usepackage[margin=2.5cm]{geometry}
\numberwithin{equation}{section}
\usepackage{hyperref}

\begin{document}
\begin{flushright}
CERN-TH-2021-123\\
\today
\end{flushright}
\vspace{4mm}
\begin{center}
{
\Large {\bf Geodesics in the extended K\"ahler cone of Calabi-Yau threefolds}\\[12pt]
\vspace{1cm}
\normalsize
{\bf{Callum R.~Brodie}$^{a,}$\footnote{callum.brodie@ipht.fr}},  
{\bf{Andrei Constantin$^{b,}$}\footnote{andrei.constantin@physics.ox.ac.uk}},
{\bf{Andre Lukas$^{b,}$}\footnote{andre.lukas@physics.ox.ac.uk}},
{\bf{Fabian Ruehle$^{c,b,}$}\footnote{fabian.ruehle@cern.ch}}
\bigskip}\\[0pt]
\vspace{0.23cm}
${}^a$ {\it 
Institut de Physique Th\'eorique, Universit\'e Paris Saclay, CEA, CNRS\\
Orme des Merisiers, 91191 Gif-sur-Yvette CEDEX, France
}\\[2ex]
${}^b$ {\it 
Rudolf Peierls Centre for Theoretical Physics, University of Oxford\\
Parks Road, Oxford OX1 3PU, UK
}\\[2ex]
${}^c$ {\it 
CERN, Theoretical Physics Department\\
1 Esplanade des Particules, Geneva 23, CH-1211, Switzerland
}
\end{center}
\vspace{0.5cm}

\begin{abstract}\noindent
We present a detailed study of the effective cones of Calabi-Yau threefolds with $h^{1,1}=2$, including the possible types of walls bounding the K\"ahler cone and a classification of the intersection forms arising in the geometrical phases. For all three normal forms in the classification we explicitly solve the geodesic equation and use this to study the evolution near K\"ahler cone walls and across flop transitions in the context of M-theory compactifications. In the case where the geometric regime ends at a wall beyond which the effective cone continues, the geodesics ``crash'' into the wall, signaling a breakdown of the M-theory supergravity approximation. For illustration, we characterise the structure of the extended K\"ahler and effective cones of all $h^{1,1}=2$ threefolds from the CICY and Kreuzer-Skarke lists, providing a rich set of examples for studying topology change in string theory. These examples show that all three cases of intersection form are realised and suggest that isomorphic flops and infinite flop sequences are common phenomena. 
\end{abstract}

\setcounter{footnote}{0}
\setcounter{tocdepth}{2}
\clearpage
\tableofcontents

\section{Introduction}
Topology change is an intriguing feature of string theory, and possibly of quantum gravity more generally, which was first discovered and studied some time ago~\cite{Aspinwall:1993yb, Witten:1993yc, Aspinwall:1993nu, Strominger:1995cz, Greene:1995hu, Greene:1996cy, Witten:1996qb}. It is an interesting question how recent attempts to extract general features of low-energy theories from quantum gravity, in the context of the swampland programme~\cite{Vafa:2005ui, Ooguri:2006in, Brennan:2017rbf, Palti:2019pca} relate to topology change and whether topology change itself might perhaps give rise to such general features.\\[2mm]
In the present paper, we will study the extended K\"ahler moduli space of Calabi-Yau (CY) threefolds, including the boundary structure and the flop transitions connecting K\"ahler subcones, in relation to geodesic motion. Our main context is M-theory compactifications on CY threefolds to five-dimensional $N=1$ supergravity, although some of our results can also be directly applied to type II compactifications. To keep the discussion explicit we will focus on CY threefolds $X$ with Picard number $h^{1,1}(X)=2$, the simplest case where we can expect a non-trivial K\"ahler cone structure.\\[2mm]
We review the different types of K\"ahler cone walls, namely flop walls, walls along which a divisor collapses to a curve or a point, and effective cone walls on which the volume of the CY goes to zero, and show how they can be identified from basic topological CY data, as well as discuss the properties of the moduli space metric near these walls. Some emphasis is placed on isomorphic flops, that is, flop transitions of a CY manifold to another version of the same topological type, which can give rise to infinite sequences of flop transitions. A classification of intersection forms for $h^{1,1}(X)=2$ CY threefolds is presented which exhibits three different cases.\\[2mm]
To substantiate our discussion we have compiled a detailed dataset which contains the cone and wall structure of the extended and effective cones for all $h^{1,1}(X)=2$ manifolds within the complete intersection CYs (CICYs)~\cite{Candelas:1987kf} and the CY hypersurfaces in toric fourfolds (THCYs)~\cite{Kreuzer:2000xy}. This data shows that the K\"ahler moduli space structure is quite rich, even at the level of Picard number two, and it should provide a useful resource for future studies of topology change in string theory.\\[2mm] 
Our main goal is the study of geodesics and we will show that, thanks to the classification of intersection forms, the geodesic equation can be explicitly solved for all $h^{1,1}(X)=2$ CYs. This allows us to follow geodesic motion near K\"ahler cone walls and across flop transitions. Our results further substantiate a recent discussion of  how geodesic motion across the extended K\"ahler moduli space relates to the distance conjecture~\cite{Brodie:2021ain}. In particular, we argue that the existence of infinite-length geodesics for CYs with infinite flop sequences does not contradict the distance conjecture.\\[2mm]
The plan of the paper is as follows. In the next section, we review the structure of the CY K\"ahler moduli space, its possible walls, isomorphic and non-isomorphic flop transitions as well as infinite flop sequences. In Section~\ref{sec:geometrymodspace}, we introduce the metric and the geodesic equation on K\"ahler moduli space, discuss how it relates to the cone and wall structure and prepare the ground for solving the geodesic equation by carrying out a classification of intersection forms. The explicit CY data, covering the $h^{1,1}(X)=2$ manifolds from the CICY and THCY lists, will be introduced in Section~\ref{sec:CYdata}, with the detailed information provided in Appendices~\ref{appA} and \ref{appB}. In Section~\ref{sec:geodesic} we study the geodesic equation in detail and show that it can be explicitly solved for all $h^{1,1}(X)=2$ CYs. Several explicit examples of these solutions will be presented in Section~\ref{sec:examples}. We conclude in Section~\ref{sec:conclusion}.

\section{K\"ahler moduli space}
\label{sec:KahlerCones}
In this introductory section, we review a number of relevant features of the K\"ahler moduli space of Calabi-Yau (CY) threefolds, to prepare for our later study of geodesics. Some of these have been well-known for some time, but we will also discuss certain features emphasised more recently, including the possibility of isomorphic flop transitions  and infinite chains of flops. We will focus on Picard number two manifolds and, in particular, present a classification of intersection forms for this case.

\subsection{K\"ahler cones}
\begin{figure}
\centering
\includegraphics[width=.4\textwidth]{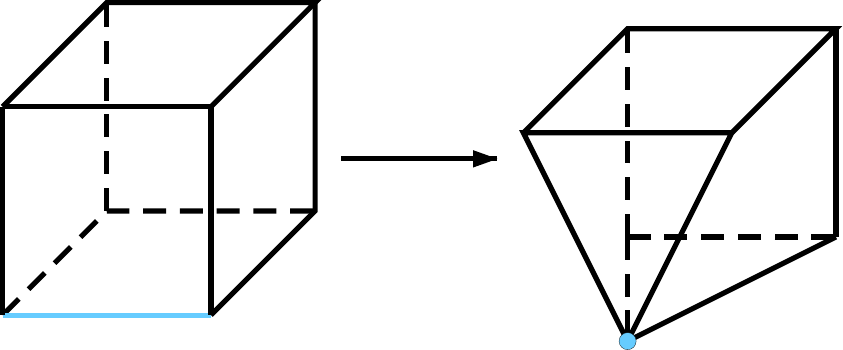}\qquad\qquad
\includegraphics[width=.4\textwidth]{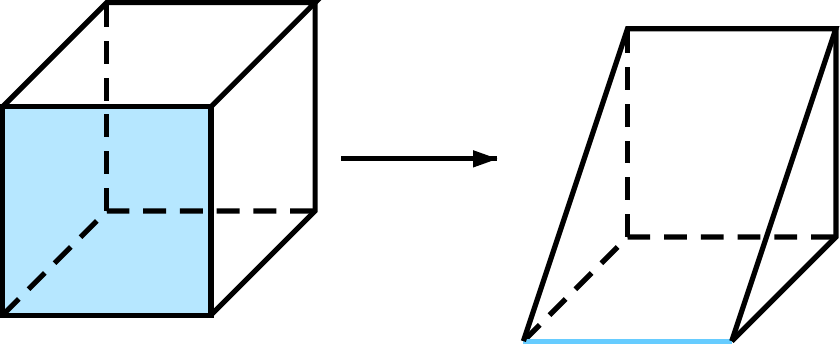}\\
\includegraphics[width=.4\textwidth]{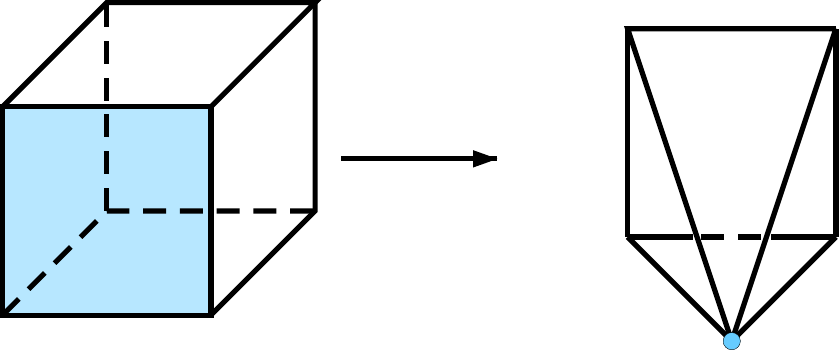}\qquad\qquad
\includegraphics[width=.4\textwidth]{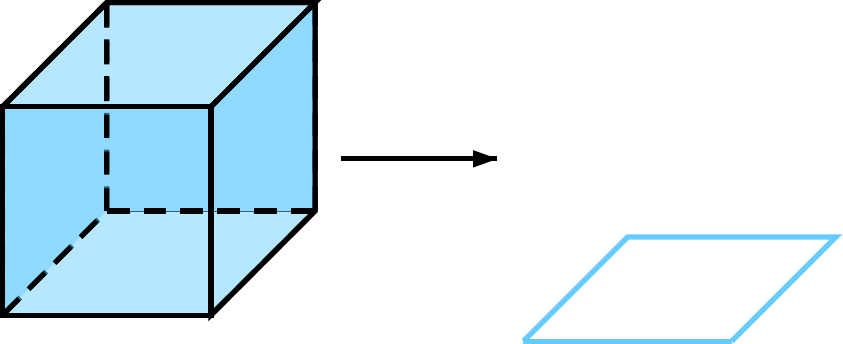}
\caption{(Top left:) Collapse of a curve. (Top right:) Collapse of a divisor to a curve. (Bottom left:) Collapse of a divisor to a point. (Bottom right:) Collapse of the CY.}
\label{fig:WallTypes}
\end{figure}

We are interested in CY threefolds $X$ and their K\"ahler cone $\mathcal{K}(X)$. The K\"ahler cone consists of all closed $(1,1)$-forms $J$ on $X$ which satisfy
\begin{equation}\label{eq:volumes}
  {\rm vol}(C):=\int_CJ>0\;,\qquad {\rm vol}(D):=\frac{1}{2}\int_D J^2>0\;,\qquad {\rm vol}(X):=\frac{1}{6}\int_XJ^3>0\; ,
\end{equation}
for all holomorphic curves $C\subset X$ and all effective divisors $D\subset X$. Its dimension equals $h = h^{1,1}(X)$ and it is usually parametrised as $J=t^iJ_i$, where $t=(t^1,\ldots ,t^h)$ are the K\"ahler moduli and $(J_1,\ldots ,J_h)$ forms a suitably chosen basis of the second cohomology of $X$, with a Poincar\'e dual basis $(D_1,\ldots ,D_h)$ of divisor classes. We also choose an integral basis $(C^1,\ldots ,C^h)$ of dual curve classes, so that
\begin{equation}
\int_{C^i}J_k=\delta^i_k\;,\qquad t^i={\rm vol}(C^i)=\int_{C^i}J\; .
\end{equation}
and introduce the triple intersection numbers $d_{ijk}$ and the pre-potential $\kappa$ by
\begin{equation}\label{eq:isecdef}
d_{ijk}=(D_i,D_j,D_k)=\int_XJ_i\wedge J_j\wedge J_k\;,\qquad \kappa=6\,{\rm vol}(X)=d_{ijk}t^it^jt^k\; ,
\end{equation}
where $(\cdot,\cdot,\cdot)$ is the triple intersection form on $X$. At the boundary of the K\"ahler cone the volume of a sub-manifold vanishes and there are three qualitatively different scenarios of what can happen at a given boundary point, depending on which type of integral in Eq.~\eqref{eq:volumes} approaches zero (see for example Refs.~\cite{Wilson1992,Witten:1996qb} for more information),
\begin{enumerate}
\item[(1)] \textbf{Flop wall:} The volume of a curve in $X$ goes to zero, while the volumes of divisors and the volume of $X$ remain finite. 
\item[(2)] \textbf{Zariski wall:} The volume of a divisor $D$ in $X$ goes to zero, while the volume of $X$ remains finite. There are two sub-cases.\\
$~~~~$(a) The divisor $D$ collapses to a curve.\\
$~~~~$(b) The divisor $D$ collapses to a point.
\item[(3)]\textbf{Effective cone wall:} The volume of $X$ goes to zero, together with the volumes of some divisors and curves. 
\end{enumerate}
A cartoon of these three possibilities is shown in Figure~\ref{fig:WallTypes}.\\[2mm]
When a flop wall is encountered, the geometric moduli space continues beyond the wall into an adjacent K\"ahler cone of another CY manifold $X'$, birationally equivalent to $X$. The volume of the collapsing curves becomes formally negative from the perspective of the original CY $X$, but they are replaced by a new set of holomorphic curves on $X'$. The manifolds $X$ and $X'$ can be isomorphic or non-isomorphic and we will refer to isomorphic and non-isomorphic flops accordingly. We will see later that isomorphic flops are, in fact, rather common. If the new CY $X'$ allows for a flop other than the one leading back to $X$ the process can be continued. By exhausting all possible flops in this way the extended K\"ahler cone, $\mathcal{K}_{\rm ext}(X)$, of $X$ is produced. It has been conjectured that the extended K\"ahler cone only contains a finite number of non-isomorphic CY manifolds, in a statement known as the Kawamata-Morrison conjecture \cite{1994alg.geom..7007M,Kawamata1997OnTC} (see also Ref.~\cite{Brodie:2021ain} for a recent connection with the infinite distance conjecture). On the other hand, as we will see, the extended K\"ahler cone can contain a countably infinite number of isomorphic CY manifolds.\\[2mm]
When a Zariski wall is encountered, a divisor collapses. While the CY volume stays finite, the theory cannot be continued (in the geometric regime) to a new CY. However, the Zariski wall does not mark the end of the effective cone of $X$, denoted by ${\rm Eff}(X)$. Rather, adjacent to the Zariski wall is a Zariski cone containing effective divisors, or equivalently line bundles with global holomorphic sections.\\[2mm]
Finally, when an effective cone wall is encountered, a divisor collapses in such a way that the CY volume goes to zero. In such cases, the geometric interpretation is lost and the CY cannot be continued beyond this wall in a geometric setup. In contrast to the Zariski wall, there are no line bundles with global holomorphic sections beyond this wall, so in this case the wall is both a boundary of the K\"ahler cone and a boundary of the effective cone of $X$.\\[2mm]
\subsection{Picard number two manifolds}
In this paper, we are primarily interested in CY manifolds with Picard number $h^{1,1}(X)=2$ and in this case we can be slightly more explicit about what happens at the K\"ahler cone boundaries. First, the K\"ahler cone is necessarily simplicial and we can choose an integral  basis $(D_1,D_2)$ of K\"ahler cone generators so that the K\"ahler cone $\mathcal{K}(X)=\{xD_1+yD_2\,|\,x,y>0\}$ is the positive quadrant in the coordinates $t=(t^1,t^2)=(x,y)$. The effective cone ${\rm Eff}(X)$ is not fixed but has generators which we can parametrise as
\begin{equation}
 \mathcal{D}_1=v_{11}D_1+v_{12}D_2\;,\quad \mathcal{D}_2=v_{21}D_1+v_{22}D_2\;,
\end{equation}
so that ${\rm Eff}(X) = \{l^1\mathcal{D}_1+l^2\mathcal{D}_2 \,|\, l^1,l^2 > 0 \}$, and in the basis $(D_1,D_2)$ we can write these generators as vectors
\begin{equation}
 v_1=\left(\begin{array}{l}v_{11}\\v_{12}\end{array}\right)\;,\qquad
 v_2=\left(\begin{array}{l}v_{21}\\v_{22}\end{array}\right)\;.
\end{equation}
Finally, the prepotential~\eqref{eq:isecdef} 
\begin{equation}
 \kappa=d_{111}x^3+3d_{112}x^2y+3d_{122}xy^2+d_{222}y^3\
\end{equation}
depends on four intersection numbers, and for convenience we group these into the two vectors
\begin{equation}
 d_1=\left(\begin{array}{l}d_{122}\\d_{222}\end{array}\right)\;,\qquad 
 d_2=\left(\begin{array}{l}d_{111}\\d_{112}\end{array}\right)\;.
\end{equation}
The main CY data which will enters the subsequent discussion consists of these vectors $(d_1,d_2,v_1,v_2)$, defined relative to the basis in which the K\"ahler cone is the positive quadrant. For the two most important classes of CY examples (complete intersections in products of projective spaces and hypersurfaces in toric varieties), they will be explicitly determined in Section~\ref{sec:CYdata}, but for now we proceed by analysing which conclusions can be drawn in general.\\[2mm]
We would like to understand to what extent the data $(d_1,d_2,v_1,v_2)$ encodes the behaviour of the CY $X$ at its K\"ahler cone boundaries. Of course, the cycle volumes
\begin{equation}
 {\rm vol}(C^1)=x\;, \qquad {\rm vol}(C^2)=y\; ,
\end{equation}
vanish at the boundaries $x=0$ and $y=0$, respectively. To decide whether these correspond to flop boundaries we need to look at the volumes of effective divisors and the entire CY. For concreteness we will analyse this for the boundary at $x=0$ on the understanding that the analogous statements for the boundary $y=0$ are obtained by the index exchange $1\leftrightarrow 2$.\\[2mm]
First, for the volumes of $D_1$ and $D_2$ we have
\begin{equation*}
\begin{array}{rclcrcl}
 {\rm vol}(D_1)&=&\frac{1}{2}(d_{111}x^2+2d_{112}xy+d_{122}y^2)&&
 {\rm vol}(D_2)&=&\frac{1}{2}(d_{211}x^2+2d_{212}xy+d_{222}y^2)\\
 &\stackrel{x=0}{\longrightarrow}&\frac{1}{2}d_{122}y^2\;,&&&\stackrel{x=0}{\longrightarrow}&\frac{1}{2}d_{222}y^2\;.
\end{array} 
\end{equation*} 
so that the volume of an arbitrary divisor $D=k_1D_1+k_2D_2$ is given by
\begin{equation}\label{eq:volD}
 {\rm vol}(D)=\frac{1}{2}\left[(k\cdot d_2) x^2+2(k\cdot d)xy+(k\cdot d_1) y^2\right]
 \, \stackrel{x=0}{\longrightarrow} \,\frac{1}{2}(k\cdot d_1)y^2\; ,
 \end{equation}
where $k=(k_1,k_2)^T$ and $d=(d_{112},d_{122})^T$. For the total volume we have
\begin{equation}\label{eq:kappa}
 \kappa=6\,{\rm vol}(X)=d_{111}x^3+3d_{112}x^2y+3d_{122}xy^2+d_{222}y^3 \, \stackrel{x=0}{\longrightarrow} \, d_{222}y^3\; .
\end{equation}
From these expressions we can read off simple criteria, as summarised in Table~\ref{fig:walltypes}, which allow us to determine  the type of boundary wall in terms of the intersection numbers and the effective cone generators. It is worth noting that, while the intersection numbers are topological and, in particular, complex-structure independent, this is not necessarily the case for the effective cone ${\rm Eff}(X)$. New effective divisors can appear for specific complex structure choices and this can lead to a flop wall turning into a Zariski wall. We will later see an example of this phenomenon.
\begin{table}[t]
\begin{center}
\begin{tabular}{|l|c|c|}\hline
 boundary at $x=0$&condition&event\\\hline\hline
 \rule{0pt}{2.5ex}flop wall&$d_1\cdot v_1>0$ and $ d_1\cdot v_2>0$&${\rm vol}(C^1)\rightarrow 0$ \\ \hline
 \rule{0pt}{2.5ex}Zariski wall&$d_{222}\neq 0$ \text{ \& }$\exists k\in{\rm Eff}(X), k\neq \mathbf{0}: k\cdot d_1=0$&${\rm vol}(C^1),{\rm vol}(D)\rightarrow 0$, \\
\hskip 5mm type (a)&$k\cdot d\neq 0$&$D$ collapses to curve\\
\hskip 5mm type (b)&$k\cdot d=0$&$D$ collapses to point\\\hline 
\rule{0pt}{2.5ex}effective cone wall&$d_{222}=0$&${\rm vol}(C^1),{\rm vol}(D_2),{\rm vol}(X)\rightarrow 0$\\\hline
\end{tabular}
\end{center} 
\caption{Conditions for possible events at the $x=0$ boundary of the K\"ahler cone. The analogous statements at the $y=0$ boundary are obtained by the index exchange $1\leftrightarrow 2$. Here $d_1=(d_{122},d_{222})^T$, $d=(d_{112},d_{122})^T$ capture intersection number data, $v_1, v_2$ describe the generators of the effective cone, $k=(k_1,k_2)^T$, and $D=k_1D_1+k_2D_2$.}\label{fig:walltypes}
\end{table}

\subsection{Flops for Picard number two}\label{sec:flops2}
Now suppose the wall at $x=0$ is a flop transition to another, birationally equivalent CY manifold $X^1_1$ whose K\"ahler cone $\mathcal{K}(X^1_1)$ shares the boundary $x=0$ with $\mathcal{K}(X)$. To be more precise, $\mathcal{K}(X^1_1)$ is a cone in $H^2(X^1_1)$, while $\mathcal{K}(X)$ is a cone in $H^2(X)$. However, a flop induces an isomorphism $\rho: H^2(X)\rightarrow H^2(X^1_1)$ such that $\mathcal{K}(X)$ and $\rho^{-1}(\mathcal{K}(X^1_1))$ share a common wall. To simplify notation, from now on we will say that $\mathcal{K}(X)$ and $\mathcal{K}(X^1_1)$ share a common wall.

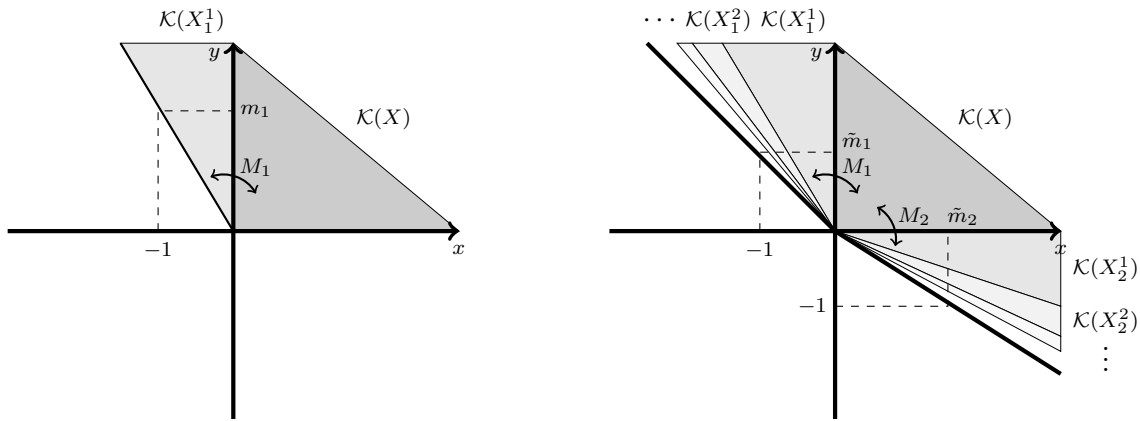
\begin{figure}[t]
\begin{center}
\begin{tikzpicture}
\draw[fill=gray!40,ultra thin] (-8,0)--(-5,0)--(-8,2.5)--(-8,0);
\draw[fill=gray!20,ultra thin] (-8,0)--(-9.5,2.5)--(-8,2.5)--(-8,0);
\draw[ultra thick,->] (-11,0)--(-5,0);
\draw[ultra thick,->] (-8,-2.5)--(-8,2.5);
\draw[thick] (-8,0)--(-9.5,2.5);
\draw[thin,dashed] (-9,0)--(-9,1.6)--(-8,1.6);
\draw [thick,<->] (-7.7,0.5) arc [radius=0.5, start angle=30, end angle= 110];
\node at (-5,-0.25) {\scriptsize $x$};
\node at (-8.25,2.35) {\scriptsize $y$};
\node at (-9,-0.25) {\scriptsize $-1$};
\node at (-7.7,1.6) {\scriptsize $m_1$};
\node at (-8.55,2.8) {\scriptsize $\mathcal{K}(X_1^1)$};
\node at (-6,1.5) {\scriptsize $\mathcal{K}(X)$};
\node at (-7.7,0.85) {\scriptsize $M_1$};
\draw[fill=gray!40,ultra thin] (0,0)--(3,0)--(0,2.5)--(0,0);
\draw[fill=gray!20,ultra thin] (0,0)--(-1.5,2.5)--(0,2.5)--(0,0);
\draw[fill=gray!10,ultra thin] (0,0)--(-1.9,2.5)--(-1.5,2.5)--(0,0);
\draw[fill=gray!5,ultra thin] (0,0)--(-2.1,2.5)--(-1.9,2.5)--(0,0);
\draw[ultra thick] (0,0)--(-2.5,2.5);
\draw[fill=gray!20,ultra thin] (0,0)--(3,-1)--(3,0)--(0,0);
\draw[fill=gray!10,ultra thin] (0,0)--(3,-1.4)--(3,-1)--(0,0);
\draw[fill=gray!5,ultra thin] (0,0)--(3,-1.6)--(3,-1.4)--(0,0);
\draw[ultra thick] (0,0)--(3,-1.9);
\draw[ultra thick,->] (-3,0)--(3,0);
\draw[ultra thick,->] (0,-2.5)--(0,2.5);
\draw [thick,<->] (0.3,0.5) arc [radius=0.5, start angle=30, end angle= 110];
\draw [thick,<->] (0.8,-0.2) arc [radius=0.5, start angle=-10, end angle= 60];
\draw[thin,dashed] (-1,0)--(-1,1.05)--(0,1.05);
\draw[thin,dashed] (0,-1)--(1.5,-1)--(1.5,0);
\node at (3,-0.25) {\scriptsize $x$};
\node at (-0.25,2.35) {\scriptsize $y$};
\node at (2,1.5) {\scriptsize $\mathcal{K}(X)$};
\node at (-0.55,2.8) {\scriptsize $\mathcal{K}(X_1^1)$};
\node at (-1.55,2.8) {\scriptsize $\mathcal{K}(X_1^2)$};
\node at (-2.3,2.8) {$\cdots$};
\node at (3.6,-0.5) {\scriptsize $\mathcal{K}(X_2^1)$};
\node at (3.6,-1.2) {\scriptsize $\mathcal{K}(X_2^2)$};
\node at (3.6,-1.6) {$\vdots$};
\node at (0.3,0.85) {\scriptsize $M_1$};
\node at (1.05,0.2) {\scriptsize $M_2$};
\node at (-1,-0.25) {\scriptsize $-1$};
\node at (-0.3,-1) {\scriptsize $-1$};
\node at (0.3,1.2) {\scriptsize $\tilde{m}_1$};
\node at (1.7,0.2) {\scriptsize $\tilde{m}_2$};
\end{tikzpicture}
\end{center}
\caption{An isomorphic flop between K\"ahler cones $\mathcal{K}(X)$ and $\mathcal{K}(X_1^1)$ with involution generated by $M_1$ (left) and an infinite flop chain, generated by $M_1$ and $M_2$ (right).}\label{fig:kahlercones}
\end{figure}

At the $x=0$ boundary $N$ curves $\mathcal{C}_\alpha$ collapse, of which $n_1$ have class $C^1$ and $n_2=N-n_1$ have class $2C^1$. The intersection forms $(\cdot,\cdot,\cdot)$ and $(\cdot,\cdot,\cdot)'$ of $X$ and $X^1_1$ are then related by
\begin{equation}
 (D,E,F)'=(D,E,F)-\sum_{\alpha=1}^N(D,{\cal C}^\alpha)(E,{\cal C}^\alpha)(F,{\cal C}^\alpha)\; ,
\end{equation}
where $D,E,F$ are divisors. This implies the triple intersection numbers $d_{ijk}=(D_i,D_j,D_k)$ and $d'_{ijk}=(D_i,D_j,D_k)'$ satisfy
\begin{equation}\label{eq:isecrel}
 d_{ijk}'=d_{ijk}-n\,\delta_{1i}\delta_{1j}\delta_{1k}\; ,
\end{equation}
where $n=n_1+8n_2$. These two sets of triple intersection numbers can be basis-transformed into each other by
\begin{equation}\label{eq:M1}
 d'_{ijk}=d_{abc}{M^a_1}_i{M^b_1}_j{M^c_1}_k\;,\qquad M_1=\left(\begin{array}{cc}-1&0\\m_1&1\end{array}\right)\; ,
\end{equation}
provided that
\begin{equation}\label{eq:m1n}
 m_1=\frac{2d_{122}}{d_{222}}\; ,\qquad n=2d_{111}-3m_1d_{112}+m_1^2d_{122}\; .
\end{equation} 
If $m_1$ is fractional then $X$ and $X^1_1$ are different CY manifolds and the above equation for $n$ is typically not satisfied since $n$ needs to be integer. Hence, the triple intersection forms are not related by the matrix $M_1$ and we have a non-isomorphic flop. On the other hand, for $m_1$ integer the transformation $M_1$ is integral and $X$ and $X_1^1$ may be isomorphic. In this case, the second Eq.~\eqref{eq:m1n} is a formula for a specific combination, $n$, of Gromov-Witten invariants in terms of the intersection numbers. In either case, it can happen that $X^1_1$ has another flop boundary, different from the one it shares with $X$, which connects it to a further manifold $X_1^2$ and so forth. It is conjectured that only a finite number of non-isomorphic manifolds can arise in this way. However, as we will see below, there exist infinite chains of isomorphic flops.\\[2mm]
If $X$ and $X_1^1$ are non-isomorphic we are not aware of a general method to read off the K\"ahler cone of the flopped space $X^1_1$ from simple data on $X$. It seems $\mathcal{K}(X^1_1)$ has to be determined from an explicit construction of $X^1_1$ or from line bundle cohomology data on $X$. We will rely on the latter method for our example manifolds. The situation is much simpler if $X$ and $X^1_1$ are isomorphic in which case the K\"ahler cone generators for $X^1_1$ are $D_1'=-D_1+m_1D_2$ and $D_2'=D_2$ and the K\"ahler cones are given by
\begin{equation}
 \mathcal{K}(X)=\{x D_1+yD_2\,|\, x,y>0\}\;,\qquad  \mathcal{K}(X^1_1)=\{x D_1+yD_2\,|\, x<0,\; m_1x+y>0\}\; .
\end{equation}
Then, the matrix $M_1$ in Eq.~\eqref{eq:M1} generates an involution which exchanges $\mathcal{K}(X)$ with $\mathcal{K}(X^1_1)$. The involution should be regarded as a map between CY threefolds of the same type, but with different K\"ahler classes and different complex structures \cite{Katz:1996ht}.  Under this involution, a divisor (curve) class $D=\beta^iD_i$ ($C=\gamma_iC^i$) on $X$ and its equivalent $D'={\beta^i}'D_i$ ($C'={\gamma_i}'C^i$) on $X'_1$ are related by
\begin{equation}
 {\beta^i}'={M_1^i}_j\beta^j\;,\qquad \gamma_i'={M_{1i}}^j\gamma_j\; .
\end{equation}
All this is illustrated on the left in Fig.~\ref{fig:kahlercones}. It has been argued in Ref.~\cite{Brodie:2021ain} that the involution generated by $M_1$ is gauged and that the Kahler cones $\mathcal{K}(X)$ and $\mathcal{K}(X_1^1)$ should, hence, be identified.
At first sight, it might seem that flops between isomorphic CYs are a somewhat exotic and rare phenomenon. On the other hand, the ratio of intersection numbers in Eq.~\eqref{eq:m1n} being integer does not appear to be a particularly strong constraint. Indeed, as we will see in the next section when we discuss examples, isomorphic flops are quite common.\\[2mm]
\subsection{Infinite flop chains}
The previous discussion about flops of course applies equally to the boundary at $y=0$, subject to the index exchange $1\leftrightarrow 2$. In particular, an isomorphic flop at $y=0$ leads to an involution generated by 
\begin{equation}\label{eq:M2}
 M_2=\left(\begin{array}{cc}1&m_2\\0&-1\end{array}\right)\;,\qquad m_2=\frac{2d_{211}}{d_{111}}\; ,
\end{equation}
which exchanges the K\"ahler cone $\mathcal{K}(X)$ with
\begin{equation}
 \mathcal{K}(X_2^1)=\{xD_1+yD_2\,|\, x+m_2y>0,\; y<0\}\; .
\end{equation} 
The matrices $M_1$ and $M_2$ in Eqs.~\eqref{eq:M1} and \eqref{eq:M2} do not commute and they generate a certain discrete group $G$. To determine this group we consider the product of the two involutions,
\begin{equation}
 M=M_1M_2=\left(\begin{array}{cc}-1&-m_2\\m_1&-1+m_1m_2\end{array}\right)\; ,
\end{equation} 
which is of finite order, $s$, if $m_1m_2<4$ and generates a group isomorphic to $\mathbb{Z}$ for $m_1m_2\geq 4$. The elements of $G$ can be written in the unique form $M_1^qM^k$, where $q\in\{0,1\}$ and either $k\in\{0,\ldots ,s-1\}$ when $m_1m_2<4$ or $k\in\mathbb{Z}$ if $m_1m_2\geq 4$. In the former case, $G$ is a finite group whose structure is indicated in the table below.
\begin{center}
\begin{tabular}{|c||c|c|c|}\hline
$(m_1,m_2)$&$(1,1)$&$(1,2)$&$(1,3)$\\\hline
$G\cong$&$\mathbbm{Z}_2\ltimes\mathbbm{Z}_3$&$\mathbbm{Z}_2\ltimes\mathbbm{Z}_4$&$\mathbbm{Z}_2\ltimes\mathbbm{Z}_6$\\\hline
\end{tabular}
\end{center}
For $m_1m_2\geq 4$, on the other hand, $G$ is an infinite group isomorphic to $\mathbbm{Z}_2\ltimes\mathbb{Z}$.\\[2mm]
The extended K\"ahler cone $\mathcal{K}_{\rm ext}(X)$ is generated by acting with $G$ on $\mathcal{K}(X)$. For the finite cases with $m_1m_2<4$ the image of $\mathcal{K}(X)$ under $G$ covers the $x$-$y$ plane minus the negative quadrant and is, therefore, not actually a cone. The conclusion is that $h^{1,1}(X)=2$ CY manifolds with intersections numbers that would lead to $m_1m_2<4$ do not exist.\\[2mm]
For $m_1m_2\geq 4$ the extended K\"ahler cone contains an infinite number of cones, $\mathcal{K}(X_1^k)$ and $\mathcal{K}(X_2^k)$, where $k=1,2,\ldots $, on either side of the original cone $\mathcal{K}(X)$, as indicated on the right in Fig.~\ref{fig:kahlercones}. These infinite sequences of isomorphic flops converge to limits which mark the boundary of the extended cone
\begin{equation}
 \mathcal{K}_{\rm ext}(X)=\{xD_1+yD_2\,|\, \tilde{m}_1x+y>0,\; x +\tilde{m}_2y>0\}
\end{equation}
where
\begin{equation}
 \tilde{m}_i=\frac{m_i}{2}\left(1+\sqrt{1-\frac{4}{m_1m_2}}\right)\; .
\end{equation}
This cone is rational for $m_1=m_2=2$ and irrational for all other cases, and in particular for $m_1m_2>4$. Just as for the case of a single isomorphic flop, it has been argued~\cite{Brodie:2021ain} that $G$ is a gauge symmetry and should be divided out, so that all the cones $\mathcal{K}(X_1^k)$ and $\mathcal{K}(X_2^k)$ are identified with $\mathcal{K}(X)$.

\section{Geometry on K\"ahler moduli space}\label{sec:geometrymodspace}
We will now introduce the metric on K\"ahler moduli space and discuss to what extent it contains information about the wall and cone structure of this moduli space. Basic properties of the associated geodesic equation will be discussed and we present a classification of intersection forms for the case $h^{1,1}(X)=2$ which facilitates solving the geodesic equation. Finally, we review M-theory compactifications on threefolds to five-dimensional $N=1$ supergravity, the low-energy context within which we prefer to consider the geodesic equation.

\subsection{Moduli space metric}
The K\"ahler cone $\mathcal{K}(X)$ is equipped with the moduli space metric
\begin{align}\label{eq:Gdef}
G_{ij}=-\frac13\partial_i\partial_j\ln\kappa=-2\left(\frac{\kappa_{ij}}{\kappa}-\frac32\frac{\kappa_i\kappa_j}{\kappa^2}\right)\;,
\end{align}
where
\begin{align}
\kappa_i=\frac13\partial_i\kappa=d_{ijk}t^jt^k\,,\qquad \kappa_{ij}=\frac16 \partial_i\partial_j \kappa = d_{ijk}t^k\; .
\end{align}
This metric is of course positive-definite on $\mathcal{K}(X)$. It can be used to define the contravariant coordinates $t_i=G_{ij}t^j$ which satisfy the useful relations
\begin{equation}
 t_i=\frac{\kappa_i}{\kappa}\; ,\qquad t_it^i=1\; .
\end{equation} 
The associated Levi-Civita connection reads explicitly
\begin{align}\label{eq:LC}
\Gamma_{ijk}=\frac12\partial_iG_{jk}=-\frac{d_{ijk}}{\kappa}+9\frac{\kappa_{(ij}\kappa_{k)}}{\kappa^2}-9\frac{\kappa_i\kappa_j\kappa_k}{\kappa^3}\,,\qquad \Gamma^i_{jk}=G^{il}\Gamma_{ljk}\; .
\end{align}
Using that $G_{ij}$ is a homogeneous function of degree $-2$ in the coordinates $t^i$ and applying Euler's theorem, it follows that $\partial_i G_{jk}t^i=-2G_{jk}$ or, equivalently,
\begin{align}\label{eq:gammares}
\Gamma_{ijk}t^k=-G_{ij}\,,\qquad \Gamma^i_{jk}t^k=-\delta^i_j\; .
\end{align}
Given our focus on geodesics, we would like to understand to what extent the type of the K\"ahler cone wall, as classified in Table~\ref{fig:walltypes}, is encoded in the behaviour of $G_{ij}$ at or near the wall\footnote{See also Ref.~\cite{Mayer2004} for a discussion of the regularity of the metric at boundaries of the K\"ahler cone.}. As before, we focus on the boundary at $x=0$, with the understanding that results for the $y=0$ boundary are obtained by the index exchange $1\leftrightarrow 2$. At this boundary, the metric becomes
\begin{equation}
 G|_{x=0}=\frac{1}{d_{222}^2y^2}\left(\begin{array}{cc}3d_{122}^2-2d_{222}d_{112}&d_{122}d_{222}\\
                                                                                            d_{122}d_{222}&d_{222}^2\end{array}\right)\;,\quad
{\rm det}(G|_{x=0})=\frac{2(d_{122}^2-d_{112}d_{222})}{d_{222}^2y^4}\; .                                                                                            
\end{equation}
Clearly, the metric diverges at $x=0$ if $d_{222}=0$ and this coincides with the condition for an effective cone wall in Table~\ref{fig:walltypes}. Now assuming that $d_{222}\neq 0$, so that the metric does not diverge for $x=0$, can we distinguish the other wall types? The above expression for the determinant shows that the metric becomes singular at $x=0$ iff
\begin{equation}\label{eq:detisec}
 d_{112}d_{222}-d_{122}^2=0\; .
\end{equation}
If this condition is satisfied $G$ develops one zero eigenvalue, so its rank reduces to one. We will now show that such a singularity in the metric cannot arise in the case that $x=0$ is a flop wall. To this end, we assume that $G$ is non-divergent but singular at $x=0$, so $d_{222}\neq 0$ but $d_{122}^2=d_{112}d_{222}$, and assume that a flop to a manifold $X'$ arises at $x=0$. We will show that these assumptions lead to a contradiction by studying the behaviour of $G$ on either side of the boundary $x=0$. For $x\geq 0$, so inside the K\"ahler cone of $X$, the determinant to first order in $x$ is given by
\begin{equation}
 {\rm det}(G|_{x\geq 0})=2\frac{d_{122}^3-d_{111}d_{222}^2}{d_{222}^3y^5}x+{\cal O}(x^2)=
2\frac{d_{112}d_{122}-d_{111}d_{222}}{d_{222}^2y^5}x+{\cal O}(x^2) \; .
\end{equation} 
The numerator $d_{112}d_{122}-d_{111}d_{222}$ cannot be zero as this would imply the vanishing of two minors of the matrix in Eq.~\eqref{eq:phidef} below and, hence, a reduction in the rank of $\varphi$, which as we will see is excluded. Hence since $G$ is positive definite for $x> 0$ we require that $d_{122}^3-d_{111}d_{222}^2>0$. For $x\leq 0$ we can carry out the same calculation as above but we have to use the intersection numbers $d'_{ijk}$ of $X'$ given in Eq.~\eqref{eq:isecrel}. Since $d_{111}$ is, in fact, the only intersection number which changes, we have
\begin{equation}
 {\rm det}(G|_{x\leq 0})=2\frac{d_{122}^3-d'_{111}d_{222}^2}{d_{222}^3y^5}x+{\cal O}(x^2)=
2\frac{d_{122}^3-d_{111}d_{222}^2+nd_{222}^2}{d_{222}^3y^5}x+{\cal O}(x^2)
\end{equation} 
Evidently, the numerator of this expression remains positive (as $n\geq 0$) but, since $x\leq 0$, it follows that ${\rm det}(G|_{x<0})<0$ near the wall which contradicts positive definiteness of the metric. This shows that the metric $G$ is necessarily non-singular at a flop boundary and, conversely, that a non-divergent but singular metric indicates a Zariski wall.\\[2mm]
We can be even more specific and decide which type of Zariski wall a singular metric corresponds to. From Table~\ref{fig:walltypes} the condition for a Zariski wall at $x=0$ is $k\cdot d_1=0$ for some effective divisor $k \in {\rm Eff}(X)$. For a type (b) Zariski wall (divisor collapses to a point) we need in addition that $k\cdot d=0$. Together these imply that the matrix
\begin{equation}
 (d,d_1)=\left(\begin{array}{ll}d_{112}&d_{122}\\d_{122}&d_{222}\end{array}\right)
\end{equation}
is singular, which is the case iff the determinant in Eq.~\eqref{eq:detisec} vanishes. Conversely, if this matrix is singular, so that $d \propto d_1$, there must exist an effective $k$ satisfying $k\cdot d_1=0$ and $k\cdot d =0$, since we know from Table~\ref{fig:walltypes} that at a Zariski wall either $d_1 \cdot v_1 \leq 0$ or $d_1 \cdot v_2 < 0$, where $v_1$ and $v_2$ describe the generators of the effective cone. The conclusion is that a non-divergent but singular metric indicates a type (b) Zariski wall, where a divisor collapses to a point.\\[2mm]
The above statements, relating the behaviour of the metric to the type of the boundary wall, are summarised in Table~\ref{tab:Gwallrel}.
\begin{table}[t]
\begin{center}
\begin{tabular}{|c|c|c|}\hline
metric behaviour&condition&boundary type\\\hline\hline
\rule{0pt}{2.5ex}non-divergent, non-singular&$d_{222}\neq 0$, $d_{112}d_{222}-d_{122}^2\neq 0$&flop or type (a) Zariski wall\\\hline
\rule{0pt}{2.5ex}non-divergent, singular&$d_{222}\neq 0$, $d_{112}d_{222}-d_{122}^2= 0$&type (b) Zariski wall\\\hline
\rule{0pt}{2.5ex}divergent&$d_{222}=0$&effective cone wall\\\hline
\end{tabular}
\end{center}
\caption{Relation between metric behaviour at $x=0$ and the type of K\"ahler cone boundary.}\label{tab:Gwallrel}
\end{table}
The metric behaviour is apparently not sufficient to distinguish between flop and type (a) Zariski walls. Indeed, one and the same K\"ahler cone boundary can switch between the two types, depending on complex structure choice, but without any change of the metric $G_{ij}$. An example for this phenomenon will be presented in Section~\ref{sec:CYdata}.

\subsection{Classification of intersection forms}\label{sec:class}
\begin{figure}[t]
\centering
\includegraphics[width=.31\textwidth]{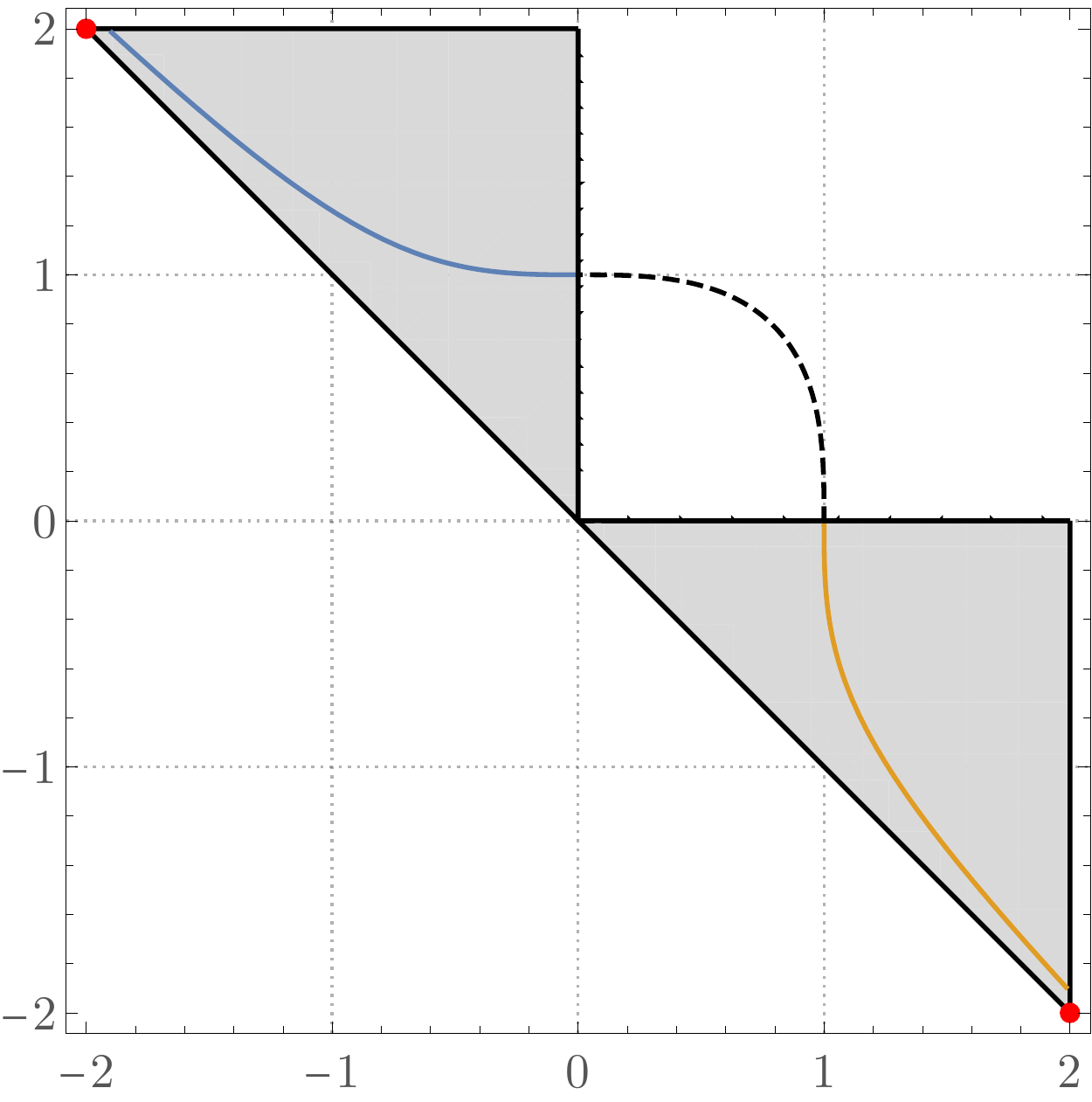}~~
\includegraphics[width=.31\textwidth]{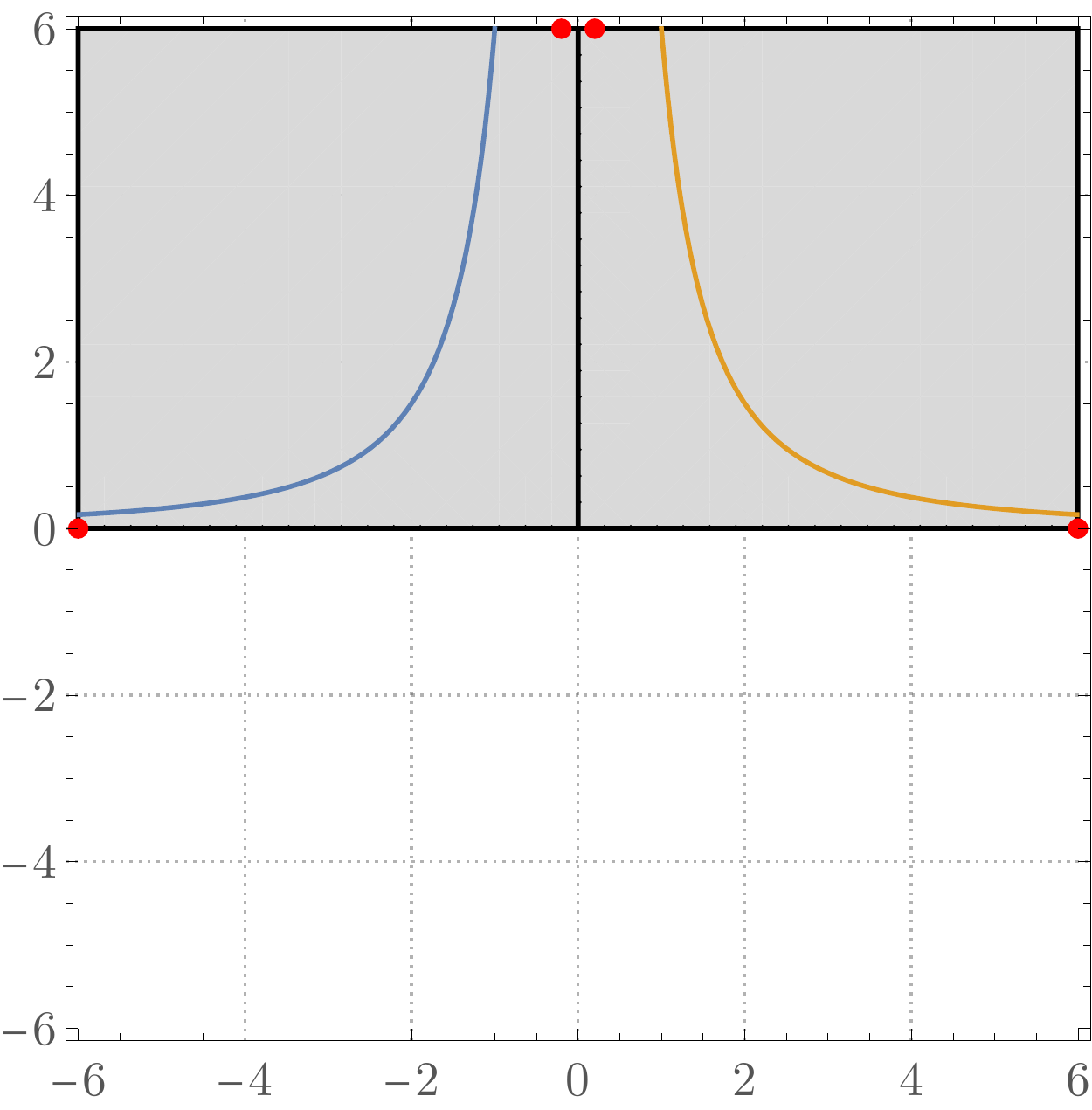}~~
\includegraphics[width=.31\textwidth]{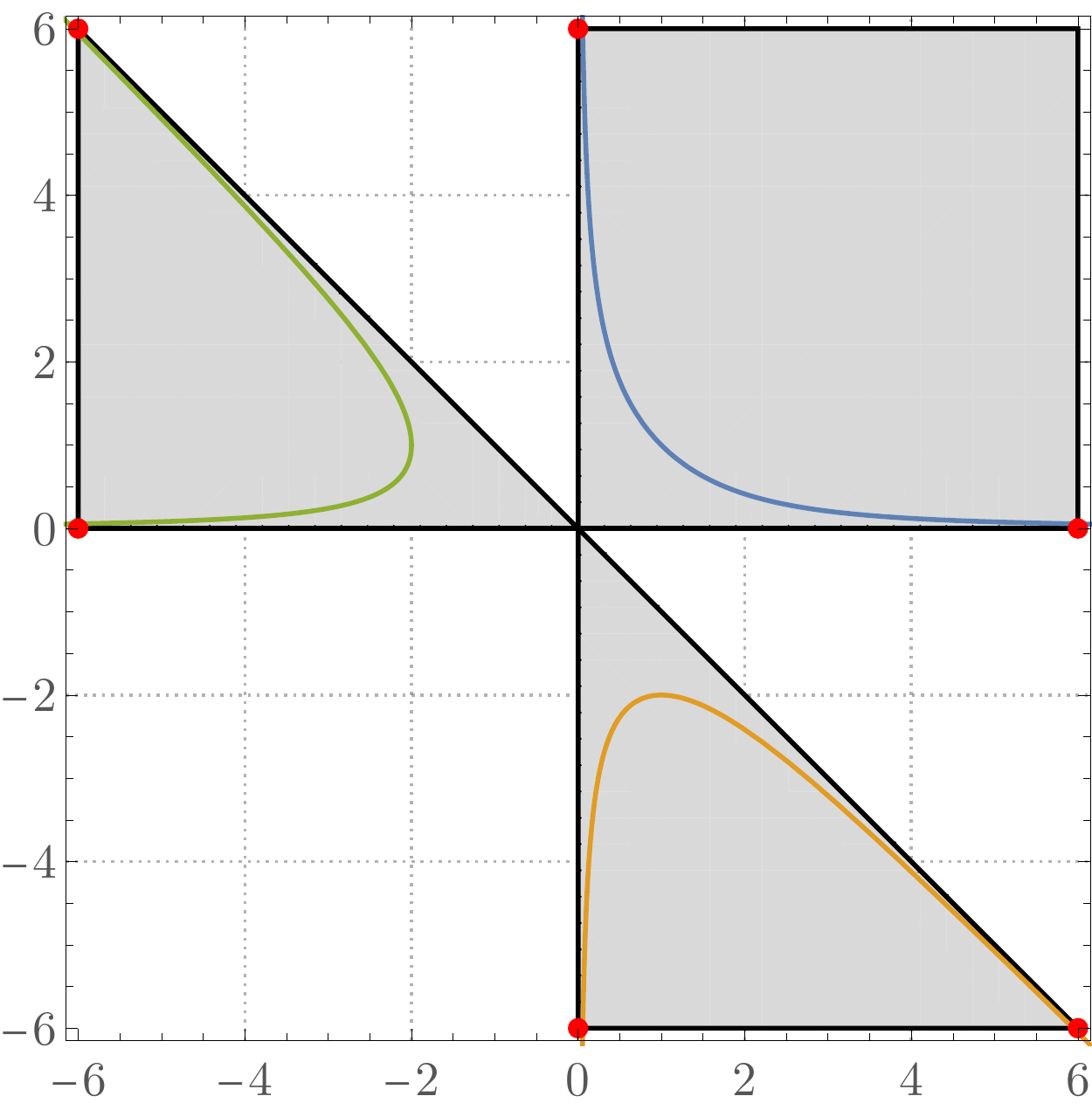}
\caption{Shaded regions indicate where $\hat{\kappa}>0$ and $\hat{G}$ is positive definite, for case 1 (left), case 2 (middle) and case 3 (right) from Table~\ref{tab:EquivalenceClasses}. The coloured lines are isochores (taking $\hat{\kappa}=6$) and the red dots indicate boundaries for which $\hat{\kappa}$ vanishes at the boundary. The dashed line in case 1 indicates the isochore outside the consistent region.}
\label{fig:Geodesics3Cases}
\end{figure}

The intersection form, together with the K\"ahler cone, are the main ingredients in discussing geodesics. For this reason it makes sense to discuss basis transformations of intersection forms and find suitably simple normal forms. To this end, we call two intersection forms 
\begin{align}
\kappa(t) = d_{ijk} t^it^jt^k\qquad\text{and}\qquad \hat{\kappa}(t)=\hat{d}_{ijk}t^it^jt^k
\end{align}
equivalent if there exists a basis transformation $P\in\text{GL}(\mathbbm{R}^h)$ such that $\hat{\kappa}(Pt)=\kappa(t)$ or, equivalently, if the intersection numbers are related by $d_{ijk}=\hat{d}_{lmn}{P^l}_i{P^m}_j{P^n}_k$~\footnote{The notation of integrality is of course typically lost under this equivalence but this will not be essential for the purpose of solving the geodesic equations.}. Evidently, this defines an equivalence relation and we are interested in finding the equivalence classes and a suitably simple normal form for each class. Class functions can be helpful to carry this out explicitly. One such class function is the rank, ${\rm rk}(\varphi)$, of the map $\varphi:S^2\mathbb{R}^h\rightarrow \mathbb{R}^h$ defined by $[\varphi(s)]_i=d_{ijk}s^{jk}$. However, by itself it is not sufficient to distinguish all classes, as we will see below. For arbitrary Picard number $h$, the  classification is a complicated problem but for $h=2$, our main case of interest, it is not too difficult to complete.\\[2mm]
First, we note that, for $h=2$, the general intersection form~\eqref{eq:kappa} is determined by four intersection numbers so our classification is carried out in a space isomorphic to $\mathbb{R}^4$. The intersection form~\eqref{eq:kappa} also shows that the map $\varphi:S^2\mathbb{R}^2\rightarrow\mathbb{R}^2$ is represented by the matrix 
\begin{align}\label{eq:phidef}
\varphi\sim\left(\begin{array}{ccc}d_{111}&d_{112}&d_{122}\\ d_{112}&d_{122}&d_{222}\end{array}\right)\qquad\Rightarrow\qquad\text{rk}(\varphi)\in\{0,1,2\}\; .
\end{align}
As already mentioned, this rank by itself is not sufficient to distinguish all classes. Another class function can be constructed as follows. Seen as a cubic in $\mathbbm{RP}^1$, $\kappa$ has either one, two, or three distinct zeros, and we define the class of $\kappa$ to be this number, that is
\begin{align}\label{eq:classdef}
  \text{cl}(\kappa)=\left|\left\{[x:y]\in\mathbbm{RP}^1~|~\kappa(x,y)=0\right\}\right|~\in~\{1,2,3\}\; .
\end{align}
Taken together, the rank, $\text{rk}(\varphi)$, and the class, $\text{cl}(\kappa)$, are sufficient to characterise the equivalence classes of intersection forms, as can be shown by explicitly carrying out basis transformations. It turns out there are four classes (apart from the trivial class which consists of the zero polynomial) which are summarised in Table~\ref{tab:EquivalenceClasses}, together with suitable normal forms.
\begin{table}[t]
\centering
\begin{tabular}{|c@{~}|c@{~}|c@{~}||c@{~}|c@{~}|}
\hline
case&normal form $\hat{\kappa}$&(cl($\kappa$),rk($\varphi$))& metric $\hat{G}$&$\mathcal{K}(X)$ contained in\\
\hline
\hline
\rule{0pt}{4ex}0&$x^3$&(1,1)&$\frac{1}{x^2}\left(\begin{array}{rr}1&0\\0&0\end{array}\right)$&$\{\}$\\\hline
\rule{0pt}{4ex}1&$x^3+y^3$&(1,2)&
$\frac{1}{\kappa^2}\left(
\begin{array}{cc}
 x^4-2 x y^3 & 3 x^2 y^2 \\
 3 x^2 y^2 & y^4-2 x^3 y \\
\end{array}
\right)$&$\begin{array}{c}\{x<0,x+y>0\}\\\cup\,\{y<0,x+y>0\}\end{array}$\\\hline
\rule{0pt}{4ex}2&$x^2y$&(2,2)&
$\left(
\begin{array}{cc}
 \frac{2}{3x^2} & 0 \\
 0 & \frac{1}{3y^2} \\
\end{array}
\right)$
&$\{x\neq0,y>0\}$\\\hline
\rule{0pt}{4ex}3&$x^2y+xy^2$	&(3,2)&
$\frac{1}{3}\left(
\begin{array}{cc}
\frac{1}{x^2}+\frac{1}{(x+y)^2} & \frac{1}{(x+y)^2} \\
 \frac{1}{(x+y)^2} &\frac{1}{y^2}+\frac{1}{(x+y)^2} \\
\end{array}
\right)$
&$\begin{array}{c}\{x>0,y>0\}\\\cup\,\{x>0,x+y<0\}\\\cup\,\{y>0,x+y<0\}\end{array}$\\\hline
\end{tabular}
\caption{Classification of intersection forms for CYs $X$ with $h^{1,1}(X)=2$. The metric $\hat{G}$ has been computed from the prepotential $\hat{\kappa}$, using Eq.~\eqref{eq:Gdef}. The regions where $\hat{\kappa}>0$ and $\hat{G}$ is positive definite are given in the last column and are plotted in Figure~\ref{fig:Geodesics3Cases}.}
\label{tab:EquivalenceClasses}
\end{table}
The regions in the last column of Table~\ref{tab:EquivalenceClasses} are the ones where $\hat{\kappa}>0$ and $\hat{G}$ is positive definite, so they indicate the maximal K\"ahler cones possible in each case. For case 0 the set is empty since the metric $\hat{G}$ is singular everywhere, so this case is irrelevant for the discussion of CY intersection forms. For the remaining three cases, the maximal K\"ahler cones have been plotted in Fig.~\ref{fig:Geodesics3Cases}.
The boundaries $x=0$ and $y=0$ of the grey region in case 1 are special in that the volume $\hat{\kappa}$ remains finite (away from the origin) but the metric $\hat{G}$ becomes singular. On all other boundaries of the grey regions in Fig.~\ref{fig:Geodesics3Cases} the volume $\hat{\kappa}$ vanishes.\\[2mm]
How does this classification of intersection forms relate to our previous discussion of the K\"ahler moduli space and its boundaries? For starters, throughout the K\"ahler cone $\mathcal{K}(X)$ of a CY threefold $X$ the volume is positive and the metric is non-singular so such a moduli space must map to a cone within one of the maximal cones in Fig.~\ref{fig:Geodesics3Cases}. The price we pay for choosing coordinates where the intersection form becomes one of the normal forms is a more complicated K\"ahler cone in those coordinates, not necessarily equal to the positive quadrant. More precisely, if $P$ transforms the intersection form $\kappa(t)=d_{ijk}t^it^jt^k$ with associated K\"ahler cone $\mathcal{K}(X)=\{xD_1+yD_2\,|\, x,y>0\}$ into one of the normal forms $\hat{\kappa}$ from Table~\ref{tab:EquivalenceClasses}, so that $\hat{\kappa}(Pt)=\kappa(t)$, then the K\"ahler cone $\hat{\mathcal{K}}(X)$ in normal form coordinates is the image
\begin{equation}
 \hat{\mathcal{K}}(X)=P\left(\{(x,y)\,|\,x,y>0\}\right)
\end{equation}
of the positive quadrant under $P$. For cases 2 and 3 we can always choose $P$ such that this cone is contained in the positive quadrant, but it may not take up the entire quadrant. For case 1 we can ensure it lies within the upper cone and, again, it may or may not fill out the entire region.\\[2mm]
There are three qualitatively different possibilities for how a boundary of the mapped K\"ahler cone $\hat{\mathcal{K}}(X)$ can relate to the maximal cones in Fig.~\ref{fig:Geodesics3Cases}. First, if the boundary in question is in the interior of a maximal cone the metric is non-divergent and non-singular, so from Table~\ref{tab:Gwallrel} this must correspond to a flop or type (a) Zariski wall. If a boundary of $\hat{\mathcal{K}}(X)$ coincides with one of the maximal cone boundaries where the volume vanishes (all but the $x=0$ and $y=0$ boundaries of case 1) this must be an effective cone wall. Finally, if a boundary of $\hat{\mathcal{K}}(X)$ coincides with the $x=0$ (or $y=0$) boundary for case 1 then the volume remains finite and the metric is singular so that, from Table~\ref{tab:Gwallrel}, we have a type (b) Zariski wall. This means that type (b) Zariski walls can only ever occur for case 1 intersection forms. The structure of the cones in Fig.~\ref{fig:Geodesics3Cases} also shows that a type~(b) Zariski wall can only arise for at most one of the two boundaries of the K\"ahler cone.

\subsection{M-theory on threefolds}
The physics associated with the CY K\"ahler cone and its boundaries depends somewhat on the string compactification considered. In type II string compactifications on a CY threefold $X$ the K\"ahler moduli $t^i$ reside in $h^{1,1}(X)$ four-dimensional $N=2$ vector multiplets but there are a number of complications: (i) The K\"ahler moduli $t^i$ are complexified by axions. (ii) Instanton effects correct the prepotential $\kappa$ and, hence, the metric $G_{ij}$. These effects become important for small curve volumes. (iii) Non-geometric phases, described by abstract conformal field theories, may arise beyond certain boundaries of the K\"ahler cone.\\[2mm]
In this paper, we focus on M-theory compactifications on CY threefolds instead, where these complications are largely absent. These compactifications lead to five-dimensional $N=1$ supergravity theories with $h^{1,1}(X)-1$ vector multiplets and $h^{2,1}(X)+1$ hypermultiplets. The vector multiplet scalars are the ``shape moduli"
\begin{equation}
 b^i=\frac{t^i}{a}\;,\qquad a^3:={\rm vol}(X)=\frac{\kappa}{6}\;,
\end{equation} 
while the overall volume modulus $a$ resides in the hypermultiplet sector. Evidently, the prepotential $\tilde{\kappa}$ as a function of the shape moduli satisfies a constant volume contraint
\begin{align}
\tilde{\kappa}=d_{ijk}b^ib^jb^k\qquad\Rightarrow\qquad \tilde{\kappa}=6\; ,
\end{align}
which accounts for the reduction from $h^{1,1}(X)$ to $h^{1,1}(X)-1$, as required by the number of vector multiplets. The metric and connection on the five-dimensional vector moduli space are obtained from $\tilde{\kappa}$ by the analogue of Eqs.~\eqref{eq:Gdef} and \eqref{eq:LC} and these quantities will be denoted by $\tilde{G}_{ij}$ and $\tilde{\Gamma}_{ij}^k$. From homogeneity of these functions it follows that
\begin{align}\label{eq:Gtildedef}
\tilde{\kappa}=a^{-3}\kappa\,,\qquad \tilde{G}_{ij}=a^2 G_{ij}\,,\qquad \tilde{\Gamma}^i_{jk}=a\Gamma^{i}_{jk}\,.
\end{align}
The detailed five-dimensional effective action in the Einstein frame can, for example, be found in Ref.~\cite{Lukas:1998tt}. For our purposes it is sufficient to know that geodesics in vector moduli space are governed by the metric $\tilde{G}_{ij}$ and that geodesic distances are measured in units of the five-dimensional Planck length.\\[2mm]
As is clear from the above discussion, the vector multiplet scalars $b^i$ are not complexified, so this type II complication is absent. Further, instanton effects are proportional to $\exp(-R)$, where $R$ is the M-theory radius (or the radius between four- and five-dimensional effective theories), so they vanish in the five-dimensional limit $R\rightarrow\infty$. Hence, $\tilde{\kappa}$ and $\tilde{G}_{ij}$ do not receive instanton corrections. Finally, non-geometric phases are absent in the five-dimensional theory~\cite{Witten:1996qb}. Having said this, our results can be straightforwardly applied to IIA, provided it can be argued that the above complications can be neglected. Of course, the results only apply to the sub-sector of IIA where axion dynamics have been switched off (which can be done consistently). Further, we need to require that cycles either retain large volumes or, if they become small, they do not give rise to instanton corrections and that the evolution stays away from non-geometric phases. Of course these conditions have to be checked case by case.\\[2mm]
What is the physics of those M-theory compactifications to five-dimensions which correspond to the different types of K\"ahler cone boundaries listed in Table~\ref{fig:walltypes}? At a flop wall, a number, $N$, of curves in the CY manifold $X$ shrink to zero size and the five-dimensional effective theory acquires $N$ additional hypermultiplets which become massless at the flop. They originate from membranes wrapping the collapsing cycles. These hypermultiplets can be explicitly incorporated into the five-dimensional supergravity~\cite{Brandle:2002fa}. As one passes through the flop wall, from $X$ to a birationally equivalent CY $X'$, one-loop corrections due to these hypermultiplets change the intersection numbers $d_{ijk}$ of $X$ which appear in the five-dimensional theory to the intersection numbers $d'_{ijk}$ of $X'$ \cite{Witten:1996qb}. With this adjustment of intersection numbers understood we can think of the five-dimensional theory as a theory on the extended K\"ahler moduli space of $X$.\\[2mm]
For a type (a) Zariski wall we have a divisor which shrinks to a curve. This marks the end of the five-dimensional vector multiplet moduli space with an SU$(2)$ gauge theory from membranes wrapping the developing $A_1$ singularity appearing at the boundary~\cite{Katz:1996ht}.\footnote{The five-dimensional effective theory including this SU$(2)$ gauge theory has been constructed in Ref.~\cite{Mohaupt2001} for the special case that there are no massless charged hypermultiplets.} For a type (b) Zariski wall a divisor collapses to a point and this is the end of the moduli space. An infinite tower of states arises from membranes wrapping the curves within the shrinking divisor as well as from a five-brane wrapping the entire divisor (leading to a tensionless string in five dimensions).\\[2mm]
Finally, an effective cone wall combines the collapse of a divisor and the entire CY and also marks the end of the vector moduli space. An infinite tower of light states appears from the collapsing divisor or from Kaluza-Klein modes. 

\section{Calabi-Yau constructions}\label{sec:CYdata}
To add substance to the discussion, we now introduce two classes of CY manifolds with $h^{1,1}(X)=2$ and explicitly determine their K\"ahler cone structure. The first set consists of the $h^{1,1}(X)=2$ manifolds from the list\footnote{The CICY table is available at \href{http://www-thphys.physics.ox.ac.uk/projects/CalabiYau/cicylist/}{http://www-thphys.physics.ox.ac.uk/projects/CalabiYau/cicylist/}} of complete intersection CYs in product of projective spaces (CICYs)~\cite{Candelas:1987kf} and the second is the subset of manifolds with $h^{1,1}(X)=2$ among all CYs defined as hypersurfaces in toric ambient spaces\footnote{The Kreuzer-Skarke list is available at multiple places, e.g.\ at \href{http://hep.itp.tuwien.ac.at/~kreuzer/CY/}{http://hep.itp.tuwien.ac.at/~kreuzer/CY/}. We use the SAGE package polytopes\_db\_4d.}  (THCYs) as captured in the Kreuzer-Skarke list~\cite{Kreuzer:2000xy}. As it happens, both of these data sets contains $36$ topological types of CY manifolds each.\\[2mm]
The manifolds in either data set are constructed in an ambient space $\mathcal{A}$. For the CICYs this is a product of projective spaces with the CY defined as the common zero locus of sections whose associated line bundles form a nef partition of the anti-canonical bundle of $\mathcal{A}$. For THCYs the ambient space is a toric fourfold and the CY is defined as the zero locus of a section with associated first Chern class equal to that of the anti-canonical bundle of $\mathcal{A}$. While the K\"ahler cone of the ambient space $\mathcal{A}$ can be easily determined in either case, finding the K\"ahler cone of the CY manifolds $X\subset\mathcal{A}$ is, in general, not straightforward. First of all, the second cohomology of $X$ might not entirely descend from the second cohomology of $\mathcal{A}$. If it does the CY is called {\it favourable}. It turns out that all $h^{1,1}(X)=2$ manifolds in either data set are favourable in this sense. Secondly, even in a favourable case, the K\"ahler cone of $X$ can be larger than the cone obtained from the K\"ahler cone of $\mathcal{A}$ by restriction. When the two cones are equal the CY is called {\it K\"ahler-favourable}. It turns out that all $h^{1,1}(X)=2$ CICYs are K\"ahler-favourable. On the other hand, while many of the $h^{1,1}(X)=2$ THCYs are K\"ahler-favourable as well, this data set also contains some K\"ahler non-favourable cases. The overlap between the CICY and THCY lists for $h^{1,1}(X)=2$ consists of precisely two manifolds, which are the two $h^{1,1}(X)=2$ CICYs (with numbers 7884 and 7887 in the standard list of Ref.~\cite{Candelas:1987kf}) defined as hypersurfaces  in the four-dimensional ambient spaces $\mathcal{A}=\mathbb{P}^2\times\mathbb{P}^2$ and $\mathcal{A}=\mathbb{P}^1\times\mathbb{P}^3$.\\[2mm]
Before presenting our examples in more detail it is worth commenting on some general features of the data in Appendices~\ref{appA} and \ref{appB}. First, the structure of the extended and effective cones is typically quite rich, even for our relatively simple examples with $h^{1,1}(X)=2$. We note that most of the flop transitions (indeed all flop transitions in the CICY case) in our examples are different from the traditional type where a flop in a toric ambient space descends to a flop on the CY. For our examples, there is no ambient space flop and the topological transition only arises on the CY itself. Somewhat unexpectedly, isomorphic flops are rather common and perhaps even more surprisingly, multiple isomorphic flops leading to an infinite flop sequence are by no means rare. We also learn that the three non-trivial classes of intersection forms in Table~\ref{tab:EquivalenceClasses} are all realised by CY manifolds, although the CICY examples fall either into case 2 or case 3. What happens to the class of the intersection form under flop transitions? Of course isomorphic flops preserve the class but we do not have a general statement for non-isomorphic flops. The examples show that non-isomorphic flops typically do change the class, usually between $3\leftrightarrow 1$ and $2\leftrightarrow 1$. This is often to account for a secondary cone with a type (b) Zariski wall which, as we know from our classification, can only arise for case 1 intersection forms. 

\subsection[The \texorpdfstring{$h^{1,1}(X)=2$}{h11=2} CICYs]{The \texorpdfstring{$\boldsymbol{h^{1,1}(X)=2}$}{h11=2} CICYs}\label{sec:cicyex}
The $h^{1,1}(X)=2$ CICYs are embedded in an ambient space $\mathcal{A}=\mathbb{P}^{d_1}\times\mathbb{P}^{d_2}$ and are defined as the common zero locus of $K=d_1+d_2-3$ polynomials, homogeneous in coordinates of each projective space factor. The classification of CICYs in Ref.~\cite{Candelas:1987kf} lists $36$ topological types, each specified by a configuration matrix which contains the bi-degrees of the defining polynomials. Since all of these manifolds are K\"ahler-favourable a suitable basis of K\"ahler cone generators is obtained by restricting the standard  K\"ahler forms on the projective space factors to $X$ \cite{Anderson:2017aux}. This gives rise to forms $J_1$, $J_2$ and Poincar\'e dual divisor classes $D_1$ and $D_2$. Below we will specify divisor classes as two-dimensional numerical vectors relative to the basis $(D_1,D_2)$. Curve classes will be represented relative to the basis $(C^1,C^2)$ dual to $(D_1,D_2)$. The triple intersection numbers can be computed by standard methods, and the intersection form case in Table~\ref{tab:EquivalenceClasses} is determined by computing the class from Eq.~\eqref{eq:classdef}. Finding the structure of the various cones and the nature of the cone boundaries is more difficult, but can be accomplished by computing line bundle cohomology on $X$, following Ref.~\cite{Brodie:2020fiq}. The resulting information has been compiled in the table in Appendix~\ref{appA}.\\[2mm]
As an example for how to read this information, we consider from the table in Appendix~\ref{appA} the CICY \#7885  with configuration matrix
\begin{equation}
 X\in\left[\begin{array}{c|cc}\mathbb{P}^1&1&1\\\mathbb{P}^4&4&1\end{array}\right]^{2,86}_{-168}\; .
\end{equation}
This notation denotes a manifold defined in the ambient space $\mathcal{A}=\mathbb{P}^1\times\mathbb{P}^4$ as the common zero locus of two polynomials with bi-degrees $(1,4)$ and $(1,1)$. The Hodge numbers, attached as superscripts, are $h^{1,1}(X)=2$ and $h^{2,1}(X)=86$ and the Euler number, attached as a subscript, is $\chi(X)=-168$. The triple intersection numbers can be straightforwardly computed from this description to be
\begin{equation}
 \left(\begin{array}{ll}d_{122}&d_{222}\\d_{112}&d_{111}\end{array}\right)= \left(\begin{array}{ll}4&5\\0&0\end{array}\right)\qquad\Rightarrow\qquad
 \kappa=12xy^2+5y^3=y^2(12x+5y)\; .
\end{equation}
This intersection form has two zeros, $[1,0]$ and $[-5,12]$, in $\mathbb{RP}^1$ and is, hence, a realisation of case 2 from Table~\ref{tab:EquivalenceClasses}. More generally, the table in Appendix~\ref{appA} shows that CICYs realise case 2 and case 3 intersection forms, but there is no case 1 example~\footnote{However, some of the CYs obtained from CICYs via non-isomorphic flops have case 1 intersection forms.}. As we will see in the next subsection, case 1 intersection forms can be obtained from THCYs.

Returning to our example, one finds that the generators $v_i$ of the effective cone, and the further subdivision of this cone by flop walls and Zariski walls, as discussed in Section~\ref{sec:KahlerCones}, are given by
\begin{equation}
 v_1=\left(\begin{array}{r}-1\\1\end{array}\right)\;,\qquad v_2=\left(\begin{array}{r}1\\0\end{array}\right)\;,\qquad
 \left(\begin{array}{l}Z\\K_1\\K_2\end{array}
\begin{array}{|rr|}
 -1 & 1 \\
 -1 & 4 \\
 0 & 1 \\
 1 & 0 \\
\end{array}
\begin{array}{l}E\\Z\\F_{16,0}\\E\end{array}\right) \;.
\end{equation}
In the matrix on the right, the numerical rows in the middle provide the generators of the various cone boundaries. The first and the last of these generators are evidently the effective cone boundaries and are, hence, marked by an $E$ on the right. The boundary generated by $(-1,4)^T$ is a Zariski boundary marked by a $Z$ while the boundary generated by $(0,1)^T$ is a non-isomorphic flop boundary, marked by $F_{16,0}$. The subscripts are the Gromov-Witten invariants $n_1=16$, $n_2=0$ which determine the change in intersection numbers. The symbols on the left of the matrix indicate the nature of the cones. The cone $\langle (-1,1)^T,(-1,4)^T\rangle$ is a Zariski cone, marked as $Z$. The cone $\langle (-1,4)^T,(0,1)^T\rangle$ is a K\"ahler cone $\mathcal{K}(X')$, marked by $K_1$, where the subscript indicates that it has a case 1 intersection form. Its intersection form $\kappa'$ can be determined from Eq.~\eqref{eq:isecrel} with $n=n_1+8n_2=16$, so that
\begin{equation}
 \kappa'=-16x^3+12xy^2+5y^3
\end{equation}
Finally, the cone $\langle (0,1)^T,(1,0)^T\rangle$ is the K\"ahler cone of the original CY, which is case 2 and accordingly denoted by $K_2$. A plot of this cone structure is presented in Fig.~\ref{fig:ConesExample3} below.\\[2mm]
Of course a manifold $X'$ obtained from a CICY $X$ via a non-isomorphic flop does not need to have a CICY realisation but is, more generally, described by a complete intersection in a toric ambient space. The explicit construction of $X'$ is not of immediate relevance for the study of geodesics and will be presented in a forthcoming paper~\cite{CICYFlops}.\\[2mm]
As another example, consider CICY \#7887 with configuration matrix
\begin{equation}
 X\in\left[\begin{array}{c|c}\mathbb{P}^1&2\\\mathbb{P}^3&4\end{array}\right]\quad
 \begin{array}{l}x=(x_0,x_1)\\y=(y_0,y_1,y_2,y_3)\end{array}
\end{equation} 
where the homogeneous coordinates of $\mathbb{P}^1[x]$ and $\mathbb{P}^3[y]$ are defined on the right. The defining polynomial $p$ has the structure
\begin{equation}\label{eq:p7887}
 p=x_0^2\, p_0(y)+x_0x_1\,p_1(y)+x_1^2\, p_2(y)\; ,
\end{equation}
where $p_0,p_1,p_2$ are quartics in $y$. The intersection numbers give rise to a case 2 intersection form
\begin{equation}
 \left(\begin{array}{ll}d_{122}&d_{222}\\d_{112}&d_{111}\end{array}\right)= \left(\begin{array}{ll}4&2\\0&0\end{array}\right)\qquad\Rightarrow\qquad
 \kappa=12xy^2+2y^3=y^2(12x+2y)\; .
\end{equation}
and the cone structure is specified by
\begin{equation}
 v_1=\left(\begin{array}{r}-1\\4\end{array}\right)\;,\qquad v_2=\left(\begin{array}{r}1\\0\end{array}\right)\;,\qquad
 \left(\begin{array}{l}K_2\\K_2\end{array}
\begin{array}{|rr|}
 -1 & 4 \\
 0 & 1 \\
 1 & 0 \\
\end{array}
\begin{array}{l}E\\I_{64,0}^4\\E\end{array}\right)\; .
\end{equation}
The matrix denotes that the primary K\"ahler cone $\langle(0,1)^T,(1,0)^T\rangle$ of $X$ is connected by an isomorphic flop along the boundary generated by $(0,1)^T$, denoted $I_{64,0}^4$, to the K\"ahler cone of an isomorphic CY $X'$. Here the subscripts are the Gromov-Witten invariants $n_1=64$, $n_2=0$ while the superscript $m=4$ determines the generator for the involution of K\"ahler cones in Eq.~\eqref{eq:M1}.\\[2mm]
It is important to note that this result is valid for generic choices of complex structure, that is, for generic polynomials of the form~\eqref{eq:p7887}. For non-generic choices, previously non-effective divisors can become effective. If these divisors collapse when the K\"ahler form reaches a wall, this can change the wall type as well as the type of the adjacent cone. To illustrate this, consider a special defining polynomial with $p_1=0$, so that
\begin{equation}\label{eq:p7887ng}
 p=x_0^2\, p_0(y)+x_1^2\, p_2(y)\; .
\end{equation}
For this choice the generically non-effective divisor $D=-2D_1+4D_2$ becomes effective, as can be seen by computing $h^0(X,{\cal O}_X(D))$. Comparison with Eq.~\eqref{eq:volD} shows that its volume vanishes along the $x=0$ boundary. Hence, this boundary, which is generically an isomorphic flop wall, turns into a Zariski wall for non-generic defining polynomials of the form~\eqref{eq:p7887ng}. Since we are dealing with a case 2 intersection form this must necessarily be a type (a) Zariski wall, as type (b) Zariski walls can only arise for case 1 intersection forms.

\subsection[The \texorpdfstring{$h^{1,1}(X)=2$}{h11=2} THCYs]{The \texorpdfstring{$\boldsymbol{h^{1,1}(X)=2}$}{h11=2} THCYs}\label{sec:tcyex}
A (generic) CY hypersurface in a four-dimensional toric variety is specified by a choice of a four-dimensional reflexive polytope and a triangulation of its faces. The toric variety is constructed from the fan over the chosen triangulation, while the monomials that contribute to the defining polynomial are in one-to-one correspondence with the lattice points of the dual polytope. 

The polytope can be specified either directly, through its vertices, or indirectly, using homogeneous coordinates. It is the latter approach that we use in Appendix~\ref{appB}. Thus if $z_0,z_2,\ldots,z_m$ denote the homogeneous coordinates on a four-dimensional toric variety, these coordinates need to be identified under $m-3$ scaling relations specified as the rows of a weight system, also known as a charge matrix in GLSM language. Each homogeneous coordinate is associated with an irreducible toric divisor $D_i$ defined by $z_i=0$ and the scaling relations translate into linear relations between the toric divisors with coefficients given by the rows of the charge matrix. In order to construct the reflexive polytope and, subsequently, the fan of the toric variety, one associates a four-dimensional vector (ray) $v_i$ to each toric divisor $D_i$ satisfying the same linear relations that hold between the toric divisors. The vectors $v_i$ then specify the vertices of the polytope. 

The triangulation of the surface of the polytope corresponds to the information about the allowed simultaneous vanishings of the homogeneous coordinates. The non-allowed simultaneous vanishings are to be taken away before quotienting the homogeneous coordinates by the scaling relations. If the generators corresponding to a number of coordinates share a common cone, then the coordinates are allowed to simultaneously vanish, otherwise they are not. The allowed vanishings are collected in the Stanley-Reisner ideal denoted in Appendix~\ref{appB} as `SRI'.

Finally, the rank of the Picard group of the toric variety is given by the number of rays minus the dimension of the fan, $4$  in our case. Since we are interested in Picard number 2 compact toric varieties associated with reflexive polytopes, we will always have six rays. 

Changing the triangulation of the polytope defining the toric variety corresponds to flopping the ambient space. This might or might not descend to a flop on the CY hypersurface. If it does, the flop leads to a flop of the CY, and the K\"ahler cones of the two manifolds glue along a common wall to form (a part of) the extended K\"ahler cone. If it does not, the ambient space flop indicates a boundary of the ambient space K\"ahler cone which is not a boundary of the CY K\"ahler cone. This means that the two ambient space K\"ahler cones can be glued together to form (at least part of) the CY K\"ahler cone. In particular, nothing special happens on the CY at the ambient space flop locus. We will see examples illustrating both situations in Section~\ref{sec:examples}. 

\section{Geodesics in K\"ahler moduli space}\label{sec:geodesic}
In this chapter, we discuss general properties of the geodesic equation in K\"ahler moduli space, both in terms of the K\"ahler moduli $t^t$ and the shape moduli $b^i$. For $h^{1,1}(X)=2$ we show that the geodesic equations for $b^i$ can be integrated and we carry this out explicitly for the three intersection normal forms in Table~\ref{tab:EquivalenceClasses}.

\subsection{Generalities}
The geodesic equation for a trajectory $t^i=t^i(s)$ in K\"ahler moduli space reads
\begin{align}
\label{eq:GeodesicKahler}
\ddot{t}^i+\Gamma^i_{jk}\dot{t}^j\dot{t}^k=0\,,
\end{align}
where the dot denotes the derivative with respect to the curve parameter $s$. Multiplying the geodesic equation with $G_{ij}\dot{t}^j$ leads to a first ``energy'' integral
\begin{align}\label{eq:tenergy}
0=G_{ij}\dot{t}^i\dot{t}^j+\frac12\partial_iG_{jk}\dot{t}^i\dot{t}^j\dot{t}^k=\frac{d}{ds}\left(\frac12G_{ij}\dot{t}^i\dot{t}^j\right)\qquad\Rightarrow\qquad \frac{1}{2}G_{ij}\dot{t}^i\dot{t}^j=E\,,
\end{align}
with $E\geq0$ constant. The geodesic length between the points $t_i(s_1)$ and $t_i(s_2)$ is simply
\begin{align}
\label{eq:GeodesicDistanceKahler}
\Delta\tau=\tau_2-\tau_1=\int_{s_1}^{s_2}ds\sqrt{G_{ij}(t(s))\dot{t}^i\dot{t}^j}=\sqrt{2E}\,(s_2-s_1)=\sqrt{2E}\,\Delta s\,.
\end{align}
We would now like to show that the geodesic equation implies a decoupled differential equation for the overall volume modulus $\kappa=6a^3$. To do this we first note that Eq.~\eqref{eq:gammares} implies the result $\dot{t}_i=-G_{ij}\dot{t}^j$ for the derivative of the contravariant K\"ahler moduli and this leads to
\begin{equation}\label{kdd}
 \dot{\kappa}=3\kappa t_i\dot{t}^i\;,\qquad \ddot{\kappa}=3\kappa t_i(\ddot{t}^i+\Gamma_{jk}^i\dot{t}^j\dot{t}^k)+\frac{\dot{\kappa}^2}{\kappa} \;.
\end{equation} 
Provided the geodesic equation~\eqref{eq:GeodesicKahler} is satisfied, the first term on the right-hand side vanishes and we find the desired equation
\begin{equation} \label{eq:aeq}
 \ddot{\kappa}=\frac{\dot{\kappa}^2}{\kappa}\qquad\Leftrightarrow\qquad \ddot{a}=\frac{\dot{a}^2}{a}\; ,
\end{equation}
for the volume modulus. The solution is
\begin{equation}\label{eq:asol}
 a(s)=a_0\, e^{\alpha s}\; ,
\end{equation}
where $a_0$ and $\alpha$ are real constants. In particular, constant volume geodesics correspond to $\alpha=0$, while $\alpha>0$ ($\alpha<0$) implies an exponentially expanding (contracting) volume.

\subsection{Geodesics in very special geometry}
\label{sec:GeodesicsVector}
Given that the volume modulus $a$ can be decoupled from the geodesic equation it is reasonable to ask if the same can be accomplished for the shape moduli $b^i=t^i/a$. A straightforward computation of the shape moduli derivatives lead to 
\begin{align}
\dot{b}^i = a^{-1}(\dot{t}^i-t_j\dot{t}^jt^i)\,,\qquad \ddot{b}^{i}=a^{-1}\left(\ddot{t}^{i} - 2 t_j\dot{t}^j \dot{t}^i - t_j\dot{t}^jt_k\dot{t}^k t^i - (t_j\ddot{t}^j-G_{jk}\dot{t}^j\dot{t}^k)t^i\right)\; ,
\end{align}
and combining these results it follows that
\begin{align}\label{eq:geodesicrel}
\ddot{b}^i+\tilde{\Gamma}^i_{jk}\dot{b}^j\dot{b}^k=\frac1a\left[\ddot{t}^i+\Gamma^i_{jk}\dot{t}^j\dot{t}^k-\frac{1}{3\kappa}\left(\ddot{\kappa}-\frac{\dot{\kappa}^2}{\kappa}\right)t^i\right]\; .
\end{align}
Evidently, if the geodesic equation~\eqref{eq:GeodesicKahler} for the K\"ahler moduli $t^i$ (and, hence,  Eq.~\eqref{eq:aeq} for the volume modulus) is satisfied then the right-hand side of this equation vanishes and we find that the shape moduli satisfy the geodesic equation 
\begin{equation}
\label{eq:GeodesicVector}
\ddot{b}^i+\tilde{\Gamma}^i_{jk}\dot{b}^j\dot{b}^k=0\; ,
\end{equation}
where $\tilde{\Gamma}^i_{jk}$ is the Levi-Civita connection in terms of the shape moduli, as in Eq.~\eqref{eq:Gtildedef}. Conversely, satisfying the geodesic equation~\eqref{eq:GeodesicVector} for the shape moduli and for the overall volume modulus~\eqref{eq:aeq} implies the geodesic equation~\eqref{eq:GeodesicKahler} for $t^i$, as Eq.~\eqref{eq:geodesicrel} shows.\\[2mm]
All this is consistent with the structure of five-dimensional $N=1$ supergravity, where the shape moduli $b^i$ are the vector multiplet moduli. As explained earlier, we will focus on M-theory compactifications to such five-dimensional supergravity theories and, therefore, study the geodesic equation~\eqref{eq:GeodesicVector} for $b^i$. However, a solution $b^i(s)$ of this geodesic equation can be multiplied with a solution~\eqref{eq:asol} for $a(s)$ to produce a solution $t^i(s)=a(s)b^i(s)$ of the geodesic equation~\eqref{eq:GeodesicKahler} for the K\"ahler moduli $t^i$. This can be interpreted as an M-theory solution with evolving vector multiplet moduli and evolving volume modulus (which resides in the hypermultiplet sector) or as a IIA solution with the same properties. In the latter case we have to assume that the complicating effects which arise in IIA, that is, possible axion dynamics, instanton effects and the presence of non-geometric phases, can be neglected.\\[2mm]
Of course the geodesic equation~\eqref{eq:GeodesicVector} also has a first energy integral
\begin{equation}\label{eq:Ebint}
 \frac{1}{2}\tilde{G}_{ij}b^ib^j=\tilde{E}\; ,
\end{equation}
where $\tilde{E}\geq 0$ is a constant. The constants $\tilde{E}$ and $E$ in Eq.~\eqref{eq:tenergy} and the expansion coefficient $\alpha$ of the volume modulus in Eq.~\eqref{eq:asol} are related via
\begin{align}
\frac{1}{2}G_{ij}\dot{t}^i\dot{t}^j=\frac{1}{2}\tilde{G}_{ij}\dot{b}^i\dot{b}^j+\frac{\dot{a}^2}{2a^2}\qquad\Rightarrow\qquad E=\tilde{E}+\frac{1}{2}\alpha^2 \;.
\end{align}
This means the geodesic length~\eqref{eq:GeodesicDistanceKahler} becomes
\begin{align}\label{eq:dtau}
\Delta\tau = \sqrt{2\tilde{E}+\alpha^2}\,\Delta s\; .
\end{align}
A solution $t^i(s)$ with varying volume is of the form 
\begin{equation}\label{eq:tsol}
 t^i(s)=a(s) b^i(s)=a_0e^{\alpha s}b^i(s)\; ,
\end{equation} 
where $b^i(s)$ is a solution to the geodesic equation~\eqref{eq:GeodesicVector}. For constant volume, $a={\rm const}$, the case we focus on, we should set $\alpha=0$ in this equation. More generally, in the context of M-theory, we can think of $\alpha$ as the contribution of the entire hypermultiplet moduli sector. Then equation~\eqref{eq:dtau} shows that the $\alpha=0$ case, where all hypermultiplet moduli are kept constant, provides a lower bound  on the geodesic length.\\[2mm]
For our purposes it is important to note that geodesics $b^i(s)$ for the shape moduli satisfy the two equations
\begin{equation}\label{eq:2eqs}
 d_{ijk}b^ib^jb^k=6\;,\qquad d_{ijk}b^i\dot{b}^j\dot{b}^k=-\tilde{E}\; .
\end{equation}
The first of these is simply the constant volume constraint, $\tilde{\kappa}=6$, and the second one follows from the energy integral~\eqref{eq:Ebint}, given that $\tilde{\kappa}=6$ implies $\tilde{\kappa}_i\dot{b}^i=0$. In general, these two equations are of course not sufficient to find a solution, but they are for $h^{1,1}(X)=2$, our main case of interest. 

\subsection{Geodesics and K\"ahler cone walls}
Next we discuss how geodesics behave near walls of the K\"ahler cone, covering the different types of walls listed in Table~\ref{fig:walltypes}.\\[2mm]
At flop walls of the K\"ahler cone $\mathcal{K}(X)$, the metric $G_{ij}$ remains finite and non-singular. Hence near a flop wall, geodesics behave regularly, qualitatively no different to how they behave in the interior of the K\"ahler cone, and flop walls appear at finite geodesic distance. What happens at the flop wall depends on the dynamics of the additional hypermultiplets which become massless. If these hypermultiplets do not evolve, as is plausible since they are massive away from the flop wall, the evolution continues through the wall and into the adjacent K\"ahler cone $\mathcal{K}(X')$. We must therefore discuss how to match geodesics across a flop wall. Recall from Eq.~\eqref{eq:Gdef} that the metric $G_{ij}$ on $\mathcal{K}(X)$ depends on the intersection numbers $d_{ijk}$. The metric $G'_{ij}$ on $\mathcal{K}(X')$ is given by the same general expression but now in terms of the intersection numbers $d'_{ijk}$ of $X'$. From Eq.~\eqref{eq:isecrel}, we can formally incorporate this change into the five-dimensional theory by thinking of the metric as a function
\begin{equation}
 G=G(b,d_{ijk}(b))\;,\qquad d_{ijk}(b)=d_{ijk}-\theta(-b^1)n \delta_{1i}\delta_{1j}\delta_{1k}\; ,
\end{equation} 
of the shape moduli $b^i$ and the moduli-dependent intersection numbers $d_{ijk}(b)$. Here, we are assuming that the flop arises at $b^1=0$, and $b^1>0$ ($b^1<0$) corresponds to the K\"ahler cone of $X$ ($X'$). As we compute the Levi-Civita connection~\eqref{eq:LC} from this metric we would expect additional terms proportional to $\delta(b^1)$. However, these terms are always multiplied by factors of $b^1$. Hence, the connection has a discontinuity (due to the theta function which arises in the first term in Eq.~\eqref{eq:LC}) across the flop wall but it does not have a delta function singularity. The conclusion is that geodesics should be matched across a flop wall such that $b^i$ and $\dot{b}^i$ are continuous. In particular this means a geodesic in $\mathcal{K}(X)$ uniquely determines its continuation into the adjacent K\"ahler cone $\mathcal{K}(X')$.\\[2mm]
The above discussion has interesting implications for isomorphic flops. Suppose a K\"ahler cone $\mathcal{K}(X)$ within the region $b^1>0$ and an adjacent K\"ahler cone $\mathcal{K}(X')$ within the region $b^1<0$ are related by an isomorphic flop at $b^1=0$. Consider a geodesic $b^i(s)$ across both cones with $b^i(s)>0$ for $s<0$ and $b^i(s)<0$ for $s>0$, so that the flop arises for parameter value $s=0$. The two K\"ahler cones are related by an involution, generated by $M$ (see Eqs.~\eqref{eq:M1}) which leaves the flop locus $b^1=0$ unchanged. We can use this involution to map the geodesic $b^i(s)$ for $s>0$ which is inside $\mathcal{K}(X')$ back into the primary K\"ahler cone $\mathcal{K}(X)$, that is, onto ${M^i}_jb^j(s)$ for $s\geq 0$. This is of course in line with the idea that the involution is a gauge symmetry and that the K\"ahler cones $\mathcal{K}(X)$ and $\mathcal{K}(X')$ should be identified. But the geodesics $b^i(s)$ for $s\leq 0$ and ${M^i}_jb^j(s)$ for $s\geq 0$ have the same initial conditions at the flop locus $s=0$ so they represent the same trajectories
\begin{equation}
 {M^i}_jb^j(s)=b^i(-s)
\end{equation}
for $s\geq 0$. Hence, in a ``downstairs" description where we identify the extended K\"ahler moduli space with $\mathcal{K}(X)$, geodesics ``bounce" back from a flop wall and ``re-trace" their original path. If we have two isomorphic flop walls, giving rise to an infinite flop sequence, then the geodesic motion in the downstairs picture corresponds to an ``oscillation" between these two walls, along the same trajectory.\\[2mm]
At a type (a) Zariski wall the metric also remains finite and non-singular just like for a flop, so in the naive effective theory they arise at finite geodesic distance. In fact, we have seen that it is possible to switch between flop and type (a) Zariski walls by a choice of complex structure, without changing the metric $G_{ij}$. However, unlike flop walls, type (a) Zariski walls mark the end of the moduli space. This is not apparent from the naive effective theory, in which geodesic evolution proceeds just as for a flop, that is through to the other side of the wall. Of course, for a correct description of the physics near a type (a) Zariski wall the SU$(2)$ gauge theory which appears there has to be included in the low-energy theory. We note that this modified geodesic evolution has been determined in a simplified setting in Ref.~\cite{Mohaupt2001}.\\[2mm]
For type (b) Zariski walls the metric is non-divergent but singular at the wall. This means the wall can be reached at finite geodesic distance (at least within the naive effective theory) but one combination of ``velocities" $\dot{b}^i$ is not bounded by the energy condition~\eqref{eq:Ebint} and can, hence, diverge. This is indeed what happens near a type (b) Zariski wall as our explicit solution in the next subsection will show. Of course, as the velocity grows beyond the string or compactification scale the low-energy theory becomes invalidated. Hence, the naive low-energy theory retains some memory of the new physics which is expected at a type (b) Zariski wall, quite unlike for a type (a) Zariski wall. For a correct low-energy description the infinite towers of light states expected at a type (b) Zariski wall have to be included and they will modify the geodesic evolution near the wall.\\[2mm]
Finally, for an effective cone wall the metric diverges, so the energy condition~\eqref{eq:Ebint} implies that velocities $\dot{b}^i$ go to zero. This suggests effective cone walls are at infinite geodesic distance and this is shown explicitly by our solutions in the next subsection.

\subsection[Geodesics for \texorpdfstring{$h^{1,1}(X)=2$}{h11=2}]{Geodesics for \texorpdfstring{$\boldsymbol{h^{1,1}(X)=2}$}{h11=2}}
For $h^{1,1}(X)=2$ CYs the two equations~\eqref{eq:2eqs} completely determine the geodesic solutions. In fact, the actual geodesic curve is already determined by the first equation, the constant volume constraint $\tilde{\kappa}=6$, but in order to find the evolution and geodesic distance we have to solve both equations. In principle this is straightforward by solving the first Eq.~\eqref{eq:2eqs} for one of the shape moduli in terms of the other, inserting into the second Eq.~\eqref{eq:2eqs} and solving the resulting first-order non-linear differential equation. In practice, this is difficult to carry out for a generic intersection form. However, the classification of intersection forms in Section~\eqref{sec:class} shows that we only need to do this for the four normal forms in Table~\ref{tab:EquivalenceClasses}. In fact, case 0 is irrelevant since the metric is never positive definite, so only three interesting cases remain. We will now show that analytic solutions for the geodesics can be obtained for these three cases. As before, we denote the two shape moduli by $(x,y)=(b^1,b^2)$. 

\subsubsection*{Case 1}
The case 1 normal form from Table~\ref{tab:EquivalenceClasses} (multiplied by a convenient overall factor of $6$) is $\hat{\kappa}=6x^3+6y^3$ and inserting this into Eqs.~\eqref{eq:2eqs} gives
\begin{align}
x^3+y^3=1\,,\qquad x\dot{x}^2+y\dot{y}^2=-\frac16\tilde{E}\,.
\end{align}
Solving the first equation for $y$ and inserting into the second gives
\begin{align}
 y=(1-x^3)^{1/3}\,,\qquad \frac{x \dot{x}^2}{1-x^3}=-\frac16\tilde{E}=:-\frac16\varepsilon^2\,.
\end{align}
The differential equation can be solved analytically which leads to
\begin{align}
\label{eq:SolutionCase1}
x=-\sqrt[3]{\sinh^{2}(r_k(s))}\,,\qquad y=\sqrt[3]{\cosh^{2}(r_k(s))}\,,\qquad r_k(s)=\sqrt{\frac38}\, \varepsilon s + k\,,
\end{align}
where $k$ is an integration constant and we are choosing the branch such that the solution is real. This solution satisfies the first K\"ahler cone condition ($x<0,y>-x$) in Table~\ref{tab:EquivalenceClasses}, so it resides in the upper cone in Fig.~\ref{fig:Geodesics3Cases} on the left. The other solution, which resides in the lower cone, is obtained by exchanging $x$ and $y$. Note that the constant volume constraint $\hat{\kappa}=6$ is a result of the identity $\cosh^2x-\sinh^2x=1$. Both branches of the geodesic have been plotted in the left Fig.~\ref{fig:Geodesics3Cases}. They are evidently equivalent and we can focus on the $y>0$ branch for simplicity.\\[2mm]
The maximal (upper) cone in the left Fig.~\eqref{fig:Geodesics3Cases} has two boundaries. We know that the boundary at $x=-y$ corresponds to an effective cone wall while the boundary at $x=0$ is a type (b) Zariski wall. Let us discuss the behaviour of the geodesic~\eqref{eq:SolutionCase1} at both of these boundaries. The geodesic asymptotes to the boundary at $x=-y$ as $\sigma\rightarrow \infty$ which shows that it is at infinite geodesic distance, as expected. The type (b) Zariski wall at $x=0$ is approached by the geodesic as $r_k(s)\rightarrow 0$ and it is evident from Eq.~\eqref{eq:SolutionCase1} that $\dot{x}$ diverges in this case. As discussed above, this will eventually lead to a break-down of the naive effective theory near the type (b) Zariski wall.
\\[2mm]
Of course the K\"ahler cone $\hat{\mathcal{K}}(X)$ of an actual CY with case 1 intersection form can fill out the entire maximal cone, only one of its boundaries may coincide with a maximal cone boundary, or it may be entirely in the interior of the maximal cone. Any boundary of $\hat{\mathcal{K}}(X)$ in the interior must be either a flop or a type (a) Zariski wall.\\[1mm]
It is interesting to note that, from our classification, we can exclude certain combinations of K\"ahler cone wall types for CYs with case 1 intersection forms. For example, it is impossible for both K\"ahler cone boundaries to be type (b) Zariski walls or for both to be effective cone walls.

\subsubsection*{Case 2}
From Table~\ref{tab:EquivalenceClasses}, the case 2 normal form is given by $\hat{\kappa}=x^2 y$, so that the Eqs.~\eqref{eq:2eqs} become
\begin{equation}
x^2y=6\;,\qquad \dot{x}(y\dot{x}+2x\dot{y})=-3\tilde{E}\; .
\end{equation}
Solving the first equation for $y$ and inserting into the second, the solution is easily found to be
\begin{equation}
\label{eq:SolutionCase2}
x=ke^{\varepsilon s}\,,\quad y=\frac{6}{k^2}e^{-2\varepsilon s}\,,
\end{equation}
where $k$ is an integration constant and $\varepsilon^2=\tilde{E}/3$. Depending on the choice for the sign of $k$, we get solutions which reside in either of the two maximal cones and both of these branches are shown in the middle Fig.~\ref{fig:Geodesics3Cases}. They are equivalent and we can focus on the maximal cone which coincides with the positive quadrant.\\[2mm]
Both boundaries of the maximal cone are effective cone walls. If $\varepsilon>0$, the geodesic~\eqref{eq:SolutionCase2} asymptotes towards the $x=0$ boundary as $s\rightarrow -\infty$ and towards the $y=0$ boundary as $s\rightarrow\infty$, so either of these walls is at infinite geodesic distance, as expected. As before, the K\"ahler cone $\hat{\mathcal{K}}(X)$ of an actual CY need not fill out the entire maximal cone, and flop and type (a) Zariski walls arise at boundaries of $\hat{\mathcal{K}}(X)$ in the interior of the maximal cone. But just as for case 1, certain wall types are excluded for CYs with case 2 intersection forms. Specifically, any combination of wall types which involves a type (b) Zariski wall is ruled out.

\subsubsection*{Case 3}
From Table~\ref{tab:EquivalenceClasses}, the case 3 normal form (with a convenient factor of $3$) is $\hat{\kappa}=3x^2y+3xy^2$, so that Eqs.~\eqref{eq:2eqs} become
\begin{align}
x^2y+x y^2=2\;,\qquad \dot{x}^2y+2(x+y)\dot{x}\dot{y}+x\dot{y}^2=-\tilde{E}\; .
\end{align}
Solving the first equation for $y$ and inserting into the second equation gives
\begin{align}
\label{eq:ODECase3}
y_{\pm}=\frac{-x^2\pm\sqrt{8x+x^4}}{2x}\,,\qquad \frac{2+x^3}{x^2(x^3+8)}\dot{x}^2=\frac{\tilde{E}}{6}=:\varepsilon^2\,.
\end{align}
Note that both roots $y_\pm$ lead to the same differential equation for $x$. The square root (and hence $y_{\pm}$) is real for (i) $x\leq-2$ or (ii) $x\geq0$. In case (i), $y_{\pm}$ is positive for either of the solutions, and hence satisfies the first K\"ahler cone condition in Table~\ref{tab:EquivalenceClasses}. In case (ii), $y_+$ is negative and satisfies the second K\"ahler cone condition, while $y_-$ is positive and satisfies the third K\"ahler cone condition.  An analytic solution to~\eqref{eq:ODECase3} is given by
\begin{align}
\label{eq:SolutionCase3}
x=\sqrt[3]{-2+6\cosh(3\varepsilon s)~\sin\left(\frac13\arccos\left(\tanh(-3\varepsilon s)\right)\right)}\,,\qquad y_{\pm}=\frac{-x^2\pm\sqrt{8x+x^4}}{2x}\,,
\end{align}
where we have fixed the integration constant such that $x(0)=1$. Note that $w_1=\tanh(-3s\varepsilon)\in(-1,1)$ and $w_2=\arccos(w_1)\in(0,\pi)$, such that $\sin(w_2/3)\in(0,\sqrt{3}/2)$. The three branches of the geodesic solution are shown in Fig.~\ref{fig:Geodesics3Cases}, on the right. \\[2mm]
In fact, the intersection form $\hat{\kappa}=3(x^2y+xy^2)$ has a symmetry group $H=\langle h_1,h_2\rangle$ generated by
\begin{equation}
 h_1=\left(\begin{array}{rr}-1&-1\\0&1\end{array}\right)\;,\qquad  h_2=\left(\begin{array}{rr}0&1\\1&0\end{array}\right)\; ,
\end{equation}
which is of order $6$ and isomorphic to $S_3$. This group permutes the three maximal cones in Fig.~\ref{fig:Geodesics3Cases} along with the corresponding geodesic branches which are, hence, equivalent. We can, therefore, focus on the maximal cone which fills out the positive quadrant.\\[2mm]
This cone has two boundaries at $x=0$ and $y=0$, both of which are effective cone walls. They are approached asymptotically as $s\rightarrow \pm\infty$ and are, therefore, at infinite geodesic distance, as expected. As before, the Kahler cone $\hat{\mathcal{K}}(X)$ may not fill out the entire maximal cone and its boundaries in the interior of the maximal cone are flop or type (a) Zariski walls. Note that type (b) Zariski walls are excluded for CYs with case 3 intersection forms, just as they were for case 2.

Let us summarise these results. We have obtained explicit geodesic solutions  for the three non-trivial normal forms in Table~\ref{tab:EquivalenceClasses}. These solution prove that effective cone walls are always at infinite distance. Flop walls are reached in finite geodesic distance and the same is true for Zariski walls when considered within the naive effective theory as described by the standard geodesic equation. However, for both type (a) and type (b) Zariski walls new massless states arise at the wall which can be expected to modify the geodesic evolution near the wall. We have also seen that type (b) Zariski walls are, in a certain sense, rare. They can only arise for CYs with a case 1 intersection form and then for at most one of the K\"ahler cone boundaries. 

\section{Examples}\label{sec:examples}
In this section, we illustrate our general results with a number of examples taken from the collections of $h^{1,1}(X)=2$ CICYs and THCYs, whose data is presented in Appendices~\ref{appA} and \ref{appB}. From this data we know the structure of the extended K\"ahler cone. We can work out the intersection form for each constituent K\"ahler cone, determine its case in the classification of Table~\ref{tab:EquivalenceClasses} and the coordinate transformation to the corresponding normal form. The geodesic solution for an intersection normal form is given by one of the explicit solutions given in the previous section. Transforming these back to the original field basis for each K\"ahler cone and matching $b^i$ and $\dot{b}^i$ continuously at all flop walls gives the geodesic in the entire extended K\"ahler cone.

\subsection{Example for case 1 - flop to one isomorphic CY}
As an example for case 1, we consider a THCY $X_0$ with a flop to an isomorphic CY $X_1$, listed as example number $(4,\cdot)$ in the long table of Appendix~\ref{appB}. This flop is inherited from a flop of the ambient toric variety, which is perhaps the most well-known situation. The two toric varieties $\mathcal{A}_0$ and $\mathcal{A}_1$ are described by a GLSM charge matrix, and a Stanley-Reisner ideal which specifies the fine regular star triangulation\footnote{Here and below the two charge matrices are equivalent, but we choose them to match the two K\"ahler cone bases.},
\begin{align}
\begin{split}
X_0 \subset \mathcal{A}_0 &\sim
\begin{tabular}{|cccccc|}
\hline
$z_0$&$z_1$&$z_2$&$z_3$&$z_4$&$z_5$\\
\hline
-1&2&1&1&3&0\\
1&-1&0&0&-1&1\\
\hline
\end{tabular}
\,,\qquad
X_1 \subset \mathcal{A}_1 \sim
\begin{tabular}{|cccccc|}
\hline
$z_0$&$z_1$&$z_2$&$z_3$&$z_4$&$z_5$\\
\hline
2&-1&1&1&0&3\\
-1&1&0&0&1&-1\\
\hline
\end{tabular}\,.\\
&\hspace{.57cm}\text{SRI}_0:~~\langle z_0 z_5,~z_1 z_2 z_3 z_4 \rangle\qquad\qquad\qquad\hspace{1.28cm}\text{SRI}_1:\quad\langle z_1 z_4,~z_0 z_2 z_3 z_5\rangle
\end{split}
\end{align}
The K\"ahler cone generators are $\frac13(z_4 + z_5)$ and $z_5$ or $z_4$ for the CYs $X_0$ and $X_1$, respectively. For the following discussion, all coordinates are given relative to the basis $(D_1=z_5,D_2=\frac13(z_4 + z_5))$ of the K\"ahler cone generators of $X_0$.
\begin{figure}[t]
\centering
\hspace{50pt}\includegraphics[width=.5\textwidth]{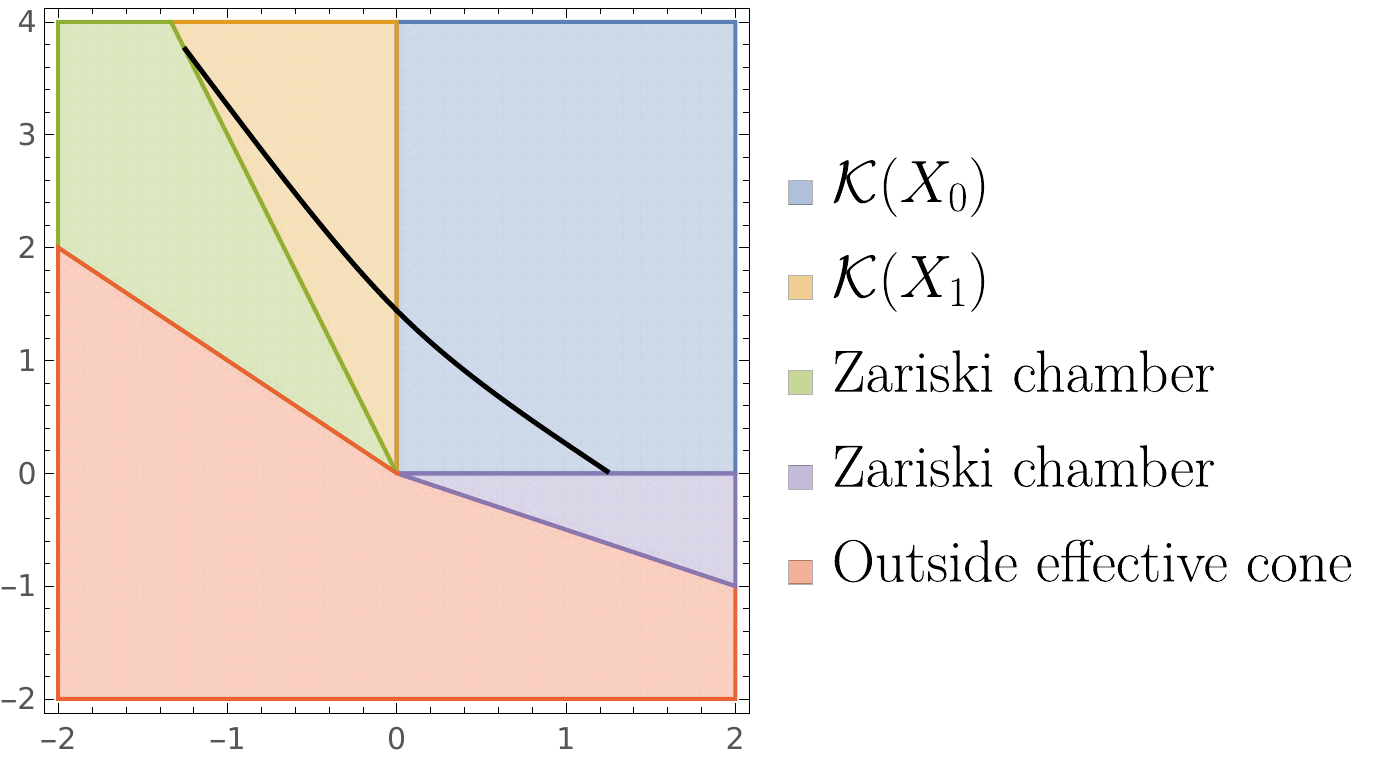}
\caption{Effective cone of the THCY example $(4,\cdot)$, showing K\"ahler cones of CYs $X_0$ and $X_1$, as well as two adjacent Zariski chambers, with a geodesic curve superimposed.}
\label{fig:ConesExample1}
\end{figure}
\noindent The structure of the effective cone is presented in Figure~\ref{fig:ConesExample1}. 

There are two K\"ahler cones corresponding to isomorphic CYs, and two Zariski chambers with type (b) Zariski walls that bound the extended K\"ahler cone on either side. The K\"ahler cones of $X_0$ and $X_1$ are given by
\begin{align}
\mathcal{K}(X_0)=\{x_0>0\,,~y_0>0\}\,,\qquad \mathcal{K}(X_1)=\{x_1<0\,,3x_1+y_1>0\}\,.
\end{align}
The involution that maps these two K\"ahler cones onto one another is given by
\begin{align}
M_1=\left(\begin{array}{rr}-1&0\\3&1\end{array}\right)\,,
\end{align}
and the intersection polynomials of $X_0$ and $X_1$ in the K\"ahler cone divisor basis of $X_0$ are given by
\begin{align}
\kappa_0=3 x^3_0 + 9 x^2_0 y_0 + 9 x_0 y^2_0 + 2 y^3_0\,,\qquad
\kappa_1=-3 x^3_1 + 9 x^2_1 y_1 + 9 x_1 y^2_1 + 2 y^3_1\,.
\end{align}
We see that the flop changes the triple intersection number $d_{111}$ by 6 and leaves all other intersection numbers invariant. These two intersection forms can be brought into the (rescaled) normal form $6x^3+6y^3$ of case 1 using the transformations
\begin{equation}
\left(\begin{array}{c}x_i\\y_i\end{array}\right)=P_i\left(\begin{array}{c}x\\y\end{array}\right)\;,\qquad 
P_0=\sqrt[3]{6}\left(\begin{array}{rc}\!\!\!1&\frac{1}{\sqrt[3]{3}}\\-1&0\end{array}\right)\,,\qquad
P_1=\sqrt[3]{6}\left(\begin{array}{rc}\!\!\!-1&-\frac{1}{\sqrt[3]{3}}\\2&(\sqrt[3]{3})^2 \end{array}\right)\,.
\end{equation}
There are two choices for these maps, corresponding to the two maximal cones in the left Fig.~\ref{fig:Geodesics3Cases} and we have selected the one which maps to the upper cone. The normal form solution in terms of the variables $(x,y)$ is given in~\eqref{eq:SolutionCase1}. Using the above maps $P_0$ and $P_1$, we can transform this solution back to the original coordinates $(x_0,y_0)$ of $X_0$ and $(x_1,y_1)$ of $X_1$. We fix the integration constants in Eq.~\eqref{eq:SolutionCase1} by demanding that the flop occurs at $s=0$, explicitly $k_{0,1}=\mp \text{arccosh}\big(\sqrt{3/2}\big)$, and demand that the solutions are continuous and differentiable across the flop, so that
\begin{align}
x_0(0)=x_1(0)=0\;,\quad y_0(0)=y_1(0)\;, \quad \dot{x}_0(0)=\dot{x}_1(0)\;,\quad \dot{y}_0(0)=\dot{y}_1(0)\; .
\end{align}
Setting $\tilde{E}=1$, $\alpha=0$ (i.e.\ fixing the volume of the CY along the geodesic), and $k_{0,1}=\mp \text{arccosh}\big(\sqrt{3/2}\big)$, the geodesics are
\begin{equation}
\begin{array}{rclcl}
\left(\begin{array}{c}x_0(s)\\y_0(s)\end{array}\right)&=&{\scriptsize\left(\begin{array}{c}\sqrt[3]{2\cosh^2[r_{k_0}(s)]}-\sqrt[3]{6\sinh^2[r_{k_0}(s)]}\\\sqrt[3]{6\sinh^2[r_{k_0}(s)]})\end{array}\right)}&&s\leq0~~~(\text{cone}~\mathcal{K}_0)\\
\left(\begin{array}{c}x_1(s)\\y_1(s)\end{array}\right)&=&{\scriptsize\left(\begin{array}{c}-\sqrt[3]{2\cosh^2[r_{k_1}(s)]}+\sqrt[3]{6\sinh^2[r_{k_1}(s)]}\\
3\sqrt[3]{2\cosh^2[r_{k_1}(s,k)]}-2\sqrt[3]{6\sinh^2[r_{k_1}(s)]}))\end{array}\right)}&&s\geq 0~~~(\text{cone}~\mathcal{K}_1)
\end{array}
\end{equation}
We have plotted the solutions in Figure~\ref{fig:ConesExample1}. We can see the high-speed escape/crash at the Zariski wall and the continuous smooth transition across the flop wall. Since $X_0$ and $X_1$ are equivalent we should identify the two cones $\mathcal{K}_0$ and $\mathcal{K}_1$, dividing out by the involution. In the downstairs picture we can then describe the entire evolution within the cone $\mathcal{K}(X_0)$. It amounts to a high-speed escape from the Zariski wall, a motion towards and a bounce back from the flop wall  and retracing the same trajectory which ends in a high-speed crash at the Zariski wall. Since the flop arises at parameter value $\varepsilon s=0$ and the Zariski wall at $\varepsilon s=-\sqrt{8/3}\;\text{arccosh}(\sqrt{3/2})\approx1.08$, it takes a geodesic distance of
\begin{equation}\label{eq:exgeolength}
 \Delta \tau=\sqrt{2}\,\varepsilon\Delta s\simeq 1.52
\end{equation} 
to traverse the K\"ahler cone $\mathcal{K}(X_0)$ along an isochore.\\[2mm]
The above discussion has focused on isochores, but solutions with varying volume are easily obtained from Eq.~\eqref{eq:tsol}. For such solutions, with a volume expansion rate $\alpha$, the geodesic distance to traverse the K\"ahler cone $\mathcal{K}(X_0)$ is
\begin{equation}
 \Delta \tau=\sqrt{2\varepsilon^2+\alpha^2}\;\Delta s\simeq\sqrt{1+\frac{\alpha^2}{2\varepsilon^2}}\, 1.52\; ,
\end{equation} 
which is bounded from below by the isochore result~\eqref{eq:exgeolength}.

\subsection{Example for case 1 - ambient space flop without CY flop}
\begin{figure}[t]
\centering
\hspace{50pt}
\begin{tikzpicture}
   \tikzstyle{every node}=[font=\fontsize{10.25}{10.25}\selectfont]
   \node[anchor=south west,inner sep=0] at (0,0) {\includegraphics[width=.5\textwidth]{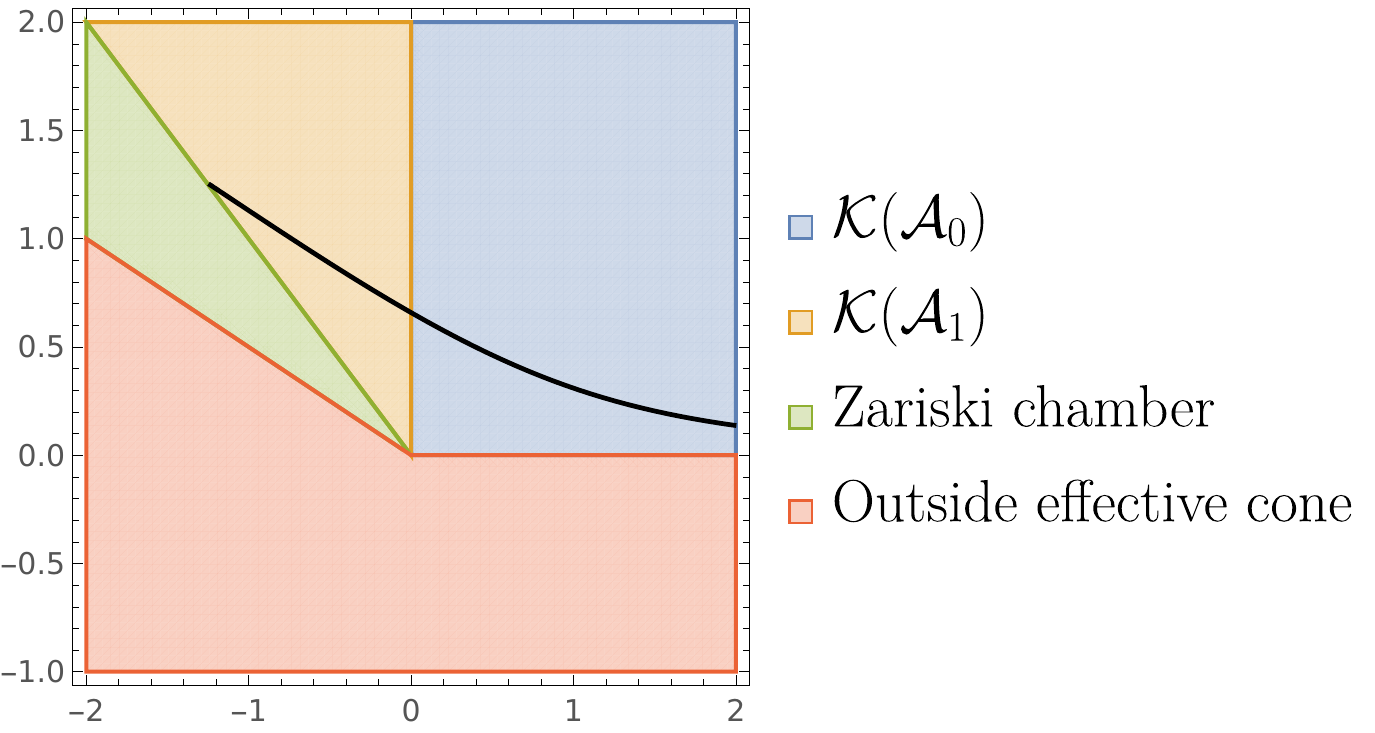}};
   \node[anchor=south west,inner sep=0] at (5.7,2.36) {$\left. \begin{array}{c} \, \\ \, \end{array} \right\} \mathcal{K}(X_0) $};
\end{tikzpicture}
\caption{Effective cone of THCY example $(17,\cdot)$, showing K\"ahler cone of CY $X_0$, which is a union of ambient space K\"ahler cones for two triangulations, as well as an adjacent Zariski chamber, with a geodesic curve superimposed.}
\label{fig:ConesExample2}
\end{figure}

A flop in the toric ambient space does not always descend to the CY hypersurface. If it does not, the geodesic evolution across such an ambient flop locus takes place within a single CY K\"ahler cone and, hence, leaves the CY topology unchanged. An explicit example is realised by the hypersurfaces $X_0$ and $X_1$ in the two four-dimensional toric varieties $\mathcal{A}_0$ and $\mathcal{A}_1$ given by the following two fine regular star triangulations
\begin{align}
\begin{split}
X_0\subset \mathcal{A}_0 &\sim
\begin{tabular}{|cccccc|}
\hline
$z_0$&$z_1$&$z_2$&$z_3$&$z_4$&$z_5$\\
\hline
1&0&0&0&1&1\\
-1&1&1&1&-2&0\\
\hline
\end{tabular}\,,\qquad
X_1\subset\mathcal{A}_1 \sim
\begin{tabular}{|cccccc|}
\hline
$z_0$&$z_1$&$z_2$&$z_3$&$z_4$&$z_5$\\
\hline
1&-1&-1&-1&2&0\\
0&1&1&1&-1&1\\
\hline
\end{tabular}\,,\\
&\hspace{.57cm}\text{SRI}_0:~~\langle z_0 z_4 z_5,~z_1 z_2 z_3 \rangle\qquad\qquad\qquad\hspace{1.28cm}\text{SRI}_1:\quad\langle z_0 z_4,~z_1 z_2 z_3 z_5 \rangle
\end{split}
\end{align}
which are related by a flop. This is example number $(17,\cdot)$ in the long table of Appendix~\ref{appB}, with $\mathcal{A}_0$ and $\mathcal{A}_1$ corresponding to triangulations $2$ and $1$ respectively. The K\"ahler cones of $\mathcal{A}_0$ and $\mathcal{A}_1$ share a common generator $z_5$, and the other generator is $\frac12(z_5\mp z_4)$. We choose the basis $(D_1=\frac12(z_5 - z_4),D_2=z_5)$ of the $\mathcal{A}_0$ K\"ahler cone generators to present our results.
Then, we find the K\"ahler cones of $\mathcal{A}_0$ and $\mathcal{A}_1$ are given by
\begin{align}
\mathcal{K}(\mathcal{A}_0)=\{x_0>0\,,~y_0>0\}\,,\qquad \mathcal{K}(\mathcal{A}_1)=\{x_1<0\,,~x_1+y_1>0\}\,,
\end{align}
and the intersection forms
$
\kappa_0=\kappa_1=9 x^2_0y_0 + 27 x_0 y^2_0 + 21 y^3_0 \,
$
on the CY hypersurfaces are, in fact, equal. The fact that the intersection form does not change across the ambient space flop locus illustrates that the CY K\"ahler cone $\mathcal{K}(X_0)$ is just given by the union of the K\"ahler cones $\mathcal{K}(\mathcal{A}_0)$ and $\mathcal{K}(\mathcal{A}_1)$. In particular, while there is an ambient space flop across $x_0=0$ the CY itself does not flop. The cone structure for this example is illustrated in Figure~\ref{fig:ConesExample2}. There is a single K\"ahler cone $\mathcal{K}(\mathcal{A}_0)\cup\mathcal{K}(\mathcal{A}_1)\cup \{x_0=0\,,~y_0>0\}$, bounded at $y_0=0$ by an effective cone wall and at $x_0+y_0=0$ by a type (b) Zariski wall.\\[2mm]
The transformation which converts the intersection form into the (rescaled) case 1 normal form is given by
\begin{equation}
 \left(\begin{array}{c}x_0\\y_0\end{array}\right)=P \left(\begin{array}{c}x\\y\end{array}\right)\;,\qquad
P=\sqrt[3]{2}\left(\begin{array}{rr}-2&-1\\1&1\end{array}\right)\,.
\end{equation}
We can fix the integration constants such that the ambient space flop is at $s=0$, which implies $k=\text{arccosh}(\sqrt{8/7})$ and the resulting solution is plotted in Fig.~\ref{fig:ConesExample2}. There is a high-speed escape/crash at the Zariski wall and the geodesic asymptotes to the effective cone wall which is at infinite geodesic distance. As before, non-isochore solutions can be obtained from Eq.~\eqref{eq:tsol}. 

\subsection{Example for case 2 - flop to a non-isomorphic CY}
\begin{figure}[t]
\centering
\hspace{50pt}\includegraphics[width=.5\textwidth]{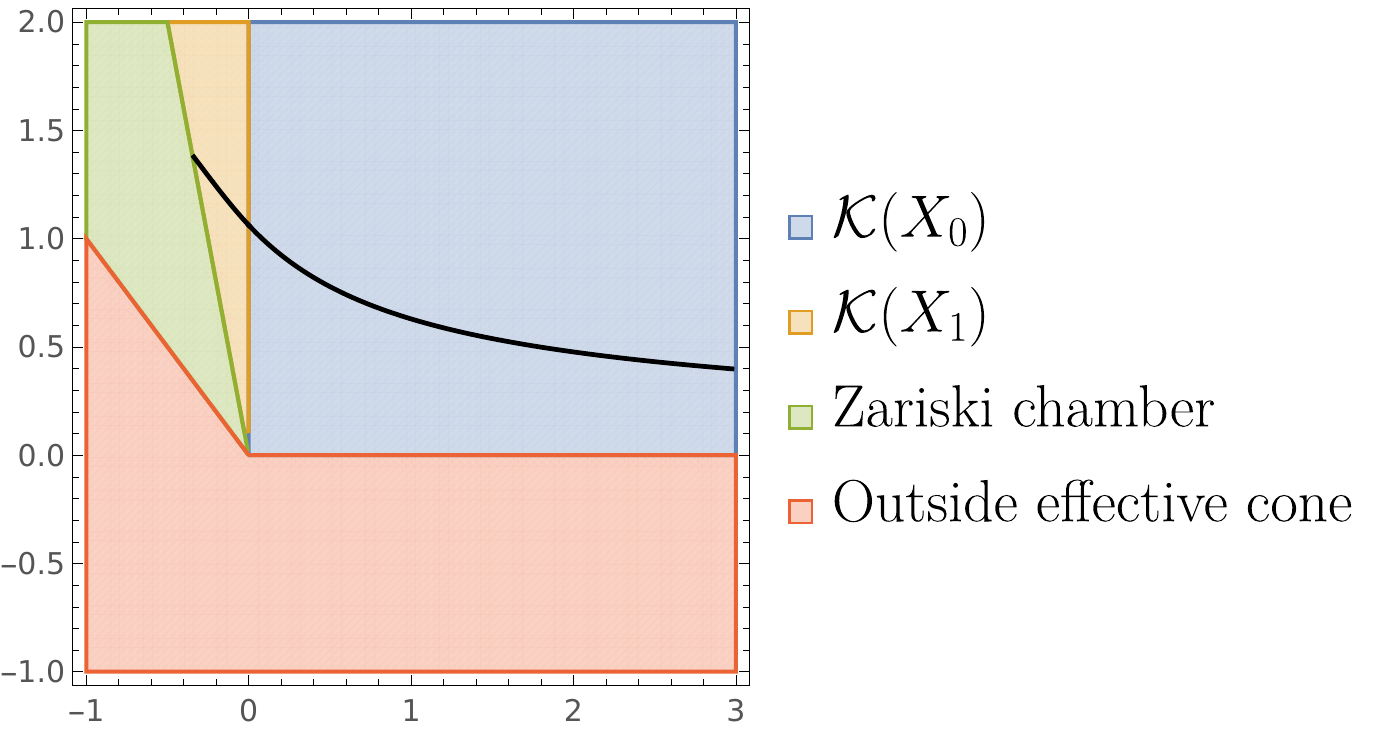}
\caption{Effective cone of CICY \#7885, showing K\"ahler cones of CYs $X_0$  and $X_1$, as well as an adjacent Zariski chamber, with a geodesic curve superimposed.}
\label{fig:ConesExample3}
\end{figure}

As an example for a flop between two non-isomorphic CYs with case 1 and case 2 intersection forms, we consider CICY \#7885 with configuration matrix
\begin{align}
X_0\sim\left[
\begin{array}{c|ccc}
\mathbbm{P}^1&1&1\\
\mathbbm{P}^4&4&1
\end{array}
\right]\,.
\end{align}
As a basis we use the divisors $(D_1,D_2)$ which generate the K\"ahler cone of $X_0$ and are dual to the restrictions $J_1,J_2$ of the standard ambient K\"ahler forms of the two projective space factors. The relevant data for this manifold is contained in the table in Appendix~\ref{appA} and has already been discussed in Section~\ref{sec:cicyex}. The K\"ahler cones of the CICY $X_0$ and the flopped CY $X_1$ are
\begin{align}
\mathcal{K}(X_0)=\{x_0>0\,,~y_0>0\}\,,\qquad \mathcal{K}(X_1)=\{x_1<0\,,~4x_1+y_1>0\}\,.
\end{align}
and they share a flop wall at $x_0=0$. The other boundaries of $\mathcal{K}(X_0)$ and $\mathcal{K}(X_1)$ are an effective cone wall and a type (b) Zariski wall, respectively, as shown in Figure~\ref{fig:ConesExample3}.

\noindent The two intersection forms
\begin{align}
\kappa_0=12x_0 y^2_0+5y^3_0\,,\qquad\kappa_1=-16x^3_1+12x_1 y^2_1+5 y^3_1\,,
\end{align}
are case 2 and case 1, respectively, and can be related to the corresponding normal forms in Table~\ref{tab:EquivalenceClasses} via the transformations
\begin{equation}
 \left(\begin{array}{c}x_i\\y_i\end{array}\right)=P_i \left(\begin{array}{c}x\\y\end{array}\right)\;,\qquad
P_0=\left(\begin{array}{cc}-\frac{5c}{12}&\frac{1}{12c^2}\\c&0\end{array}
\right)\,,\qquad
P_1=\frac{1}{\sqrt[3]{3}}\left(\begin{array}{rr}-\frac12&-\sqrt[3]{2}\\2&\sqrt[3]{2}\end{array}\right)\; ,
\end{equation}
where $c$ is a non-zero real constant. The choice of $c$ does not affect the physics and it can, in fact, be absorbed into the integration constant $k$ of the solution~\ref{eq:SolutionCase2}. Its choice also does not affect the image of the K\"ahler cone $\mathcal{K}(X_1)$ in normal form coordinates.
For the sake of being explicit, we set $c=12/5$. We use these transformations to convert the case 2 and case 1 normal form solutions in Eqs.~\eqref{eq:SolutionCase2} and \eqref{eq:SolutionCase1} back into the original coordinates.\footnote{We use here the case 1 solution in the fourth quadrant, since this allows for a more natural choice for the branch cuts used in Mathematica.} We fix integration constants such that the flop occurs at $s=0$ and require that $b^i$ and $\dot{b}^i$ are continuous across the flop. For the integration constants $\tilde{E}_0$, $k_0$ of the case 2 solution~\eqref{eq:SolutionCase2} and the integration constant $\tilde{E}_1$, $k_1$ of the case 1 solution~\eqref{eq:SolutionCase1} this leads to
\begin{align}
\tilde{E}_0=\sqrt{2\tilde{E}_1}\,,\qquad k_0=\frac12 \left(\frac56\right)^{2/3}\,,\qquad k_1=\frac12 \log\left(\frac53\right)\; .
\end{align}
The resulting geodesic is shown in Fig.~\ref{fig:ConesExample3}. Again, solutions along non-isochores can be obtained from Eq.~\eqref{eq:tsol}.

\subsection{Example for case 3 - infinitely many flops}
\label{sec:Example7863}
\begin{figure}[t]
\centering
\hspace{50pt}\includegraphics[width=.5\textwidth]{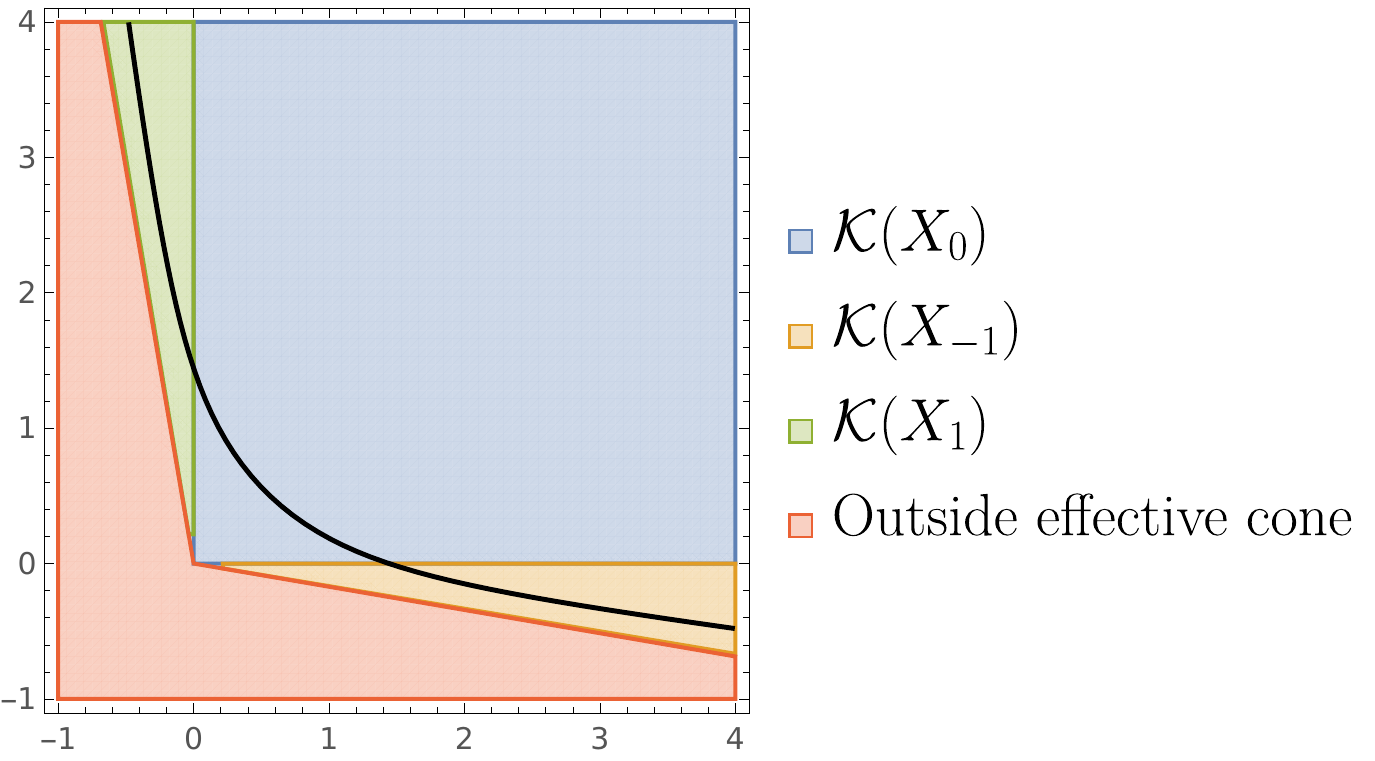}
\caption{Effective cone of CICY \#7863, showing K\"ahler cone of CY $X$, with infinitely many adjacent K\"ahler cones of isomorphic CYs (only two of which can be resolved), with a geodesic curve superimposed. The boundaries of the extended K\"ahler cone are generated by $(-1,3+2\sqrt{2})$ and $(3+2\sqrt{2},-1)$.}
\label{fig:ConesExample4}
\end{figure}

Let us finally discuss an example which combines a case 3 intersection form with an infinite flop chain. We choose CICY \#7863, which is given by the configuration matrix
\begin{align}
X\sim\left[
\begin{array}{c|ccc}
\mathbbm{P}^3&2&1&1\\
\mathbbm{P}^3&2&1&1
\end{array}
\right] \,,
\end{align}
with K\"ahler cone and intersection form given by
\begin{align}
\mathcal{K}(X_0) = \{x_0 > 0\,,~y_0 > 0\}\;, \qquad \kappa_0=2x_0^3+18x_0^2y_0+18x_0y_0^2+2y_0^3\; .
\end{align}
The space has an isomorphic flop wall at both boundaries, $y_0=0$ and $x_0=0$, leading to an infinite flop chain whose associated symmetry group is generated by
\begin{align}\label{eq:M1M2ex}
M_1=\left(\begin{array}{rr}-1&0\\6&1\end{array}\right)\,,\qquad M_2=\left(\begin{array}{cc}1&6\\0&-1\end{array}\right)\,.
\end{align}
The cone structure is plotted in Fig.~\ref{fig:ConesExample4}. There are infinitely many K\"ahler cones on either side of the primary cone, as schematically indicated in Fig.~\ref{fig:kahlercones} on the right, but those cones become so narrow that only one on each side can be resolved in Fig.~\ref{fig:ConesExample4}. The boundaries of this infinite flop sequence are along $(-1,3+2\sqrt{2})$ and $(3+2\sqrt{2},-1)$ and these are the generators of the effective cone. 

\noindent There are several choices for the map which converts the intersection form $\kappa_0$ into the case 3 normal form and for concreteness we will work with
\begin{equation}
\left(\begin{array}{c}x_0\\y_0\end{array}\right)=P\left(\begin{array}{c}x\\y\end{array}\right)\;,\qquad
P=\frac{1}{c_+^2-c_-^2}\left(
\begin{array}{rr}
c_+&-c_-\\
-c_-&c_+
\end{array}
\right)\,,
\qquad
P^{-1}=\left(
\begin{array}{cc}
c_+&c_-\\
c_-&c_+
\end{array}
\right)\; ,
\end{equation}
where
\begin{equation}
c_{\pm}=\left(7\pm9\sqrt{\frac35}\right)^{1/3}\simeq\left\{\begin{array}{ll}1.67&~~+\\0.21&~~-\end{array}\right.\,.
\end{equation}
The image $\hat{\mathcal{K}}(X_0)$ of the K\"ahler cone under the map $P^{-1}$ lies within the positive quadrant in Fig.~\ref{fig:Geodesics3Cases} on the right. However, since the boundaries of $\hat{\mathcal{K}}(X_0)$ are flop walls this image cannot possibly take up the entire positive quadrant. In fact, the generators of $\hat{\mathcal{K}}(X_0)$ are $(c_+,c_-)$ and $(c_-,c_+)$. To obtain the geodesics we proceed as before and map the case 3 solution~\eqref{eq:SolutionCase3} to the $(x_0,y_0)$ coordinates using $P$. For our choice of $P$, the solution $(x,y_+)$ will be mapped into the CY K\"ahler cone.  The resulting geodesic is plotted in Figure~\ref{fig:ConesExample4}.\\[2mm]
The geodesic length between the two flops we can worked out in either the normal form coordinates $(x,y)$ or the CY coordinates $(x_0,y_0)$. Using the former, we find that the intersections of the geodesic with the flop lines happens when 
\begin{align}
x(s_\pm)/y(s_\pm) = c_\pm/c_\mp\qquad\Leftrightarrow\qquad \varepsilon s_\pm(x)\simeq\pm0.62 \,.
\end{align}
This means the geodesic length between the two flops along the geodesic is
\begin{align}
\Delta\tau=2\sqrt{12}\, \varepsilon s_+\simeq 4.30
\end{align}
in 5D Planck units. More generally, non-isochore geodesics can be obtained from Eq.~\eqref{eq:tsol} and the geodesic distance for such solutions, with volume expansion rate $\alpha$ is
\begin{equation}
 \Delta\tau= \sqrt{1+\frac{\alpha^2}{12\varepsilon^2}}\;2\sqrt{12}\varepsilon s_+\simeq  \sqrt{1+\frac{\alpha^2}{12\varepsilon^2}}\; 4.30 \,.
\end{equation} 
This provides us with the solution in the primary K\"ahler cone $X_0$. The solution in all other K\"ahler cones is obtained\footnote{We note that, as described in Ref.~\cite{Mohaupt2001}, for the distinct case of a type (a) Zariski wall, at which an SU$(2)$ gauge theory appears, there is also a notion of continuing the geodesic evolution into the K\"ahler cone of an isomorphic CY, based on the existence of an elementary transformation, which is discussed in Ref.~\cite{Katz:1996ht}.} by acting with the elements of the symmetry group $G\cong\mathbb{Z}_2\ltimes\mathbb{Z}$, generated by the matrices in Eq.~\eqref{eq:M1M2ex}. In this way we obtain the geodesic across the entire infinite sequence of isomorphic flops. Equivalently, we can describe the situation in the downstairs picture where we use the primary cone $\mathcal{K}(X_0)$ only. In this case, we should think of the isochore geodesic as ``oscillating" along the same trajectory between the two flop-boundaries. Either way, if $k$ K\"ahler cones have been crossed the corresponding isochore geodesic distance is of course $\simeq 4.30 k $ so the geodesic length is unbounded. On the other hand, the physical theory remains unchanged from cone to cone since all CYs involved are isomorphic. The apparent contradiction with the distance conjecture is resolved because $G$ is a gauge symmetry and has to be divided out. Hence, all K\"ahler cones should be identified so that the shortest geodesic between a point and its equivalent in another cone is not, in fact, the above geodesic but the constant geodesic.

For all our examples of infinite flop sequences all CYs which appear are isomorphic. Even if a finite number of non-isomorphic CYs were involved there does not appear to be a conflict with the distance conjecture. However, an infinite flop sequence with an infinite number of non-isomorphic CYs is likely to cause a problem~\footnote{This could perhaps be avoided if the intersection numbers conspire to keep the geodesic distance finite. However, this seems to require a somewhat contrived  ``fine-tuning" of intersection numbers.}. However, such examples would be excluded by the Kawamata-Morrison conjecture, as discussed in more detail in Ref.~\cite{Brodie:2021ain}.

\section{Conclusion}\label{sec:conclusion}
In this paper, we have studied the K\"ahler moduli space of CY threefolds in relation to geodesic motion, focusing on $h^{1,1}(X)=2$ manifolds and compactifications of M-theory to 5d supergravity.\\[2mm]
We have presented a classification of intersection forms for $h^{1,1}(X)=2$ CY threefolds which shows that they fall into three different cases (see Table~\ref{tab:EquivalenceClasses}). These cases constrain the possible wall types of the K\"ahler cone. For example, type (b) Zariski walls (where divisors collapse to a point) can only arise for case 1 intersection forms and even then for at most one of the two K\"ahler cone boundaries.\\[2mm]
We have compiled detailed data on the K\"ahler moduli space structure for all $h^{1,1}(X)=2$ CICY and THCY manifolds. This should be a valuable resource for studying topology change in string theory. While the manifolds under consideration are relatively simple and easily constructed they show a remarkably rich structure of walls and cones. For some of these examples, the flops arise only on the CY itself but not on the ambient space which makes them different from the cases traditionally studied in string theory. Isomorphic flops and infinite flop sequences are much more common than perhaps naively expected. The data also shows that all three cases of intersection form are realised by actual CY manifolds, although CICY manifolds only realise two of the cases.\\[2mm]
Using the intersection normal forms (Table~\ref{tab:EquivalenceClasses}) we can find explicit solutions for the geodesics of all $h^{1,1}(X)=2$ CYs. These can be used to study geodesic motion near walls and across flop transitions. Effective cone walls are identified by a diverging moduli space metric and they are located at infinite geodesic distance. However, geodesic evolution, as described by the simple geodesic equation, does not distinguish between flops and type (a) Zariski walls - both are associated with a non-divergent and non-singular metric and are located at finite geodesic distance. Formally, the geodesic can be continued beyond the wall in either case, however, only in the flop case does this correspond to a physical evolution which takes place within the adjacent K\"ahler cone of a birationally equivalent CY manifold. Type (a) Zariski walls (where a divisor collapses to a curve) mark the end of the moduli space and the inclusion of states becoming light at the wall (in the form of an SU$(2)$ gauge theory) is expected to curtail the evolution. The choice of complex structure can switch between flop and type (a) Zariski walls and we presented an explicit example for this phenomenon. It is, therefore, expected that simple geodesic motion in K\"ahler moduli space does not distinguish between those two cases. Finally, type (b) Zariski walls (where a divisor collapses to a point) are associated to non-divergent but singular metrics. They are reached within finite geodesic distance but the effective theory breaks down close to the wall due to diverging modular velocities. As for type (a) Zariski walls, the inclusion of additional light states (indeed an infinite tower of light states) near the wall is required for a correct low-energy description.\\[2mm]
We have constructed geodesics across flop walls for both isomorphic and non-isomorphic flops as well as for infinite flop sequences. In the latter case this leads to infinite length geodesics between equivalent theories. A contradiction with the distance conjecture is avoided thanks to an infinite discrete gauge symmetry which identifies all K\"ahler cones of the infinite flop sequence. Flop sequences which contain an infinite number of inequivalent CYs would likely be a problem for the distance conjecture but they are excluded provided the Kawamata-Morrison conjecture holds.\\[2mm]
There are various extensions of the present work. First, we expect many of our results to generalise to CYs with $h^{1,1}(X)>2$. Preliminary investigation shows that complicated cone structures, isomorphic flops and infinite flop sequences are common features among CYs with $h^{1,1}(X)=3,4$. It would also be desirable to include the additional light states which appear at K\"ahler cone walls into the discussion of geodesics. This has been accomplished, to some extent, for flops~\cite{Brandle:2002fa,Jarv:2003qx,Jarv:2003qy} where the additional light hypermultiplets can be shown not to significantly affect the dynamics, at least under certain plausible assumptions. However, as we have argued, the additional light states must have a non-trivial effect on the geodesics near Zariski walls and it would be interesting to study this. Finally, it is desirable to have explicit constructions for the manifolds obtained from CICYs and THCYs via flops and work in this direction is currently underway~\cite{CICYFlops}.

\section*{Acknowledgments}
The work of CRB is supported by the John Templeton Foundation grant 61149. AC would like to thank EPSRC for grant EP/T016280/1. The authors would like to thank Antonella Grassi for useful comments and for pointing us into the direction of the Kawamata-Morrison conjecture.

\appendix
\clearpage
\section{Picard number 2 CICYs and their effective cone structure}\label{appA}
The table below lists the $36$ complete intersection CY manifolds (CICYs) with $h^{1,1}(X)=2$ and the structure of their effective cones. The meaning of the various entries is as follows. The first column contains the CICY number, which indicates its position in the standard list of Ref.~\cite{Candelas:1987kf}. Column two provides the configuration matrix
\begin{equation}
 \left(\begin{array}{cccc}q_1^1&q_2^1&\cdots&q_K^1\\q_1^2&q_2^2&\cdots&q_K^2\end{array}\right)^{h^{1,1}(X),\;h^{2,1}(X)}_{\chi(X)}
\end{equation}
where each column of this matrix specifies the bi-degree of a defining polynomial in the ambient space $\mathcal{A}=\mathbb{P}^{d_1}\times\mathbb{P}^{d_2}$ where $d_i=\sum_{a=1}^Kq_a^i-1$, and where the Hodge numbers and the Euler number $\chi(X)$ are included as superscripts and subscript respectively. The standard K\"ahler forms of the two projective space factors, restricted to $X$, are denoted by $J_1$ and $J_2$, with Poincar\'e dual divisors $D_1$ and $D_2$. All generators in the table are given relative to the basis $(D_1,D_2)$.\\[2mm]
Column three lists the four intersection numbers of $X$ and column four indicates the case of the intersection form on $X$, as defined in Table~\ref{fig:Geodesics3Cases}. Column five provides the two generators $v_1=(v_{11},v_{12})^T$ and $v_2=(v_{21},v_{22})^T$ of the effective cone. When these generators are irrational $X$ has an infinite flop sequence.\\[2mm]
The matrices in the last column summarise the information on the cone structure of $X$. Each numerical row in the centre of the matrix is the generator of a cone boundary. Note that the generators $(0,1)$, $(1,0)$ always appear as the generators of the K\"ahler cone of $X$. In case of an infinite flop sequence, only the boundaries of the primary cone and the two adjacent cones are listed. The nature of each boundary is indicated to the right of the generator, with $E$ for an effective cone wall, $Z$ for a Zariski wall, $F_{n_1,n_2}$ for a non-isomorphic flop wall, and $I_{n_1,n_2}^m$ for an isomorphic flop wall. Here $n_1,n_2$ are the Gromov-Witten invariants which determine the change in intersection numbers, as explained in Section~\ref{sec:flops2}. Further, $m$ determines the generator of the involution for an isomorphic flop and equals $m_1$ in Eq.~\eqref{eq:M1} if the boundary is above $(1,0)$ and $m_2$ in Eq.~\eqref{eq:M2} if it is below $(0,1)$. Finally, to the left of the generators the nature of the cones is indicated, with $Z$ for a Zariski cone and $K_c$ for a K\"ahler cone with a case $c$ intersection form.\\[2mm]
The information on effective cones and the nature of cones and cone boundaries in the table has been determined for generic complex structure by studying line bundle cohomology formulae, along the lines of Ref.~\cite{Brodie:2020fiq} and the earlier Refs.~\cite{Constantin:2018hvl, Klaewer:2018sfl, Larfors:2019sie, Brodie:2019pnz, Brodie:2019dfx, Brodie:2019ozt, Brodie:2020wkd}. Line bundle cohomology has been computed using the CICY package \cite{cicypackage} (see also the pyCICY package \cite{cicytoolkit}).
For special choices of complex structure new effective divisors can appear, altering the structure of the effective cone. For two manifolds we were not able to obtain enough cohomology data to identify the full structure of the effective cone, and so we have left the corresponding entries empty.

{\scriptsize
\begin{longtable}{|c|c|c|c|c|c|}\hline
\#&configuration&$\left(\begin{array}{ll}d_{122}&d_{222}\\d_{112}&d_{111}\end{array}\right)$&case&
$\left(\begin{array}{ll}v_{11}&v_{12}\\v_{21}&v_{22}\end{array}\right)$&cones\\\hline\hline
7643&$\left(
\begin{array}{cccc}
 0 & 0 & 2 & 1 \\
 2 & 2 & 1 & 1 \\
\end{array}
\right)_{-88}^{2,46}$&$\left(
\begin{array}{cc}
 12 & 8 \\
 4 & 0 \\
\end{array}
\right)$&3&$\left(
\begin{array}{cc}
 -1 & 3 \\
 1 & 0 \\
\end{array}
\right)$&$\left(\begin{array}{l}K_3\\K_3\end{array}
\begin{array}{|rr|}
 -1 & 3 \\
 0 & 1 \\
 1 & 0 \\
\end{array}
\begin{array}{l}E\\I_{40,4}^3\\E\end{array}\right)$\\\hline
7644&$\left(
\begin{array}{ccccc}
 2 & 0 & 1 & 1 & 1 \\
 0 & 2 & 1 & 1 & 1 \\
\end{array}
\right)_{-88}^{2,46}$&$\left(
\begin{array}{cc}
 12 & 4 \\
 12 & 4 \\
\end{array}
\right)$&3&$\left(
\begin{array}{cc}
 -1 & 3+2 \sqrt{2} \\
 3+2 \sqrt{2} & -1 \\
\end{array}
\right)$&$\left(\begin{array}{l}~\vdots\\K_3\\K_3\\K_3\\~\vdots\end{array}
\begin{array}{|rr|}
 -1 & 6 \\
 0 & 1 \\
 1 & 0 \\
 6 & -1 \\
\end{array}
\begin{array}{l}~\vdots\\I_{64,20}^6\\I_{64,20}^6\\ ~\vdots\end{array}\right)$\\\hline
7668&$\left(
\begin{array}{ccc}
 0 & 2 & 1 \\
 3 & 1 & 1 \\
\end{array}
\right)_{-90}^{2,47}$&$\left(
\begin{array}{cc}
 9 & 6 \\
 3 & 0 \\
\end{array}
\right)$&3&$\left(
\begin{array}{cc}
 -1 & 3 \\
 1 & 0 \\
\end{array}
\right)$&$\left(\begin{array}{l}K_3\\K_3\end{array}
\begin{array}{|rr|}
 -1 & 3 \\
 0 & 1 \\
 1 & 0 \\
\end{array}
\begin{array}{l}E\\I_{30,3}^3\\E\end{array}\right)$\\\hline
7725&$\left(
\begin{array}{ccccc}
 0 & 0 & 1 & 1 & 1 \\
 2 & 2 & 1 & 1 & 1 \\
\end{array}
\right)_{-96}^{2,50}$&$\left(
\begin{array}{cc}
 12 & 12 \\
 4 & 0 \\
\end{array}
\right)$&3&$\left(
\begin{array}{cc}
 -1 & 2 \\
 1 & 0 \\
\end{array}
\right)$&$\left(\begin{array}{l}K_3\\K_3\end{array}
\begin{array}{|rr|}
 -1 & 2 \\
 0 & 1 \\
 1 & 0 \\
\end{array}
\begin{array}{l}E\\I_{24,0}^2\\E\end{array}\right)$\\\hline
7726&$\left(
\begin{array}{ccccc}
 0 & 1 & 1 & 1 & 1 \\
 2 & 1 & 1 & 1 & 1 \\
\end{array}
\right)_{-96}^{2,50}$&$\left(
\begin{array}{cc}
 12 & 8 \\
 8 & 2 \\
\end{array}
\right)$&3&$\left(
\begin{array}{cc}
 -1 & \frac{3}{2}+\frac{\sqrt{\frac{15}{2}}}{2} \\
 4+2 \sqrt{\frac{10}{3}} & -1 \\
\end{array}
\right)$&$\left(\begin{array}{l}~\vdots\\K_3\\K_3\\K_3\\~\vdots\end{array}
\begin{array}{|rr|}
 -1 & 3 \\
 0 & 1 \\
 1 & 0 \\
 8 & -1 \\
\end{array}
\begin{array}{l}~\vdots\\I_{40,0}^3\\I_{80,0}^8\\~\vdots\end{array}\right)$\\\hline
7758&$\left(
\begin{array}{ccc}
 0 & 2 & 1 \\
 2 & 1 & 2 \\
\end{array}
\right)_{-100}^{2,52}$&$\left(
\begin{array}{cc}
 10 & 4 \\
 4 & 0 \\
\end{array}
\right)$&3&$\left(
\begin{array}{cc}
 -1 & 5 \\
 1 & 0 \\
\end{array}
\right)$&$\left(\begin{array}{l}K_3\\K_3\end{array}
\begin{array}{|rr|}
 -1 & 5 \\
 0 & 1 \\
 1 & 0 \\
\end{array}
\begin{array}{l}E\\I_{62,16}^5\\E\end{array}\right)$\\\hline
7759&$\left(
\begin{array}{cccc}
 0 & 2 & 1 & 1 \\
 2 & 1 & 1 & 1 \\
\end{array}
\right)_{-100}^{2,52}$&$\left(
\begin{array}{cc}
 10 & 4 \\
 8 & 2 \\
\end{array}
\right)$&3&$\left(
\begin{array}{cc}
 -1 & \frac{5}{2}+\frac{3 \sqrt{\frac{5}{2}}}{2} \\
 4+6 \sqrt{\frac{2}{5}} & -1 \\
\end{array}
\right)$&$\left(\begin{array}{l}~\vdots\\K_3\\K_3\\K_3\\~\vdots\end{array}
\begin{array}{|rr|}
 -1 & 5 \\
 0 & 1 \\
 1 & 0 \\
 8 & -1 \\
\end{array}
\begin{array}{l}~\vdots\\I_{70,8}^5\\I_{72,26}^8\\~\vdots\end{array}\right)$\\\hline
7761&$\left(
\begin{array}{ccccc}
 1 & 1 & 1 & 1 & 1 \\
 1 & 1 & 1 & 1 & 1 \\
\end{array}
\right)_{-100}^{2,52}$&$\left(
\begin{array}{cc}
 10 & 5 \\
 10 & 5 \\
\end{array}
\right)$&3&$\left(
\begin{array}{cc}
 -1 & 2+\sqrt{3} \\
 2+\sqrt{3} & -1 \\
\end{array}
\right)$&$\left(\begin{array}{l}~\vdots\\K_3\\K_3\\K_3\\~\vdots\end{array}
\begin{array}{|rr|}
 -1 & 4 \\
 0 & 1 \\
 1 & 0 \\
 4 & -1 \\
\end{array}
\begin{array}{l}~\vdots\\I_{50,0}^4\\I_{50,0}^4\\~\vdots\end{array}\right)$\\\hline
7799&$\left(
\begin{array}{ccc}
 2 & 1 & 1 \\
 1 & 2 & 1 \\
\end{array}
\right)_{-106}^{2,55}$&$\left(
\begin{array}{cc}
 7 & 2 \\
 7 & 2 \\
\end{array}
\right)$&3&$\left(
\begin{array}{cc}
 -1 & \frac{7}{2}+\frac{3 \sqrt{5}}{2} \\
 \frac{7}{2}+\frac{3 \sqrt{5}}{2} & -1 \\
\end{array}
\right)$&$\left(\begin{array}{l}~\vdots\\K_3\\K_3\\K_3\\~\vdots\end{array}
\begin{array}{|rr|}
 -1 & 7 \\
 0 & 1 \\
 1 & 0 \\
 7 & -1 \\
\end{array}
\begin{array}{l}~\vdots\\I_{80,15}^7\\I_{80,15}^7\\~\vdots\end{array}\right)$\\\hline
7806&$\left(
\begin{array}{cc}
 0 & 2 \\
 3 & 2 \\
\end{array}
\right)_{-108}^{2,56}$&$\left(
\begin{array}{cc}
 6 & 6 \\
 0 & 0 \\
\end{array}
\right)$&2&$\left(
\begin{array}{cc}
 -1 & 2 \\
 1 & 0 \\
\end{array}
\right)$&$\left(\begin{array}{l}K_2\\K_2\end{array}
\begin{array}{|rr|}
 -1 & 2 \\
 0 & 1 \\
 1 & 0 \\
\end{array}
\begin{array}{l}E\\I_{24,0}^2\\E\end{array}\right)$\\\hline
7807&$\left(
\begin{array}{cccc}
 0 & 1 & 1 & 1 \\
 2 & 2 & 1 & 1 \\
\end{array}
\right)_{-108}^{2,56}$&$\left(
\begin{array}{cc}
 10 & 8 \\
 4 & 0 \\
\end{array}
\right)$&3&$\left(
\begin{array}{cc}
 -1 & 2 \\
 1 & 0 \\
\end{array}
\right)$&$\left(\begin{array}{l}Z\\K_1\\K_3\end{array}
\begin{array}{|rr|}
 -1 & 2 \\
 -1 & 3 \\
 0 & 1 \\
 1 & 0 \\
\end{array}
\begin{array}{l}E\\Z_a\\F_{34,0}\\E\end{array}\right)$\\\hline
7808&$\left(
\begin{array}{cccc}
 0 & 1 & 1 & 1 \\
 3 & 1 & 1 & 1 \\
\end{array}
\right)_{-108}^{2,56}$&$\left(
\begin{array}{cc}
 9 & 9 \\
 3 & 0 \\
\end{array}
\right)$&3&$\left(
\begin{array}{cc}
 -1 & 2 \\
 1 & 0 \\
\end{array}
\right)$&$\left(\begin{array}{l}K_3\\K_3\end{array}
\begin{array}{|rr|}
 -1 & 2 \\
 0 & 1 \\
 1 & 0 \\
\end{array}
\begin{array}{l}E\\I_{18,0}^2\\E\end{array}\right)$\\\hline
\rule{0pt}{4.5ex}$7809^*$\rule{0pt}{4.5ex}&$\left(
\begin{array}{cccc}
 1 & 1 & 1 & 1 \\
 2 & 1 & 1 & 1 \\
\end{array}
\right)_{-108}^{2,56}$&$\left(
\begin{array}{cc}
 9 & 5 \\
 7 & 2 \\
\end{array}
\right)$&3&$\left(
\begin{array}{cc}
 -1 & 3 \\
 20 & -3 \\
\end{array}
\right)$&
$\left(\begin{array}{l}Z\\?\\K_3\\K_3\\?\\Z\end{array}
\begin{array}{|rr|}
 -1 & 3 \\
? & ?  \\
 0 & 1 \\
 1 & 0 \\
 7 & -1 \\
? &  ? \\
 20 & -3 \\
\end{array}
\begin{array}{l}E\\?\\F_{46,0}\\I_{84,10}^7\\F_{46,0}\\?\\E\end{array}\right)$
\\\hline
7816&$\left(
\begin{array}{ccc}
 0 & 1 & 1 \\
 2 & 2 & 2 \\
\end{array}
\right)_{-112}^{2,58}$&$\left(
\begin{array}{cc}
 8 & 8 \\
 0 & 0 \\
\end{array}
\right)$&2&$\left(
\begin{array}{cc}
 -1 & 2 \\
 1 & 0 \\
\end{array}
\right)$&$\left(\begin{array}{l}K_2\\K_2\end{array}
\begin{array}{|rr|}
 -1 & 2 \\
 0 & 1 \\
 1 & 0 \\
\end{array}
\begin{array}{l}E\\I_{32,0}^2\\E\end{array}\right)$\\\hline
7817&$\left(
\begin{array}{cccc}
 0 & 0 & 1 & 1 \\
 2 & 2 & 2 & 1 \\
\end{array}
\right)_{-112}^{2,58}$&$\left(
\begin{array}{cc}
 8 & 12 \\
 0 & 0 \\
\end{array}
\right)$&2&$\left(
\begin{array}{cc}
 -1 & 1 \\
 1 & 0 \\
\end{array}
\right)$&$\left(\begin{array}{l}Z\\K_1\\K_2\end{array}
\begin{array}{|rr|}
 -1 & 1 \\
 -1 & 2 \\
 0 & 1 \\
 1 & 0 \\
\end{array}
\begin{array}{l}E\\Z_b\\F_{16,0}\\E\end{array}\right)$\\\hline
7819&$\left(
\begin{array}{ccccc}
 0 & 0 & 0 & 1 & 1 \\
 2 & 2 & 2 & 1 & 1 \\
\end{array}
\right)_{-112}^{2,58}$&$\left(
\begin{array}{cc}
 8 & 16 \\
 0 & 0 \\
\end{array}
\right)$&2&$\left(
\begin{array}{cc}
 -1 & 1 \\
 1 & 0 \\
\end{array}
\right)$&$\left(\begin{array}{l}K_2\\K_2\end{array}
\begin{array}{|rr|}
 -1 & 1 \\
 0 & 1 \\
 1 & 0 \\
\end{array}
\begin{array}{l}E\\I_{8,0}^1\\E\end{array}\right)$\\\hline
\rule{0pt}{4.5ex}\!7821$^{**}$\!\rule{0pt}{4.5ex}&$\left(
\begin{array}{ccc}
 1 & 1 & 1 \\
 2 & 2 & 1 \\
\end{array}
\right)_{-112}^{2,58}$&$\left(
\begin{array}{cc}
 8 & 5 \\
 4 & 0 \\
\end{array}
\right)$&3&&\\\hline
7822&$\left(
\begin{array}{ccc}
 0 & 0 & 2 \\
 2 & 2 & 2 \\
\end{array}
\right)_{-112}^{2,58}$&$\left(
\begin{array}{cc}
 8 & 8 \\
 0 & 0 \\
\end{array}
\right)$&2&$\left(
\begin{array}{cc}
 -1 & 2 \\
 1 & 0 \\
\end{array}
\right)$&$\left(\begin{array}{l}K_2\\K_2\end{array}
\begin{array}{|rr|}
 -1 & 2 \\
 0 & 1 \\
 1 & 0 \\
\end{array}
\begin{array}{l}E\\I_{32,0}^2\\E\end{array}\right)$\\\hline
7823&$\left(
\begin{array}{cccc}
 0 & 0 & 0 & 2 \\
 2 & 2 & 2 & 1 \\
\end{array}
\right)_{-112}^{2,58}$&$\left(
\begin{array}{cc}
 8 & 16 \\
 0 & 0 \\
\end{array}
\right)$&2&$\left(
\begin{array}{cc}
 -1 & 1 \\
 1 & 0 \\
\end{array}
\right)$&$\left(\begin{array}{l}K_2\\K_2\end{array}
\begin{array}{|rr|}
 -1 & 1 \\
 0 & 1 \\
 1 & 0 \\
\end{array}
\begin{array}{l}E\\I_{8,0}^1\\E\end{array}\right)$\\\hline
7833&$\left(
\begin{array}{cc}
 2 & 1 \\
 1 & 3 \\
\end{array}
\right)_{-114}^{2,59}$&$\left(
\begin{array}{cc}
 7 & 2 \\
 3 & 0 \\
\end{array}
\right)$&3&$\left(
\begin{array}{cc}
 -1 & 7 \\
 1 & 0 \\
\end{array}
\right)$&$\left(\begin{array}{l}K_3\\K_3\end{array}
\begin{array}{|rr|}
 -1 & 7 \\
 0 & 1 \\
 1 & 0 \\
\end{array}
\begin{array}{l}E\\I_{64,27}^7\\E\end{array}\right)$\\\hline
7840&$\left(
\begin{array}{ccc}
 0 & 1 & 1 \\
 3 & 2 & 1 \\
\end{array}
\right)_{-120}^{2,62}$&$\left(
\begin{array}{cc}
 6 & 9 \\
 0 & 0 \\
\end{array}
\right)$&2&$\left(
\begin{array}{cc}
 -1 & 1 \\
 1 & 0 \\
\end{array}
\right)$&$\left(\begin{array}{l}Z\\K_1\\K_2\end{array}
\begin{array}{|rr|}
 -1 & 1 \\
 -1 & 2 \\
 0 & 1 \\
 1 & 0 \\
\end{array}
\begin{array}{l}E\\Z_b\\F_{12,0}\\E\end{array}\right)$\\\hline
7844&$\left(
\begin{array}{cc}
 2 & 1 \\
 2 & 2 \\
\end{array}
\right)_{-120}^{2,62}$&$\left(
\begin{array}{cc}
 6 & 2 \\
 4 & 0 \\
\end{array}
\right)$&3&$\left(
\begin{array}{cc}
 -1 & 6 \\
 1 & 0 \\
\end{array}
\right)$&$\left(\begin{array}{l}K_3\\K_3\end{array}
\begin{array}{|rr|}
 -1 & 6 \\
 0 & 1 \\
 1 & 0 \\
\end{array}
\begin{array}{l}E\\I_{80,8}^6\\E\end{array}\right)$\\\hline
7853&$\left(
\begin{array}{ccc}
 0 & 2 & 1 \\
 2 & 2 & 1 \\
\end{array}
\right)_{-124}^{2,64}$&$\left(
\begin{array}{cc}
 8 & 4 \\
 4 & 0 \\
\end{array}
\right)$&3&$\left(
\begin{array}{cc}
 -1 & 4 \\
 1 & 0 \\
\end{array}
\right)$&$\left(\begin{array}{l}K_3\\K_3\end{array}
\begin{array}{|rr|}
 -1 & 4 \\
 0 & 1 \\
 1 & 0 \\
\end{array}
\begin{array}{l}E\\I_{64,2}^4\\E\end{array}\right)$\\\hline
7858&$\left(
\begin{array}{cc}
 1 & 1 \\
 3 & 2 \\
\end{array}
\right)_{-128}^{2,66}$&$\left(
\begin{array}{cc}
 6 & 5 \\
 0 & 0 \\
\end{array}
\right)$&2&$\left(
\begin{array}{cc}
 -1 & 2 \\
 1 & 0 \\
\end{array}
\right)$&$\left(\begin{array}{l}Z\\K_1\\K_2\end{array}
\begin{array}{|rr|}
 -1 & 2 \\
 -1 & 3 \\
 0 & 1 \\
 1 & 0 \\
\end{array}
\begin{array}{l}E\\Z_b\\F_{36,0}\\E\end{array}\right)$\\\hline
7863&$\left(
\begin{array}{ccc}
 2 & 1 & 1 \\
 2 & 1 & 1 \\
\end{array}
\right)_{-128}^{2,66}$&$\left(
\begin{array}{cc}
 6 & 2 \\
 6 & 2 \\
\end{array}
\right)$&3&$\left(
\begin{array}{cc}
 -1 & 3+2 \sqrt{2} \\
 3+2 \sqrt{2} & -1 \\
\end{array}
\right)$&$\left(\begin{array}{l}~\vdots\\K_3\\K_3\\K_3\\~\vdots\end{array}
\begin{array}{|rr|}
 -1 & 6 \\
 0 & 1 \\
 1 & 0 \\
 6 & -1 \\
\end{array}
\begin{array}{l}~\vdots\\I_{80,4}^6\\I_{80,4}^6\\~\vdots\end{array}\right)$\\\hline
7867&$\left(
\begin{array}{cccc}
 0 & 0 & 1 & 1 \\
 3 & 2 & 1 & 1 \\
\end{array}
\right)_{-132}^{2,68}$&$\left(
\begin{array}{cc}
 6 & 12 \\
 0 & 0 \\
\end{array}
\right)$&2&$\left(
\begin{array}{cc}
 -1 & 1 \\
 1 & 0 \\
\end{array}
\right)$&$\left(\begin{array}{l}K_2\\K_2\end{array}
\begin{array}{|rr|}
 -1 & 1 \\
 0 & 1 \\
 1 & 0 \\
\end{array}
\begin{array}{l}E\\I_{6,0}^1\\E\end{array}\right)$\\\hline
7868&$\left(
\begin{array}{ccc}
 1 & 1 & 1 \\
 3 & 1 & 1 \\
\end{array}
\right)_{-132}^{2,68}$&$\left(
\begin{array}{cc}
 7 & 5 \\
 3 & 0 \\
\end{array}
\right)$&3&$\left(
\begin{array}{cc}
 -1 & 2 \\
 1 & 0 \\
\end{array}
\right)$&$\left(\begin{array}{l}Z\\K_1\\K_3\end{array}
\begin{array}{|rr|}
 -1 & 2 \\
 -1 & 4 \\
 0 & 1 \\
 1 & 0 \\
\end{array}
\begin{array}{l}E\\Z_a\\F_{34,0}\\E\end{array}\right)$\\\hline
7869&$\left(
\begin{array}{ccc}
 0 & 0 & 2 \\
 3 & 2 & 1 \\
\end{array}
\right)_{-132}^{2,68}$&$\left(
\begin{array}{cc}
 6 & 12 \\
 0 & 0 \\
\end{array}
\right)$&2&$\left(
\begin{array}{cc}
 -1 & 1 \\
 1 & 0 \\
\end{array}
\right)$&$\left(\begin{array}{l}K_2\\K_2\end{array}
\begin{array}{|rr|}
 -1 & 1 \\
 0 & 1 \\
 1 & 0 \\
\end{array}
\begin{array}{l}E\\I_{6,0}^1\\E\end{array}\right)$\\\hline
7873&$\left(
\begin{array}{ccc}
 0 & 1 & 1 \\
 2 & 3 & 1 \\
\end{array}
\right)_{-140}^{2,72}$&$\left(
\begin{array}{cc}
 6 & 8 \\
 0 & 0 \\
\end{array}
\right)$&2&$\left(
\begin{array}{cc}
 -1 & 1 \\
 1 & 0 \\
\end{array}
\right)$&$\left(\begin{array}{l}Z\\K_1\\K_2\end{array}
\begin{array}{|rr|}
 -1 & 1 \\
 -1 & 3 \\
 0 & 1 \\
 1 & 0 \\
\end{array}
\begin{array}{l}E\\Z_b\\F_{18,0}\\E\end{array}\right)$\\\hline
7882&$\left(
\begin{array}{cc}
 0 & 2 \\
 2 & 3 \\
\end{array}
\right)_{-148}^{2,76}$&$\left(
\begin{array}{cc}
 6 & 4 \\
 0 & 0 \\
\end{array}
\right)$&2&$\left(
\begin{array}{cc}
 -1 & 3 \\
 1 & 0 \\
\end{array}
\right)$&$\left(\begin{array}{l}K_2\\K_2\end{array}
\begin{array}{|rr|}
 -1 & 3 \\
 0 & 1 \\
 1 & 0 \\
\end{array}
\begin{array}{l}E\\I_{54,0}^3\\E\end{array}\right)$\\\hline
7883&$\left(
\begin{array}{cc}
 2 & 1 \\
 3 & 1 \\
\end{array}
\right)_{-150}^{2,77}$&$\left(
\begin{array}{cc}
 5 & 2 \\
 3 & 0 \\
\end{array}
\right)$&3&$\left(
\begin{array}{cc}
 -1 & 5 \\
 1 & 0 \\
\end{array}
\right)$&$\left(\begin{array}{l}K_3\\K_3\end{array}
\begin{array}{|rr|}
 -1 & 5 \\
 0 & 1 \\
 1 & 0 \\
\end{array}
\begin{array}{l}E\\I_{72,1}^5\\E\end{array}\right)$\\\hline
7884&$\left(
\begin{array}{c}
 3 \\
 3 \\
\end{array}
\right)_{-162}^{2,83}$&$\left(
\begin{array}{cc}
 3 & 0 \\
 3 & 0 \\
\end{array}
\right)$&3&$\left(
\begin{array}{cc}
 0 & 1 \\
 1 & 0 \\
\end{array}
\right)$&$\left(\begin{array}{l}K_3\end{array}
\begin{array}{|rr|}
 0 & 1 \\
 1 & 0 \\
\end{array}
\begin{array}{l}E\\E\end{array}\right)$\\\hline
7885&$\left(
\begin{array}{cc}
 1 & 1 \\
 4 & 1 \\
\end{array}
\right)_{-168}^{2,86}$&$\left(
\begin{array}{cc}
 4 & 5 \\
 0 & 0 \\
\end{array}
\right)$&2&$\left(
\begin{array}{cc}
 -1 & 1 \\
 1 & 0 \\
\end{array}
\right)$&$\left(\begin{array}{l}Z\\K_1\\K_2\end{array}
\begin{array}{|rr|}
 -1 & 1 \\
 -1 & 4 \\
 0 & 1 \\
 1 & 0 \\
\end{array}
\begin{array}{l}E\\Z_b\\F_{16,0}\\E\end{array}\right)$\\\hline
7886&$\left(
\begin{array}{ccc}
 0 & 1 & 1 \\
 4 & 1 & 1 \\
\end{array}
\right)_{-168}^{2,86}$&$\left(
\begin{array}{cc}
 4 & 8 \\
 0 & 0 \\
\end{array}
\right)$&2&$\left(
\begin{array}{cc}
 -1 & 1 \\
 1 & 0 \\
\end{array}
\right)$&$\left(\begin{array}{l}K_2\\K_2\end{array}
\begin{array}{|rr|}
 -1 & 1 \\
 0 & 1 \\
 1 & 0 \\
\end{array}
\begin{array}{l}E\\I_{4,0}^1\\E\end{array}\right)$\\\hline
7887&$\left(
\begin{array}{c}
 2 \\
 4 \\
\end{array}
\right)_{-168}^{2,86}$&$\left(
\begin{array}{cc}
 4 & 2 \\
 0 & 0 \\
\end{array}
\right)$&2&$\left(
\begin{array}{cc}
 -1 & 4 \\
 1 & 0 \\
\end{array}
\right)$&$\left(\begin{array}{l}K_2\\K_2\end{array}
\begin{array}{|rr|}
 -1 & 4 \\
 0 & 1 \\
 1 & 0 \\
\end{array}
\begin{array}{l}E\\I_{64,0}^4\\E\end{array}\right)$\\\hline
7888&$\left(
\begin{array}{cc}
 0 & 2 \\
 4 & 1 \\
\end{array}
\right)_{-168}^{2,86}$&$\left(
\begin{array}{cc}
 4 & 8 \\
 0 & 0 \\
\end{array}
\right)$&2&$\left(
\begin{array}{cc}
 -1 & 1 \\
 1 & 0 \\
\end{array}
\right)$&$\left(\begin{array}{l}K_2\\K_2\end{array}
\begin{array}{|rr|}
 -1 & 1 \\
 0 & 1 \\
 1 & 0 \\
\end{array}
\begin{array}{l}E\\I_{4,0}^1\\E\end{array}\right)$\\\hline
\end{longtable}}
${ }^*~$For the manifold $7809$ we had insufficient cohomology data to produce a complete description of the effective cone structure. The extended K\"ahler cone is $\langle(-1,4),(27,-4)\rangle$ and there are two rigid divisor classes $(-1,3)$ and $(20,-3)$. However, we are uncertain of how many cones (at least~2) lie between the rays $(0,1)$ and $(-1,4)$ and between the rays $(7,-1)$ and $(27,-4)$. 

${ }^{**}~$For the manifold $7821$ we were not able to identify one boundary of the effective cone, due to insufficient cohomology data. 

\section{Picard number 2 THCYs and their effective cone structure}\label{appB}
The table below lists the Picard number $2$ CYs constructed as hypersurfaces in toric varieties (THCYs) associated with the $36$ reflexive four-dimensional polytopes with six rays, and their various triangulations. We also provide the information describing the structure of the effective cones of these THCYs. The first column of the table indicates first the label of the polytope and second of the triangulation, and also includes the Hodge numbers and the Euler number $\chi(X)$ of the CY.

Column two provides the toric variety data: the weight system (charge matrix)\footnote{When there are multiple triangulations, the charge matrices are equivalent, but we choose them to match the respective K\"ahler cone bases.}, the Stanley-Reisner ideal (SRI), as well as the rays generating the ambient toric variety K\"ahler cone. The basis of $H^2(X)$ is chosen such that the positive quadrant corresponds to the restriction of the ambient space K\"ahler cone to the CY threefold $X$. In many, but not all of the cases this is also the K\"ahler cone of $X$. The K\"ahler cone of $X$ is denoted as $K_c$ where $c$ is the case number included in the fourth column. Whenever the positive quadrant is a subcone of the effective cone appearing in the last column, it is also the K\"ahler cone of $X$. In all the other cases there is no danger of confusion. 

The meaning of the remaining columns remains unchanged from Appendix~\ref{appA}. As for the case of CICYs, the structure of the effective cone has been determined for generic complex structure by studying line bundle cohomology formulae computed algorithmically using the CohomCalg package~\cite{cohomCalg:Implementation} and SAGE. For several manifolds we were not able to obtain enough cohomology data to identify the structure of the effective cone. Nevertheless, we included these manifolds in the table, for the sake of completeness, while leaving the effective cone information empty.

This table can be compared with Table~11 of Ref.~\cite{AbdusSalam:2020ywo} which identifies K3-fibrations and manifolds of Swiss cheese type amongst $h^{1,1}=2$ THCYs.

{\scriptsize
\begin{longtable}{|c|c|c|c|c|c|}\hline
\!\!\!\!\!\!\!\!(\#)$^{h^{1,1},\,h^{2,1}}_\chi \!\!\!\!\!\!\!\!$&toric variety&$\left(\begin{array}{ll}d_{122}&d_{222}\\d_{112}&d_{111}\end{array}\right)$&case&
$\left(\begin{array}{ll}v_{11}&v_{12}\\v_{21}&v_{22}\end{array}\right)$&cones\\\hline\hline
$(1,1)^{2,29}_{-54}$&$\begin{array}{c}
\begin{array}{rrrrrr}
z_0&z_1&z_2&z_3&z_4&z_5\\
\hline
1&0&0&0&1&1\\
0&1&1&1&0&0\\
\end{array}\\[4pt]
\text{SRI}=\langle z_0 z_4 z_5, z_1 z_2 z_3\rangle\\
\text{KC gen}=\{ 
z_5,
z_3
\}
\end{array}$
&$\left(
\begin{array}{cc}
 1 & 0 \\
 1 & 0 \\
\end{array}
\right)$&3&$\left(
\begin{array}{cc}
 0 & 1 \\
 1 & 0 \\
\end{array}
\right)$&$\left(\begin{array}{l}K_3\end{array}
\begin{array}{|rr|}
 0 & 1 \\
 1 & 0 \\
\end{array}
\begin{array}{l}E\\E\end{array}\right)$\\\hline
$(2,1)^{2,38}_{-72}$&$\begin{array}{c}
\begin{array}{rrrrrr}
z_0&z_1&z_2&z_3&z_4&z_5\\
\hline
0&0&1&1&0&1\\
1&1&0&0&1&-3
\end{array}\\[4pt]
\text{SRI}=\langle{z_0 z_1 z_4, z_2 z_3 z_5}\rangle\\
\text{KC gen}=\{
3 z_4 + z_5,
z_4
\}
\end{array}$
\vspace{2pt}
&$\left(
\begin{array}{cc} 1&9 \\3&9 \\
\end{array} \right)$&1
&$\left(
\begin{array}{rr} 0&1\\1&-3\\
\end{array} \right)$
&$\left(\begin{array}{l}K_1\\Z\end{array}
\begin{array}{|rr|} 0&1\\1&0\\1&-3\\\end{array}
\begin{array}{l}E\\Z_b\\E\end{array}\right)$\\
\hline
$(3,1)^{2,74}_{-144}$&$\begin{array}{c}
\begin{array}{rrrrrr}
z_0&z_1&z_2&z_3&z_4&z_5\\
\hline
-1&1&1&0&2&3\\
1&0&0&1&0&-2
\end{array}\\[4pt]
\text{SRI}=\langle{z_0 z_3, z_1 z_2 z_4 z_5}\rangle\\
\text{KC gen}=\{
\frac{z_4}{2},
\frac{3 z_4}{4}-\frac{z_5}{2}
\}
\end{array}$
\vspace{2pt}
&$\left(
\begin{array}{cc} 3&3 \\3&2 \\
\end{array} \right)$&1
&$\left(
\begin{array}{rr} -1&1\\3&-2\\
\end{array} \right)$
&$\left(\begin{array}{l}Z\\K_1\\Z\end{array}
\begin{array}{|rr|} -1
&1\\0
&1\\1
&0\\3
&-2\\\end{array}
\begin{array}{l}E\\Z_b\\Z_a\\E\end{array}\right)$\\
\hline
$(4,1)^{2,74}_{-144}$&$\begin{array}{c}
\begin{array}{rrrrrr}
z_0&z_1&z_2&z_3&z_4&z_5\\
\hline
-1&2&1&1&3&0\\
1&-1&0&0&-1&1
\end{array}\\[4pt]
\text{SRI}=\langle{z_0 z_5, z_1 z_2 z_3 z_4}\rangle\\
\text{KC gen}=\{
\frac{z_4}{3}+\frac{z_5}{3},
z_5
\}
\end{array}$
\vspace{2pt}
&$\left(
\begin{array}{cc} 3&3 \\3&2 \\
\end{array} \right)$&1
&$\left(
\begin{array}{rr} -1&1\\2&-1\\
\end{array} \right)$
&$\left(\begin{array}{l}Z\\K_1\\K_1\\Z\end{array}
\begin{array}{|rr|} -1
&1\\0
&1\\1
&0\\3
&-1\\2
&-1\\\end{array}
\begin{array}{l}E\\Z_b\\I^3_{6,0}\\Z_b\\E\end{array}\right)$\\
\hline
$(4,2)^{2,74}_{-144}$&$\begin{array}{c}
\begin{array}{rrrrrr}
z_0&z_1&z_2&z_3&z_4&z_5\\
\hline
2&-1&1&1&0&3\\
-1&1&0&0&1&-1
\end{array}\\[4pt]
\text{SRI}=\langle{z_1 z_4, z_0 z_2 z_3 z_5}\rangle\\
\text{KC gen}=\{
\frac{z_4}{3}+\frac{z_5}{3},
z_4
\}
\vspace{2pt}
\end{array}$
&$\left(
\begin{array}{cc} 3&3 \\3&2 \\
\end{array} \right)$&1
&$\left(
\begin{array}{rr} -1&1\\2&-1\\
\end{array} \right)$
&$\left(\begin{array}{l}Z\\K_1\\K_1\\Z\end{array}
\begin{array}{|rr|} -1
&1\\0
&1\\1
&0\\3
&-1\\2
&-1\\\end{array}
\begin{array}{l}E\\Z_b\\I^3_{6,0}\\Z_b\\E\end{array}\right)$\\
\hline
$(5,1)^{2,83}_{-162}$&$\begin{array}{c}
\begin{array}{rrrrrr}
z_0&z_1&z_2&z_3&z_4&z_5\\
\hline
1&0&1&0&0&1\\
0&1&0&1&1&0\\
\end{array}\\[4pt]
\text{SRI}=\langle{z_0 z_2 z_5, z_1 z_3 z_4}\rangle\\
\text{KC gen}=\{
z_5,
z_4
\}
\end{array}$
&$\left(
\begin{array}{cc} 3&0 \\3&0 \\
\end{array} \right)$&3
&$\left(
\begin{array}{rr} 0&1\\1&0\\
\end{array} \right)$
&$\left(\begin{array}{l}K_3\end{array}
\begin{array}{|rr|} 0&1\\1&0\\\end{array}
\begin{array}{l}E\\E\end{array}\right)$\\
\hline
$(6,1)^{2,84}_{-164}$&$\begin{array}{c}
\begin{array}{rrrrrr}
z_0&z_1&z_2&z_3&z_4&z_5\\
\hline
1&-1&1&1&2&0\\
0&1&0&0&-1&1
\end{array}\\[4pt]
\text{SRI}=\langle{z_1 z_5, z_0 z_2 z_3 z_4}\rangle\\
\text{KC gen}=\{
\frac{z_4}{2}+\frac{z_5}{2},
z_5
\}
\end{array}$
&$\left(
\begin{array}{cc} 5&5 \\5&3 \\
\end{array} \right)$&1
&$\left(
\begin{array}{rr} -1&1\\2&-1\\
\end{array} \right)$
&$\left(\begin{array}{l}Z\\K_1\\K_1\\Z\end{array}
\begin{array}{|rr|} -1&1\\0&1\\1&0\\5&-1\\2&-1\\\end{array}
\begin{array}{l}E\\Z_b\\F_{20,0}\\Z_b\\E\end{array}\right)$\\
\hline
$(7,1)^{2,86}_{-168}$&$\begin{array}{c}
\begin{array}{rrrrrr}
z_0&z_1&z_2&z_3&z_4&z_5\\
\hline
0&0&1&1&1&1\\
1&1&0&0&0&-2
\end{array}\\[4pt]
\text{SRI}=\langle{z_0 z_1, z_2 z_3 z_4 z_5}\rangle\\
\text{KC gen}=\{
z_4,
\frac{z_4}{2}-\frac{z_5}{2}
\}
\end{array}$
&$\left(
\begin{array}{cc} 0&0 \\4&8 \\
\end{array} \right)$&2
&$\left(
\begin{array}{rr} 0&1\\1&-2\\
\end{array} \right)$
&$\left(\begin{array}{l}K_2\\Z\end{array}
\begin{array}{|rr|} 0&1\\1&0\\1&-2\\\end{array}
\begin{array}{l}E\\Z_a\\E\end{array}\right)$\\
\hline
$(8,1)^{2,86}_{-168}$&$\begin{array}{c}
\begin{array}{rrrrrr}
z_0&z_1&z_2&z_3&z_4&z_5\\
\hline
0&1&1&0&1&1\\
1&-1&-1&1&0&0
\end{array}\\[4pt]
\text{SRI}=\langle{z_0 z_3, z_1 z_2 z_4 z_5}\rangle\\
\text{KC gen}=\{
z_5,
z_3
\}
\end{array}$
&$\left(
\begin{array}{cc} 0&0 \\4&8 \\
\end{array} \right)$&2
&$\left(
\begin{array}{rr} 0&1\\1&-1\\
\end{array} \right)$
&$\left(\begin{array}{l}K_2\\K_2\end{array}
\begin{array}{|rr|} 0&1\\1&0\\1&-1\\\end{array}
\begin{array}{l}E\\I^1_{4,0}\\E\end{array}\right)$\\
\hline
$(8,2)^{2,86}_{-168}$&$\begin{array}{c}
\begin{array}{rrrrrr}
z_0&z_1&z_2&z_3&z_4&z_5\\
\hline
1&0&0&1&1&1\\
-1&1&1&-1&0&0
\end{array}\\[4pt]
\text{SRI}=\langle{z_1 z_2, z_0 z_3 z_4 z_5}\rangle\\
\text{KC gen}=\{
z_5,
z_5-z_3
\}
\end{array}$
&$\left(
\begin{array}{cc} 0&0 \\4&8 \\
\end{array} \right)$&2
&$\left(
\begin{array}{rr} 0&1\\1&-1\\
\end{array} \right)$
&$\left(\begin{array}{l}K_2\\K_2\end{array}
\begin{array}{|rr|} 0&1\\1&0\\1&-1\\\end{array}
\begin{array}{l}E\\I^1_{4,0}\\E\end{array}\right)$\\
\hline
$(9,1)^{2,86}_{-168}$&$\begin{array}{c}
\begin{array}{rrrrrr}
z_0&z_1&z_2&z_3&z_4&z_5\\
\hline
1&0&0&0&0&1\\
0&1&1&1&1&0
\end{array}\\[4pt]
\text{SRI}=\langle{z_0 z_5, z_1 z_2 z_3 z_4}\rangle\\
\text{KC gen}=\{
z_5,
z_4
\}
\end{array}$
&$\left(
\begin{array}{cc} 4&2 \\0&0 \\
\end{array} \right)$&2
&$\left(
\begin{array}{rr} -1&4\\1&0\\
\end{array} \right)$
&$\left(\begin{array}{l}K_2\\K_2\end{array}
\begin{array}{|rr|} -1&4\\0&1\\1&0\\\end{array}
\begin{array}{l}E\\I^4_{64,0}\\E\end{array}\right)$\\
\hline
$(10,1)^{2,86}_{-168}$&$\begin{array}{c}
\begin{array}{rrrrrr}
z_0&z_1&z_2&z_3&z_4&z_5\\
\hline
1&0&0&1&0&1\\
0&1&1&0&1&-1
\end{array}\\[4pt]
\text{SRI}=\langle{z_0 z_3 z_5, z_1 z_2 z_4}\rangle\\
\text{KC gen}=\{
z_4+z_5,
z_4
\}
\end{array}$
&$\left(
\begin{array}{cc} 3&0 \\5&5 \\
\end{array} \right)$&3
&$\left(
\begin{array}{rr} 0&1\\1&-1\\
\end{array} \right)$
&$\left(\begin{array}{l}K_3\\Z\end{array}
\begin{array}{|rr|} 0&1\\1&0\\1&-1\\\end{array}
\begin{array}{l}E\\Z_a\\E\end{array}\right)$\\
\hline
$(11,1)^{2,86}_{-168}$&$\begin{array}{c}
\begin{array}{rrrrrr}
z_0&z_1&z_2&z_3&z_4&z_5\\
\hline
1&0&0&1&1&0\\
0&1&1&-1&-1&1
\end{array}\\[4pt]
\text{SRI}=\langle{z_0 z_3 z_4, z_1 z_2 z_5}\rangle\\
\text{KC gen}=\{
z_4+z_5,
z_5
\}
\end{array}$
&$\left(
\begin{array}{cc} 3&0 \\7&11 \\
\end{array} \right)$&3
&$\left(
\begin{array}{rr} 0&1\\1&-1\\
\end{array} \right)$
&$\left(\begin{array}{l}K_3\\K_2\end{array}
\begin{array}{|rr|} 0&1\\1&0\\1&-1\\\end{array}
\begin{array}{l}E\\F_{1,0}\\E\end{array}\right)$\\
\hline
$(12,1)^{2,86}_{-168}$&$\begin{array}{c}
\begin{array}{rrrrrr}
z_0&z_1&z_2&z_3&z_4&z_5\\
\hline
1&-1&-1&1&2&0\\
0&1&1&0&-1&1
\end{array}\\[4pt]
\text{SRI}=\langle{z_0 z_3 z_4, z_1 z_2 z_5}\rangle\\
\text{KC gen}=\{
\frac{z_4}{2}+\frac{z_5}{2},
z_5
\}
\end{array}$
&$\left(
\begin{array}{cc} 10&14 \\6&3 \\
\end{array} \right)$&3
&$\left(
\begin{array}{rr} -1&1\\2&-1\\
\end{array} \right)$
&$\left(\begin{array}{l}K_2\\K_3\\Z\end{array}
\begin{array}{|rr|} -1&1\\0&1\\1&0\\2&-1\\\end{array}
\begin{array}{l}E\\F_{1,0}\\Z_a\\E\end{array}\right)$\\
\hline
$(13,1)^{2,86}_{-168}$&$\begin{array}{c}
\begin{array}{rrrrrr}
z_0&z_1&z_2&z_3&z_4&z_5\\
\hline
1&0&1&1&0&1\\
0&1&0&0&1&-1
\end{array}\\[4pt]
\text{SRI}=\langle{z_1 z_4, z_0 z_2 z_3 z_5}\rangle\\
\text{KC gen}=\{
z_4+z_5,
z_4
\}
\end{array}$
&$\left(
\begin{array}{cc} 0&0 \\4&5 \\
\end{array} \right)$&2
&$\left(
\begin{array}{rr} 0&1\\1&-1\\
\end{array} \right)$
&$\left(\begin{array}{l}K_2\\K_1\\Z\end{array}
\begin{array}{|rr|} 0&1\\1&0\\4&-1\\1&-1\\\end{array}
\begin{array}{l}E\\F_{16,0}\\Z_b\\E\end{array}\right)$\\
\hline
$(14,1)^{2,86}_{-168}$&$\begin{array}{c}
\begin{array}{rrrrrr}
z_0&z_1&z_2&z_3&z_4&z_5\\
\hline
1&-1&-1&1&3&0\\
0&1&1&0&-2&1
\end{array}\\[4pt]
\text{SRI}=\langle{z_0 z_3 z_4, z_1 z_2 z_5}\rangle\\
\text{KC gen}=\{
\frac{z_4}{3}+\frac{2 z_5}{3},
z_5
\}
\end{array}$
&$\left(
\begin{array}{cc} 13&17 \\9&6 \\
\end{array} \right)$&1
&$\left(
\begin{array}{rr} -1&1\\3&-2\\
\end{array} \right)$
&$\left(\begin{array}{l}K_2\\K_1\\Z\end{array}
\begin{array}{|rr|} -1&1\\0&1\\1&0\\3&-2\\\end{array}
\begin{array}{l}E\\F_{1,0}\\Z_a\\E\end{array}\right)$\\
\hline
$(15,1)^{2,86}_{-168}$&$\begin{array}{c}
\begin{array}{rrrrrr}
z_0&z_1&z_2&z_3&z_4&z_5\\
\hline
1&0&0&0&0&1\\
0&1&1&2&1&-1
\end{array}\\[4pt]
\text{SRI}=\langle{z_0 z_5, z_1 z_2 z_3 z_4}\rangle\\
\text{KC gen}=\{
z_4+z_5,
z_4
\}
\end{array}$
&$\left(
\begin{array}{cc} 3&1 \\3&3 \\
\end{array} \right)$&1
&$\left(
\begin{array}{rr} -1&5\\1&-1\\
\end{array} \right)$
&$\left(\begin{array}{l}Z\\K_1\\K_1\\Z\end{array}
\begin{array}{|rr|} -1&5\\-1&6\\ 0&1\\ 1&0\\ 1&-1\\\end{array}
\begin{array}{l}E\\Z_b\\I^6_{60,0}\\Z_b\\E\end{array}\right)$\\
\hline
$(16,1)^{2,90}_{-176}$&$\begin{array}{c}
\begin{array}{rrrrrr}
z_0&z_1&z_2&z_3&z_4&z_5\\
\hline
1&0&0&0&0&1\\
0&1&1&1&1&-1
\end{array}\\[4pt]
\text{SRI}=\langle{z_0 z_5, z_1 z_2 z_3 z_4}\rangle\\
\text{KC gen}=\{
z_4+z_5,
z_4
\}
\end{array}$
&$\left(
\begin{array}{cc} 5&2 \\5&5 \\
\end{array} \right)$&1
&$\left(
\begin{array}{rr} -1&4\\1&-1\\
\end{array} \right)$
&$\left(\begin{array}{l}Z\\K_1\\K_1\\Z\end{array}
\begin{array}{|rr|} -1&4\\-1&5\\0&1\\1&0\\1&-1\\\end{array}
\begin{array}{l}E\\Z_b\\I^5_{60,0}\\Z_b\\E\end{array}\right)$\\
\hline
$(17,1)^{2,92}_{-180}$&$\begin{array}{c}
\begin{array}{rrrrrr}
z_0&z_1&z_2&z_3&z_4&z_5\\
\hline
1&-1&-1&-1&2&0\\
0&1&1&1&-1&1
\end{array}\\[4pt]
\text{SRI}=\langle{z_0 z_4, z_1 z_2 z_3 z_5}\rangle\\
\text{KC gen}=\{
\frac{z_4}{2}+\frac{z_5}{2},
z_5
\}
\end{array}$
&$\left(
\begin{array}{cc} 12&21 \\6&3 \\
\end{array} \right)$&1
&$\left(
\begin{array}{rr} -1&1\\2&-1\\
\end{array} \right)$
&$\left(\begin{array}{l}K_1\\Z\end{array}
\begin{array}{|rr|} -1&1\\ 1&0\\ 2&-1\\\end{array}
\begin{array}{l}E\\Z_b\\E\end{array}\right)$\\
\hline
$(17,2)^{2,92}_{-180}$&$\begin{array}{c}
\begin{array}{rrrrrr}
z_0&z_1&z_2&z_3&z_4&z_5\\
\hline
1&0&0&0&1&1\\
-1&1&1&1&-2&0
\end{array}\\[4pt]
\text{SRI}=\langle{z_0 z_4 z_5, z_1 z_2 z_3}\rangle\\
\text{KC gen}=\{
z_5,
\frac{z_5}{2}-\frac{z_4}{2}
\}
\end{array}$
&$\left(
\begin{array}{cc} 3&0 \\9&21 \\
\end{array} \right)$&1
&$\left(
\begin{array}{rr} 0&1\\1&-2\\
\end{array} \right)$
&$\left(\begin{array}{l}K_1\\Z\end{array}
\begin{array}{|rr|} 0&1\\1&-1\\1&-2\\\end{array}
\begin{array}{l}E\\Z_b\\E\end{array}\right)$\\
\hline
$(18,1)^{2,95}_{-186}$&$\begin{array}{c}
\begin{array}{rrrrrr}
z_0&z_1&z_2&z_3&z_4&z_5\\
\hline
1&0&0&1&1&0\\
0&1&1&0&-2&1
\end{array}\\[4pt]
\text{SRI}=\langle{z_0 z_3 z_4, z_1 z_2 z_5}\rangle\\
\text{KC gen}=\{
z_4+2 z_5,
z_5
\}
\end{array}$
&$\left(
\begin{array}{cc} 3&0 \\7&14 \\
\end{array} \right)$&1
&$\left(
\begin{array}{rr} 0&1\\1&-2\\
\end{array} \right)$
&$\left(\begin{array}{l}K_1\\Z\end{array}
\begin{array}{|rr|} 0&1\\1&0\\1&-2\\\end{array}
\begin{array}{l}E\\Z_a\\E\end{array}\right)$\\
\hline
$(19,1)^{2,102}_{-200}$&$\begin{array}{c}
\begin{array}{rrrrrr}
z_0&z_1&z_2&z_3&z_4&z_5\\
\hline
1&0&0&0&1&0\\
0&1&1&1&-2&1
\end{array}\\[4pt]
\text{SRI}=\langle{z_0 z_4, z_1 z_2 z_3 z_5}\rangle\\
\text{KC gen}=\{
z_4+2z_5,
z_5
\}
\end{array}$
&$\left(
\begin{array}{cc} 6&2 \\12&24 \\
\end{array} \right)$&1
&$\left(
\begin{array}{rr} -1&4\\1&-2\\
\end{array} \right)$
&$\left(\begin{array}{l}Z\\K_1\\K_1\\Z\end{array}
\begin{array}{|rr|} -1&4\\-1&6\\0&1\\1&0\\1&-2\\\end{array}
\begin{array}{l}E\\Z_b\\I^6_{48,0}\\Z_b\\E\end{array}\right)$\\
\hline
$(20,1)^{2,106}_{-208}$&$\begin{array}{c}
\begin{array}{rrrrrr}
z_0&z_1&z_2&z_3&z_4&z_5\\
\hline
1&1&1&0&-3&8\\
0&0&0&1&1&-2
\end{array}\\[4pt]
\text{SRI}=\langle{z_3 z_4, z_0 z_1 z_2 z_5}\rangle\\
\text{KC gen}=\{
z_4+\frac{z_5}{2},
4 z_4+\frac{3 z_5}{2}
\}
\end{array}$
&$\left(
\begin{array}{cc} 12&36 \\4&1 \\
\end{array} \right)$&1
&$\left(
\begin{array}{rr} -3&1\\5&-1\\
\end{array} \right)$
&$\left(\begin{array}{l}Z\\K_1\\K_1\\Z\end{array}
\begin{array}{|rr|} -3&1\\0&1\\1&0\\8&-1\\5&-1\\\end{array}
\begin{array}{l}E\\Z_b\\I^8_{40,0}\\Z_b\\E\end{array}\right)$\\
\hline
$(21,1)^{2,106}_{-208}$&$\begin{array}{c}
\begin{array}{rrrrrr}
z_0&z_1&z_2&z_3&z_4&z_5\\
\hline
1&-3&1&1&5&0\\
0&1&0&0&-1&1
\end{array}\\[4pt]
\text{SRI}=\langle{z_1 z_5, z_0 z_2 z_3 z_4}\rangle\\
\text{KC gen}=\{
\frac{z_4}{5}+\frac{z_5}{5},
z_5
\}
\end{array}$
&$\left(
\begin{array}{cc} 12&36 \\4&1 \\
\end{array} \right)$&1
&$\left(
\begin{array}{rr} -3&1\\5&-1\\
\end{array} \right)$
&$\left(\begin{array}{l}Z\\K_1\\K_1\\Z\end{array}
\begin{array}{|rr|} -3&1\\0&1\\1&0\\8&-1\\5&-1\\\end{array}
\begin{array}{l}E\\Z_b\\I^8_{40,0}\\Z_b\\E\end{array}\right)$\\
\hline
$(22,1)^{2,106}_{-208}$&$\begin{array}{c}
\begin{array}{rrrrrr}
z_0&z_1&z_2&z_3&z_4&z_5\\
\hline
1&0&0&0&1&0\\
0&1&1&2&-3&1
\end{array}\\[4pt]
\text{SRI}=\langle{z_0 z_4, z_1 z_2 z_3 z_5}\rangle\\
\text{KC gen}=\{
z_4+3 z_5,
z_5
\}
\end{array}$
&$\left(
\begin{array}{cc} 4&1 \\12&36 \\
\end{array} \right)$&1
&$\left(
\begin{array}{rr} -1&5\\1&-3\\
\end{array} \right)$
&$\left(\begin{array}{l}Z\\K_1\\K_1\\Z\end{array}
\begin{array}{|rr|} -1&5\\-1&8\\0&1\\1&0\\1&-3\\\end{array}
\begin{array}{l}E\\Z_b\\I^8_{40,0}\\Z_b\\E\end{array}\right)$\\
\hline
$(23,1)^{2,116}_{-228}$&$\begin{array}{c}
\begin{array}{rrrrrr}
z_0&z_1&z_2&z_3&z_4&z_5\\
\hline
-2&3&-2&-4&5&0\\
1&0&1&2&-1&3
\end{array}\\[4pt]
\text{SRI}=\langle{z_1 z_4, z_0 z_2 z_3 z_5}\rangle\\
\text{KC gen}=\{
\frac{3z_4}{5}+\frac{z_5}{5},
z_5
\}
\end{array}$
&$\left(
\begin{array}{cc} 25&98 \\5&1 \\
\end{array} \right)$&1
&$\left(
\begin{array}{rr} -2&1\\5&-1\\
\end{array} \right)$
&$\left(\begin{array}{l}K_1\\Z\end{array}
\begin{array}{|rr|} -2&1\\1&0\\5&-1\\\end{array}
\begin{array}{l}E\\Z_b\\E\end{array}\right)$\\
\hline
$(23,2)^{2,116}_{-228}$&$\begin{array}{c}
\begin{array}{rrrrrr}
z_0&z_1&z_2&z_3&z_4&z_5\\
\hline
0&1&0&0&1&2\\
2&-3&2&4&-5&0
\end{array}\\[4pt]
\text{SRI}=\langle{z_0 z_2 z_3, z_1 z_4 z_5}\rangle\\
\text{KC gen}=\{
z_5,
\frac{z_5}{5}-\frac{2 z_4}{5}
\}
\end{array}$
&$\left(
\begin{array}{cc} 2&0 \\16&98 \\
\end{array} \right)$&1
&&\\
\hline
$(24,1)^{2,120}_{-236}$&$\begin{array}{c}
\begin{array}{rrrrrr}
z_0&z_1&z_2&z_3&z_4&z_5\\
\hline
1&-1&1&2&0&3\\
-1&4&-1&-5&3&0
\end{array}\\[4pt]
\text{SRI}=\langle{z_1 z_4, z_0 z_2 z_3 z_5}\rangle\\
\text{KC gen}=\{
z_5,
z_4
\}
\end{array}$
&$\left(
\begin{array}{cc} 8&2 \\32&101 \\
\end{array} \right)$&1
&&\\
\hline
$(24,2)^{2,120}_{-236}$&$\begin{array}{c}
\begin{array}{rrrrrr}
z_0&z_1&z_2&z_3&z_4&z_5\\
\hline
1&-4&1&5&-3&0\\
0&1&0&-1&1&1
\end{array}\\[4pt]
\text{SRI}=\langle{z_0 z_2 z_3, z_1 z_4 z_5}\rangle\\
\text{KC gen}=\{
\frac{z_5}{3}-\frac{z_4}{3},
z_5
\}
\end{array}$
&$\left(
\begin{array}{cc} 23&101 \\5&1 \\
\end{array} \right)$&1
&$\left(
\begin{array}{rr} -4&1\\5&-1\\
\end{array} \right)$
&$\left(\begin{array}{l}Z\\K_1\\Z\end{array}
\begin{array}{|rr|} -4&1\\-3&1\\1&0\\5&-1\\\end{array}
\begin{array}{l}E\\Z_b\\Z_a\\E\end{array}\right)$\\
\hline
$(25,1)^{2,122}_{-240}$&$\begin{array}{c}
\begin{array}{rrrrrr}
z_0&z_1&z_2&z_3&z_4&z_5\\
\hline
-3&1&1&1&0&7\\
1&0&0&0&1&-2
\end{array}\\[4pt]
\text{SRI}=\langle{z_0 z_4, z_1 z_2 z_3 z_5}\rangle\\
\text{KC gen}=\{
\frac{2z_4}{7}+\frac{z_5}{7},
z_4
\}
\end{array}$
&$\left(
\begin{array}{cc} 21&63 \\7&2 \\
\end{array} \right)$&1
&$\left(
\begin{array}{rr} -3&1\\7&-2\\
\end{array} \right)$
&$\left(\begin{array}{l}Z\\K_1\\K_1\\Z\end{array}
\begin{array}{|rr|} -3&1\\ 0&1\\ 1&0\\7&-1\\ 7&-2\\\end{array}
\begin{array}{l}E\\Z_b\\I^7_{28,0}\\Z_b\\E\end{array}\right)$\\
\hline
$(26,1)^{2,122}_{-240}$&$\begin{array}{c}
\begin{array}{rrrrrr}
z_0&z_1&z_2&z_3&z_4&z_5\\
\hline
1&-3&1&1&4&0\\
0&1&0&0&-1&1
\end{array}\\[4pt]
\text{SRI}=\langle{z_1 z_5, z_0 z_2 z_3 z_4}\rangle\\
\text{KC gen}=\{
\frac{z_4}{4}+\frac{z_5}{4},
z_5
\}
\end{array}$
&$\left(
\begin{array}{cc} 21&63 \\7&2 \\
\end{array} \right)$&1
&$\left(
\begin{array}{rr} -3&1\\7&-2\\
\end{array} \right)$
&$\left(\begin{array}{l}Z\\K_1\\K_1\\Z\end{array}
\begin{array}{|rr|} -3&1\\0&1\\1&0\\7&-1\\7&-2\\\end{array}
\begin{array}{l}E\\Z_b\\I^7_{28,0}\\Z_b\\E\end{array}\right)$\\
\hline
$(27,1)^{2,122}_{-240}$&$\begin{array}{c}
\begin{array}{rrrrrr}
z_0&z_1&z_2&z_3&z_4&z_5\\
\hline
1&0&0&0&1&0\\
0&1&1&1&-3&1
\end{array}\\[4pt]
\text{SRI}=\langle{z_0 z_4, z_1 z_2 z_3 z_5}\rangle\\
\text{KC gen}=\{
z_4+3 z_5,
z_5
\}
\end{array}$
&$\left(
\begin{array}{cc} 7&2 \\21&63 \\
\end{array} \right)$&1
&$\left(
\begin{array}{rr} -2&7\\1&-3\\
\end{array} \right)$
&$\left(\begin{array}{l}Z\\K_1\\K_1\\Z\end{array}
\begin{array}{|rr|} -2&7\\-1&7\\0&1\\1&0\\1&-3\\\end{array}
\begin{array}{l}E\\Z_b\\I^7_{28,0}\\Z_b\\E\end{array}\right)$\\
\hline
$(28,1)^{2,128}_{-252}$&$\begin{array}{c}
\begin{array}{rrrrrr}
z_0&z_1&z_2&z_3&z_4&z_5\\
\hline
0&0&1&1&3&1\\
1&1&0&0&0&-2
\end{array}\\[4pt]
\text{SRI}=\langle{z_0 z_1, z_2 z_3 z_4 z_5}\rangle\\
\text{KC gen}=\{
\frac{z_4}{3},
\frac{z_4}{6}-\frac{z_5}{2}
\}
\end{array}$
&$\left(
\begin{array}{cc} 0&0 \\2&4 \\
\end{array} \right)$&2
&$\left(
\begin{array}{rr} 0&1\\1&-2\\
\end{array} \right)$
&$\left(\begin{array}{l}K_2\\Z\end{array}
\begin{array}{|rr|} 0&1\\1&0\\1&-2\\\end{array}
\begin{array}{l}E\\Z_a\\E\end{array}\right)$\\
\hline
$(29,1)^{2,128}_{-252}$&$\begin{array}{c}
\begin{array}{rrrrrr}
z_0&z_1&z_2&z_3&z_4&z_5\\
\hline
0&1&1&0&1&3\\
1&-1&-1&1&0&0
\end{array}\\[4pt]
\text{SRI}=\langle{z_0 z_3, z_1 z_2 z_4 z_5}\rangle\\
\text{KC gen}=\{
\frac{z_5}{3},
z_3
\}
\end{array}$
&$\left(
\begin{array}{cc} 0&0 \\2&4 \\
\end{array} \right)$&2
&$\left(
\begin{array}{rr} 0&1\\1&-1\\
\end{array} \right)$
&$\left(\begin{array}{l}K_2\\K_2\end{array}
\begin{array}{|rr|} 0&1\\1&0\\1&-1\\\end{array}
\begin{array}{l}E\\I^1_{2,0}\\E\end{array}\right)$\\
\hline
$(29,2)^{2,128}_{-252}$&$\begin{array}{c}
\begin{array}{rrrrrr}
z_0&z_1&z_2&z_3&z_4&z_5\\
\hline
1&0&0&1&1&3\\
-1&1&1&-1&0&0
\end{array}\\[4pt]
\text{SRI}=\langle{z_1 z_2, z_0 z_3 z_4 z_5}\rangle\\
\text{KC gen}=\{
\frac{z_5}{3},
\frac{z_5}{3}-z_3
\}
\end{array}$
&$\left(
\begin{array}{cc} 0&0 \\2&4 \\
\end{array} \right)$&2
&$\left(
\begin{array}{rr} 0&1\\1&-1\\
\end{array} \right)$
&$\left(\begin{array}{l}K_2\\K_2\end{array}
\begin{array}{|rr|} 0&1\\1&0\\1&-1\\\end{array}
\begin{array}{l}E\\I^1_{2,0}\\E\end{array}\right)$\\
\hline
$(30,1)^{2,128}_{-252}$&$\begin{array}{c}
\begin{array}{rrrrrr}
z_0&z_1&z_2&z_3&z_4&z_5\\
\hline
1&0&1&1&0&3\\
-2&3&-2&-2&3&0
\end{array}\\[4pt]
\text{SRI}=\langle{z_1 z_4, z_0 z_2 z_3 z_5}\rangle\\
\text{KC gen}=\{
z_5,
z_4
\}
\end{array}$
&$\left(
\begin{array}{cc} 0&0 \\18&108 \\
\end{array} \right)$&2
&&\\\hline
$(30,2)^{2,128}_{-252}$&$\begin{array}{c}
\begin{array}{rrrrrr}
z_0&z_1&z_2&z_3&z_4&z_5\\
\hline
0&1&0&0&1&2\\
2&-3&2&2&-3&0
\end{array}\\[4pt]
\text{SRI}=\langle{z_0 z_2 z_3, z_1 z_4 z_5}\rangle\\
\text{KC gen}=\{
z_5,
\frac{z_5}{3}-\frac{2 z_4}{3}
\}
\end{array}$
&$\left(
\begin{array}{cc} 4&0 \\24&108 \\
\end{array} \right)$&2
&&\\\hline
$(31,1)^{2,128}_{-252}$
&$\begin{array}{c}
\begin{array}{rrrrrr}
z_0&z_1&z_2&z_3&z_4&z_5\\
\hline
1&-1&-1&2&0&1\\
-2&3&3&-5&1&0
\end{array}\\[4pt]
\text{SRI}=\langle{z_0 z_3 z_5, z_1 z_2 z_4}\rangle\\
\text{KC gen}=\{
z_5,
z_4
\}
\end{array}$
&$\left(
\begin{array}{cc} 16&6 \\42&109 \\
\end{array} \right)$&1
&$\left(
\begin{array}{rr} -1&3\\2&-5\\
\end{array} \right)$
&$\left(\begin{array}{l}K_2\\K_1\\Z\end{array}
\begin{array}{|rr|} -1&3\\0&1\\1&-2\\2&-5\\\end{array}
\begin{array}{l}E\\F_{1,0}\\Z_b\\E\end{array}\right)$\\
\hline
$\begin{array}{c}
(31,2)^{2,128}_{-252}
\end{array}$
&$\begin{array}{c}
\begin{array}{rrrrrr}
z_0&z_1&z_2&z_3&z_4&z_5\\
\hline
2&-3&-3&5&-1&0\\
0&1&1&-1&1&2
\end{array}\\[4pt]
\text{SRI}=\langle{z_0 z_3, z_1 z_2 z_4 z_5}\rangle\\
\text{KC gen}=\{
z_5-2 z_4,
z_5
\}
\end{array}$
&$\left(
\begin{array}{cc} 25&109 \\5&1 \\
\end{array} \right)$&1
&&\\
\hline
$(32,1)^{2,128}_{-252}$&$\begin{array}{c}
\begin{array}{rrrrrr}
z_0&z_1&z_2&z_3&z_4&z_5\\
\hline
1&-3&-3&1&4&0\\
0&1&1&0&-1&1
\end{array}\\[4pt]
\text{SRI}=\langle{z_0 z_3 z_4, z_1 z_2 z_5}\rangle\\
\text{KC gen}=\{
\frac{z_4}{4}+\frac{z_5}{4},
z_5
\}
\end{array}$
&$\left(
\begin{array}{cc} 30&108 \\8&2 \\
\end{array} \right)$&2
&$\left(
\begin{array}{rr} -3&1\\4&-1\\
\end{array} \right)$
&$\left(\begin{array}{l}K_2\\Z\end{array}
\begin{array}{|rr|} -3&1\\1&0\\4&-1\\\end{array}
\begin{array}{l}E\\Z_a\\E\end{array}\right)$\\
\hline
$\begin{array}{c}
(32,2)^{2,128}_{-252}
\end{array}$
&$\begin{array}{c}
\begin{array}{rrrrrr}
z_0&z_1&z_2&z_3&z_4&z_5\\
\hline
1&0&0&1&1&3\\
-1&3&3&-1&-4&0
\end{array}\\[4pt]
\text{SRI}=\langle{z_1 z_2, z_0 z_3 z_4 z_5}\rangle\\
\text{KC gen}=\{
z_5,
\frac{z_5}{4}-\frac{3 z_4}{4}
\}
\end{array}$
&$\left(
\begin{array}{cc} 0&0 \\18&108 \\
\end{array} \right)$&2
&&\\
\hline
$(33,1)^{2,132}_{-260}$&$\begin{array}{c}
\begin{array}{rrrrrr}
z_0&z_1&z_2&z_3&z_4&z_5\\
\hline
1&-1&-1&-1&2&0\\
0&1&1&1&-1&2
\end{array}\\[4pt]
\text{SRI}=\langle{z_0 z_4, z_1 z_2 z_3 z_5}\rangle\\
\text{KC gen}=\{
\frac{z_4}{2}+\frac{z_5}{4},
\frac{z_5}{2}
\}
\end{array}$
&$\left(
\begin{array}{cc} 8&14 \\4&2 \\
\end{array} \right)$&1
&$\left(
\begin{array}{rr} -1&1\\2&-1\\
\end{array} \right)$
&$\left(\begin{array}{l}K_1\\Z\end{array}
\begin{array}{|rr|} -1&1\\1&0\\2&-1\\\end{array}
\begin{array}{l}E\\Z_b\\E\end{array}\right)$\\
\hline
$(33,2)^{2,132}_{-260}$&$\begin{array}{c}
\begin{array}{rrrrrr}
z_0&z_1&z_2&z_3&z_4&z_5\\
\hline
1&0&0&0&1&2\\
-1&1&1&1&-2&0
\end{array}\\[4pt]
\text{SRI}=\langle{z_0 z_4 z_5, z_1 z_2 z_3}\rangle\\
\text{KC gen}=\{
\frac{z_5}{2},
\frac{z_5}{4}-\frac{z_4}{2}
\}
\end{array}$
&$\left(
\begin{array}{cc} 2&0 \\6&14 \\
\end{array} \right)$&1
&$\left(
\begin{array}{rr} 0&1\\1&-2\\
\end{array} \right)$
&$\left(\begin{array}{l}K_1\\Z\end{array}
\begin{array}{|rr|} 0&1\\1&-1\\1&-2\\\end{array}
\begin{array}{l}E\\Z_b\\E\end{array}\right)$\\
\hline
$(34,1)^{2,132}_{-260}$&$\begin{array}{c}
\begin{array}{rrrrrr}
z_0&z_1&z_2&z_3&z_4&z_5\\
\hline
1&-2&-2&-4&7&0\\
0&1&1&2&-3&1
\end{array}\\[4pt]
\text{SRI}=\langle{z_0 z_4, z_1 z_2 z_3 z_5}\rangle\\
\text{KC gen}=\{
\frac{z_4}{7}+\frac{3 z_5}{7},
z_5
\}
\end{array}$
&$\left(
\begin{array}{cc} 49&114 \\21&9 \\
\end{array} \right)$&1
&$\left(
\begin{array}{rr} -2&1\\7&-3\\
\end{array} \right)$
&$\left(\begin{array}{l}K_1\\Z\end{array}
\begin{array}{|rr|} -2&1\\1&0\\7&-3\\\end{array}
\begin{array}{l}E\\Z_b\\E\end{array}\right)$\\
\hline
$\begin{array}{c}
(34,2)^{2,132}_{-260}
\end{array}$
&$\begin{array}{c}
\begin{array}{rrrrrr}
z_0&z_1&z_2&z_3&z_4&z_5\\
\hline
1&0&0&0&1&2\\
-1&2&2&4&-7&0
\end{array}\\[4pt]
\text{SRI}=\langle{z_0 z_4 z_5, z_1 z_2 z_3}\rangle\\
\text{KC gen}=\{
z_5,
\frac{z_5}{7}-\frac{2 z_4}{7}
\}
\end{array}$
&$\left(
\begin{array}{cc} 2&0 \\16&114 \\
\end{array} \right)$&1
&
&\\
\hline
$(35,1)^{2,144}_{-284}$&$\begin{array}{c}
\begin{array}{rrrrrr}
z_0&z_1&z_2&z_3&z_4&z_5\\
\hline
1&-2&-2&-2&5&0\\
0&1&1&1&-2&1
\end{array}\\[4pt]
\text{SRI}=\langle{z_0 z_4, z_1 z_2 z_3 z_5}\rangle\\
\text{KC gen}=\{
\frac{z_4}{5}+\frac{2z_5}{5},
z_5
\}
\end{array}$
&$\left(
\begin{array}{cc} 50&124 \\20&8 \\
\end{array} \right)$&1
&$\left(
\begin{array}{rr} -2&1\\5&-2\\
\end{array} \right)$
&$\left(\begin{array}{l}K_1\\Z\end{array}
\begin{array}{|rr|} -2&1\\1&0\\5&-2\\\end{array}
\begin{array}{l}E\\Z_b\\E\end{array}\right)$\\
\hline
$\begin{array}{c}
(35,2)^{2,144}_{-284}
\end{array}
$
&$\begin{array}{c}
\begin{array}{rrrrrr}
z_0&z_1&z_2&z_3&z_4&z_5\\
\hline
1&0&0&0&1&2\\
-1&2&2&2&-5&0
\end{array}\\[4pt]
\text{SRI}=\langle{z_0 z_4 z_5, z_1 z_2 z_3}\rangle\\
\text{KC gen}=\{
z_5,
\frac{z_5}{5}-\frac{2 z_4}{5}
\}
\end{array}$
&$\left(
\begin{array}{cc} 4&0 \\24&124 \\
\end{array} \right)$&1
&
&\\
\hline
$(36,1)^{2,272}_{-540}$&$\begin{array}{c}
\begin{array}{rrrrrr}
z_0&z_1&z_2&z_3&z_4&z_5\\
\hline
0&0&0&2&3&1\\
1&1&1&0&0&-3
\end{array}\\[4pt]
\text{SRI}=\langle{z_0 z_1 z_2, z_3 z_4 z_5}\rangle\\
\text{KC gen}=\{
\frac{z_4}{3},
\frac{z_4}{9}-\frac{z_5}{3}
\}
\end{array}$
&$\left(
\begin{array}{cc} 1&0 \\3&9 \\
\end{array} \right)$&1
&$\left(
\begin{array}{rr} 0&1\\1&-3\\
\end{array} \right)$
&$\left(\begin{array}{l}K_1\\Z\end{array}
\begin{array}{|rr|} 0&1\\1&0\\1&-3\\\end{array}
\begin{array}{l}E\\Z_b\\E\end{array}\right)$\\
\hline
\end{longtable}}

\providecommand{\href}[2]{#2}\begingroup\raggedright\endgroup

\end{document}